\documentclass[journal]{IEEEtran}
\usepackage{cite}%引用
\usepackage{amsmath,amssymb,amsfonts}%公式
\usepackage{algorithm}%算法浮动体环境
\usepackage{algorithmic}%算法书写
\usepackage{graphicx}%插图
\usepackage{subfigure}%插子图
\usepackage{subfig}
\usepackage{textcomp}
\usepackage{xcolor}
\usepackage{bm}
\usepackage{makecell}
\usepackage{balance}
\usepackage[hyphenbreaks]{breakurl}

\begin{document}
    \title{Towards a Systematic Survey for Carbon Neutral Data Centers}
	%\title{Towards a Systematic Survey for carbon-neutral data center}

\author{Zhiwei~Cao,
        Xin~Zhou,
        Han~Hu,
        Zhi~Wang, and
        Yonggang~Wen,~\IEEEmembership{Fellow,~IEEE}
        
\thanks{The first two authors contribute equally.}     
\thanks{Zhiwei Cao, Xin Zhou, Yonggang Wen are with the School of Computer Science and Engineering, Nanyang Technological University 639798, Singapore (e-mail: zhiwei003@e.ntu.edu.sg, zhouxin@ntu.edu.sg, ygwen@ntu.edu.sg).}% <-this % stops a space
\thanks{Han Hu is with the School of Information and Electronics, Beijing Institute of Technology, Beijing 100081, P.R.China (e-mail: hhu@bit.edu.cn)}% <-this % stops a space
\thanks{Zhi Wang is with Tsinghua Shenzhen International Graduate School, Tsinghua University, Beijing 100084, P.R.China (e-mail: wangzhi@sz.tsinghua.edu.cn)}
}
\markboth{Journal of \LaTeX\ Class Files,~Vol.~14, No.~8, August~2015}%
{Shell \MakeLowercase{\textit{\textit{\textit{et al}.}}}: Bare Demo of IEEEtran.cls for IEEE Communications Society Journals}


\maketitle

\begin{abstract}
Data centers are experiencing unprecedented growth as the fourth industrial revolution's supporting pillars and the engine for the future digitalized world. However, data centers are carbon-intensive enterprises due to their massive energy consumption, and it is estimated that data center industry will account for 8\% of global carbon emissions by 2030. Meanwhile, both technological and policy instruments for reducing or even neutralizing data center carbon emissions have not been thoroughly investigated, despite the fact that several global cloud providers including Google and Facebook, have pledged to achieve carbon neutrality in their hyperscale data centers. To bridge this gap, this survey paper proposes a roadmap towards carbon-neutral data centers that takes into account both policy instruments and technological methodologies. We begin by presenting the carbon footprint of data centers, as well as some insights into the major sources of carbon emissions. Following that, carbon neutrality plans for major global cloud providers are discussed to summarize current industrial efforts in this direction. In what follows, we introduce the carbon market as a policy instrument to explain how to offset data center carbon emissions in a cost-efficient manner. On the technological front, we propose achieving carbon-neutral data centers by increasing renewable energy penetration, improving energy efficiency, and boosting energy circulation simultaneously. A comprehensive review of existing technologies on these three topics is elaborated subsequently. Based on this, a multi-pronged approach towards carbon neutrality is envisioned and a digital twin-powered industrial artificial intelligence (AI) framework is proposed to make this solution a reality. Furthermore, three key scientific challenges for putting such a framework in place are discussed. Finally, several applications for this framework are presented to demonstrate its enormous potential.
\end{abstract}

\IEEEpeerreviewmaketitle
\section{Introduction}
\begin{figure}[t]
	\centering
	\includegraphics[width=.48\textwidth]{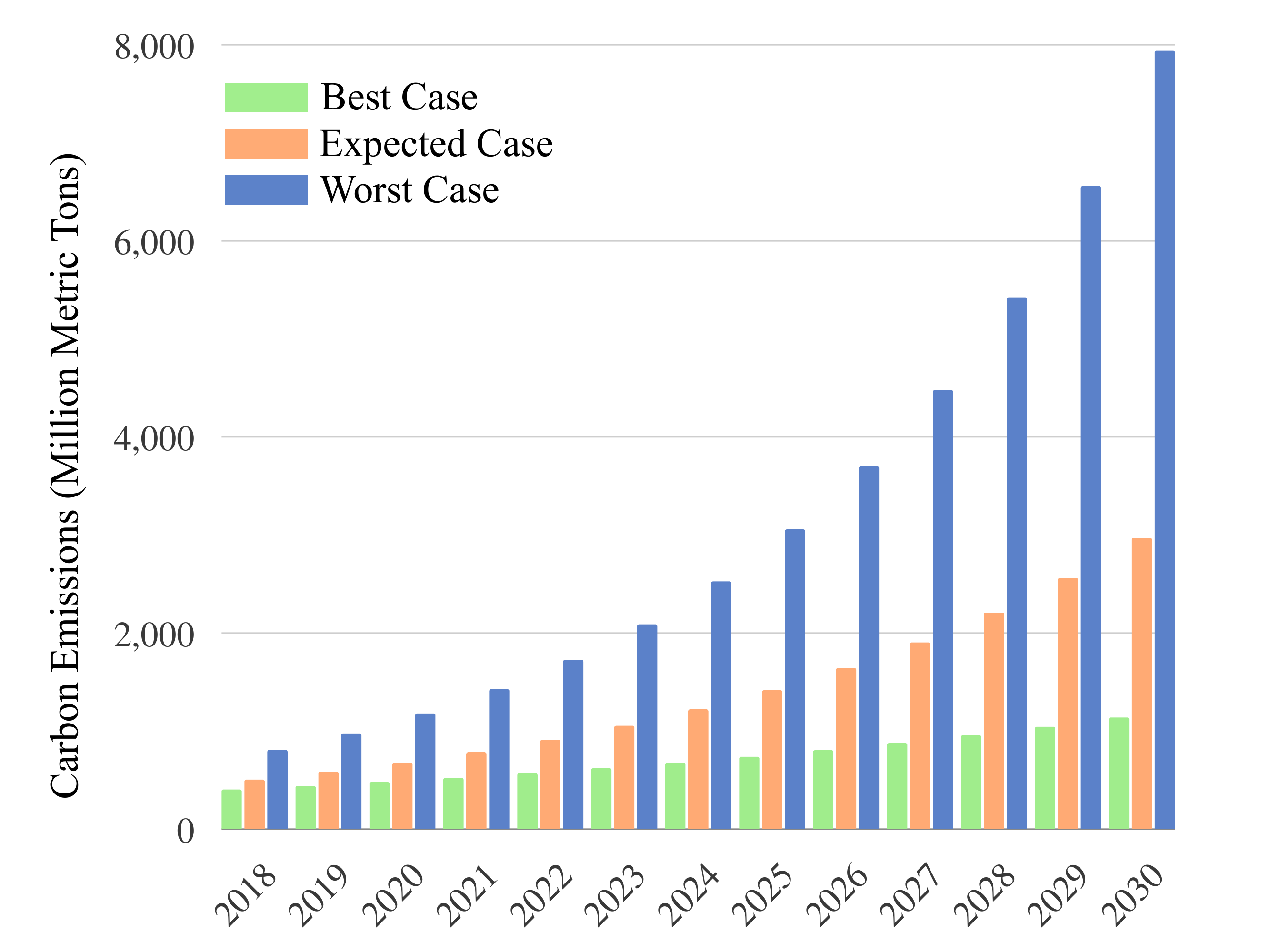}
	\caption{Estimated global carbon emissions from data centers from 2018 to 2030. A skyrocketing rising trend of global data center carbon emissions can be observed.}
	\label{GlobalCarbonEmissionsDC} 
\end{figure}

% \footnote{The first two authors contribute equally.}
\begin{table*}
    \centering
    \begin{tabular}{c!{\vrule width 1pt}c!{\vrule width 1pt}c!{\vrule width 1pt}c!{\vrule width 1pt}c}
         \noalign{\hrule height 1pt}
         Reference & Carbon Offset Mechanism & Energy Efficiency & Renewable Energy & Waste Heat Recovery \\
         \noalign{\hrule height 1pt}
         \cite{pore2015techniques} & \quad & \checkmark & \quad & \quad \\
         \hline
         \cite{zhang2016towards} & \quad & \checkmark & \quad & \quad \\
         \hline
         \cite{kong2014survey} & \checkmark & \quad & \checkmark & \quad \\
         \hline
         \cite{WasteHeatReview2014} & \quad & \quad & \quad & \checkmark \\
         \hline
         Ours & \checkmark & \checkmark & \checkmark & \checkmark \\
         \noalign{\hrule height 1pt}
    \end{tabular}
    \caption{Comparison with other survey papers.}
    \label{SurveyCompare}
\end{table*}
\IEEEPARstart{C}{arbon} emissions from cloud data centers are attracting global concerns from both governments and cloud service providers due to their skyrocketing energy demands to serve as an essential infrastructure of the global digital economy. As a noticeable massive electricity consumer, data centers consume 1.8\% of electricity in the United States \cite{USDCElectricity} and the share is expected to reach 20\% globally in 2025 as estimated by \cite{andrae2017total}. Noted that electricity generation is currently driven by carbon-intensive fossil fuels such as natural gas and coal, massive electricity consumption will result in huge carbon emissions. It is estimated that the data center industry currently contributes 0.3\% of global carbon emissions, and This growth trend will continue in the next decade \cite{DCNature}. As climate change caused by excessive carbon emissions is one of the gravest challenges the human race faces, data centers pose a major threat to the global climate change mitigation efforts. Fig. \ref{GlobalCarbonEmissionsDC} depicts the estimated global carbon emissions from data centers from 2018 to 2030\footnote{Carbon emissions are calculated by multiplying the carbon emission factor of the electrical grid with electricity consumption of data centers. Carbon emission factor of the grid can be found in https://www.iea.org/data-and-statistics. Estimation of global electricity consumption of data centers can be found in \cite{andrae2015global}.}, and an obvious escalation can be observed, indicating the urgent need to reduce data center carbon emissions. 
From a global perspective, the world has started to take steps to mitigate the environmental impact of data centers. One year after the European Green Deal adoption of the European Green Deal, leading cloud infrastructure providers and data center operators in Europe formed the Climate Neutral data center Pact \cite{EUPact}. In an unprecedented fashion, 25 companies and 17 associations have agreed to a self-regulatory initiative to make European data centers climate neutral by 2030, with ambitious measurable targets such as purchasing of carbon-free energy, water conservation, reuse and repair of servers, and heat recycling. The Chinese government has recently announced its carbon peak and carbon neutrality targets \cite{nogrady2021china}. In tandem with this, Tencent is accelerating its carbon neutrality plan, with the goal of achieving zero carbon emissions through the use of advanced Artificial Intelligence (AI) technologies. Chindata Group, a leading carrier-neutral hyperscale data center solution provider in China, has also released its roadmap to achieve carbon neutrality for all of its next-generation hyperscale data centers in China by 2030, using a 100\% renewable energy solution. The data center hyperscalers are taking even more aggressive stances in the United States~\cite{COMPUTERWORLD}. Google, for instance, has pledged to run all of their global data centres and corporate campuses on 100\% carbon-free power by 2030 \cite{GoogleReport2020}. Furthermore, Microsoft has even declared to be carbon-negative by 2030 and remove all historical carbon footprint by 2050 \cite{Microsoft}. 

To realize carbon-neutral data centers, there are two complementary ways. The first way is to reduce data center carbon emissions by improving data center energy efficiency or increasing renewable energy penetration. The other way is to balance carbon emissions through some policy instruments such as the carbon offset mechanism which reduces Green House Gas (GHG) emissions or removes carbon dioxide from the atmosphere to compensate for emissions elsewhere. In this regard, a series of surveys have been summarized on data center energy efficiency optimization \cite{pore2015techniques}\cite{zhang2016towards}, renewable energy integrated data center management \cite{kong2014survey} and data center low-grade waste heat recovery \cite{WasteHeatReview2014}. In \cite{pore2015techniques}, Pore \textit{et al.} provided tutorials on how to achieve energy-proportional data centers by managing IT devices at multiple scales. Furthermore, the environment of a data hall is complicated due to the underlying dynamics including time-varying workload and outdoor environment, complex cooling air flow, and so on. In \cite{zhang2016towards}, a comprehensive survey on joint optimization of the IT and cooling system of a data center was conducted, where the enabling techniques, modeling issues, related optimization problems and testbeds for green data centers were discussed. In terms of the management of renewable energy integrated data centers, Kong \textit{et al.} conducted an insightful survey on how to improve renewable energy utilization and several techniques as well as basic carbon market mechanisms were introduced \cite{kong2014survey}. In addition, as an new technique for reducing carbon emissions, low-quality waste heat recovery systems for data centers were reviewed in \cite{WasteHeatReview2014} where several potential data center waste heat applications were discussed in terms of technical feasibility, economic applicability, and thermal efficiency. However, none of them presented a 
comprehensive framework for reducing data center carbon emissions. Moreover, the carbon credit mechanism are ignored by the existing surveys, despite the fact that it is a critical approach for accelerating decarbonization of data centers. Table \ref{SurveyCompare} summarizes the differences between this paper and other survey papers on data center management.

This survey paper summarizes the existing literature on carbon markets as a policy instrument, data center energy efficiency improvement, data center renewable energy management, and data center waste heat recycling to present a comprehensive view of carbon-neutral data centers. We first present the carbon footprint of a data center from both the lifecycle and daily operations perspectives to shed some light on various carbon emission sources in a data center as well as their relative importance. Second, we give the definition of carbon-neutral data centers and some metrics for evaluating the carbon efficiency of a data center. In addtion, industrial efforts from global cloud providers towards carbon-neutral data centers are presented and various approaches are summarized. Following that, we introduce the carbon market as an policy instrument to achieve carbon neutrality in a cost-efficient way. Subsequently, we summarize works on the optimization of renewable energy integrated data centers based on various optimization goals, and common renewable energy options are discussed as well. We then present the works on improving data center energy efficiency from both the IT and cooling system perspectives. We also provide a brief overview of existing applications for data center low-grade waste heat recycling which reduces carbon emissions of either a data center or its waste heat consumers. In addition, we envision a digital twin-assisted industrial AI framework for carbon-neutral data centers in which the internet of digital twins solves data scarcity issues and AI serves as a powerful instrument for implementing intelligent management. With such a framework in place, we hope to shed some light on several technological trends towards carbon-neutral data centers. 

\begin{figure*}[t]
	\centering
	\includegraphics[width=.95\textwidth]{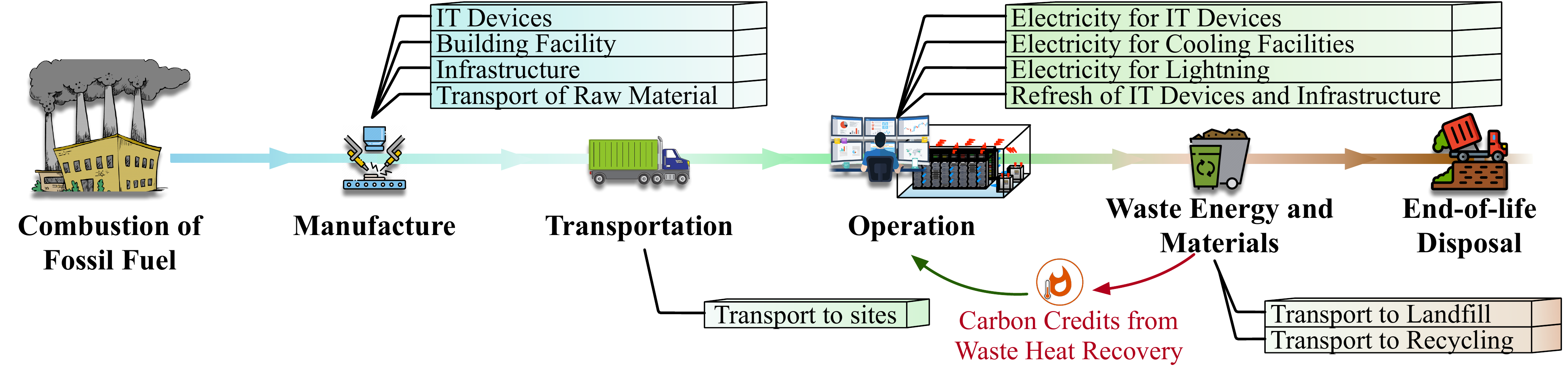}
	\caption{Life cycle carbon flow of a data center. The source of all carbon emissions is the combustion of fossil fuels. Emboddied carbon emissions include those resulted from the manufacturing of IT and non-IT components, assembling of a data center, transportation and waste disposal. Operational carbon emissions cover those incured by electricity consumption and refreshment of IT devices and infrastructure.} 
	\label{CoarseCarbonFlow} 
\end{figure*}

The rest of the paper is organized as follows. First, we present the carbon footprint of data centers in Section \ref{Footprint}. Second, definition of carbon-neutral data centers as well as some metrics to evaluate data center carbon efficiency are provided in Section \ref{CarbonNeutrailityPlans}. Furthermore, industrial efforts from data center hyperscalers and multiple approaches towards this goal are also discussed. Third, the carbon market is introduced in Section \ref{Carbon Market}. Subsequently, we provide a review of optimization problems related to renewable energy integrated data centers in Section \ref{Energy Supply}. Following that, enabling techniques and optimization approaches for improving data center energy efficiency are presented in Section \ref{Energy Utilization}. We then discuss some potential applications for low-grade data center waste heat in Section \ref{Energy Circulation}. In Section \ref{DT}, we envision a digital twin-assist industrial AI framework for carbon-neutral data centers. Finally, we summarize this paper in Section \ref{Summary}.

\section{Data Center Carbon Footprint}\label{Footprint}
As the first step towards carbon neutrality, data center operators should be aware of their data centers' carbon footprint throughout their life cycle. In this section, we first present the high-level carbon flow graph, with the goal of gaining a comprehensive understanding of the lifecycle carbon footprint. The three types of carbon emissions are then summarized in the following. Finally, current metrics for benchmarking the carbon efficiency of a data center are discussed. 

\subsection{Data Center Lifecycle Carbon Flow}
The relationship between the data center industry and carbon emissions appears to be counterintuitive, as the data center does not consume a large amount of carbon-intensive fossil fuel for electricity generation and domestic heating on its own. However, the data center sector does emit over 0.3\% equivalent $\text{CO}_2$ per year globally \cite{DCNature}. In this section, we reveal the carbon footprint during the entire lifecycle of a data center to gain insights.

In general, the carbon footprint of a data center can be coarsely categorized into three classes: carbon footprint from manufacture of the data center and its purchased equipment, carbon footprint from daily operations, and carbon footprint of end-of-life disposal of equipment and decommissioning of the data center. Once the data center is established, the carbon footprint from manufacturing can be considered constant, whereas that from daily operations is cumulative and dynamic since it is primarily derived from electricity drawn from the electrical grid. Therefore, in this section, we begin with a coarse carbon flow graph focusing on the integrity of the lifecycle carbon footprint. Subsequently, a fine-grained daily operational carbon flow will be illustrated, facilitating the understanding of the carbon flow within the data center enclosure in daily operations. 

Fig. \ref{CoarseCarbonFlow} depicts the high level carbon flow in a data center over its lifecycle. The combustion of fossil fuels for the production of electricity, which emits a large amount of $\text{CO}_2$, is the driving force for the carbon flow. The electricity is leveraged to extract useful materials from metal ores and fossils, manufacture all necessary equipment and infrastructure, power IT devices and infrastructure, and dispose of obsolete equipment. Furthermore, in case of unexpected power outages, many data centers are equipped with diesel generators to provide emergency power, which emits a certain amount of $\text{CO}_2$. In addition, carbon emissions will also occur in transportation of equipment to ensemble a data center or to landfill. Carbon emissions from material extraction and manufacturing are referred to as embodied emissions or embedded emissions, while emissions from daily operations are referred to as operational emissions. It is worth noting that because some devices can be recycled and reused, the corresponding carbon emissions should be deducted from the lifecycle carbon footprint. 

\begin{figure}[t]
	\centering
	\includegraphics[width=.46\textwidth]{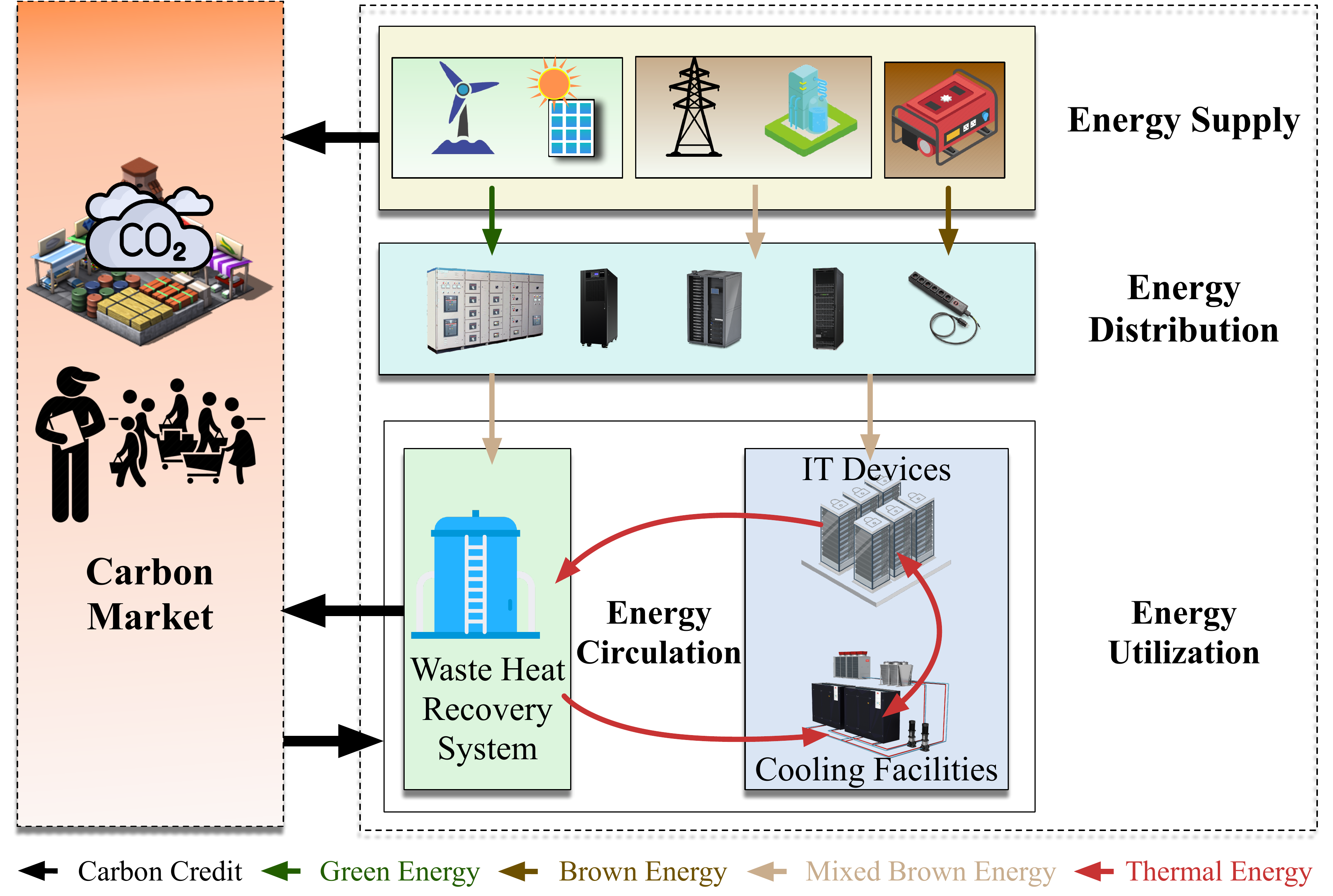}
	\caption{A data center's operational carbon flow. Electricity supply to a data center is a mix of green and brown electricity. The electricity is used for powering the IT, cooling and waste recovery system. A data center can also sell its surplus renewable energy to obtain carbon credits or purchase carbon credits from the carbon market.} 
	\label{Fine-grainedCarbonFlow} 
\end{figure}

As data center operators typically focus on optimizing their daily operations in order to reduce the environmental impact of their data centers, a deeper understanding of the carbon flow in the day-to-day operation of a data center is indispensable. Fig. \ref{Fine-grainedCarbonFlow} depicts the operational carbon flow of a data center.

According to Fig. \ref{Fine-grainedCarbonFlow}, a data center has four subsystems: IT system, cooling system, power distribution system, and waste heat recovery system. Among these four subsystems, the waste heat recovery system is gaining popularity because it can partially utilize the waste heat dissipated by servers to significantly reduce energy consumption and corresponding carbon emissions \cite{WasteHeatReview2014}. A traditional data center power distribution system consists of the substation connected to the electrical grid, the Uninterruptible Power Supply (UPS) and the diesel generator to ensure $24\times7$ operation. As more cloud providers claim to reduce their carbon footprint, the on-site renewable energy generator has emerged as a viable option for powering the data center due to its near-zero carbon footprint \cite{Google,Amazon,Facebook}. Due to the intermittent nature of renewable energy sources, an increasing number of data centers are integrating with energy storage systems to store excess energy and discharge it when an energy shortage occurs. The IT, cooling, and waste heat recovery subsystems are powered by the electricity supplied by the power delivery subsystem, which is often a mixture of the renewable and the nonrenewable sources. Renewable and the nonrenewable electricity are also referred to as green and brown electricity respectively. Green electricity incurs nearly zero carbon emission, whereas brown electricity is the major source of carbon emissions during the operation phase. It is also worth noting that the waste heat recovery system is able to reuse the waste heat for domestic heating\cite{oro2019DomesticHeat}, driving absorption chillers \cite{haywood2012AbsorptionChiller}, desalinating salty seawater \cite{sondur2018Deselination}, and even providing heating and drying for a biomass generator to produce renewable electricity \cite{sharma2010BiomassWH}. The use of waste heat allows data center operators to reduce either the carbon footprint of the data center or their customers, and to bring back carbon credits that can be used to offset their carbon emissions. In addition, with the on-site renewable energy generator, the data center operator can sell excess renewable energy back to the electrical grid via the net metering technology \cite{NetMetering}, resulting in a reduction on the net electricity consumption from the electrical grid, as well as its carbon emissions.

\subsection{Categorical Evaluation of Data Center Carbon Emissions}
From the data center operator's stand point of view, it is critical to categorize carbon emissions in the lifecycle of a data center to different types and identify the importance of different kinds of carbon emissions in order to reduce carbon emission in a cost-effective manner. In this section, we first present current adopted carbon emission classification approach. Following that, data centers carbon emissions will be categorized according to the method and the share of different carbon emission types will be presented. 

According to the United States Environmental Protection Agency (EPA) \cite{EPA}, three scopes of carbon emissions of a business entity should be considered and tracked to evaluate its carbon footprint:
\begin{itemize}
    \item \textbf{SCOPE1 Emission} refers to the \textit{direct} emission from the combustion of fossil fuels, company vehicles and any other fugitive activities such as the use of diesel generators.  
    \item \textbf{SCOPE2 Emission} is the \textit{indirect} emission from the purchase of electricity, heat, or steam from local utilities. For SCOPE2 Emission, there are two accounting methods: location-based and market-based. The location-based method is based on the carbon emission and electricity production data and averaged within a geographical boundary during a predefined time period. It is applicable everywhere and is related to the level of decarbonization of the local grid. The market-based approach refers to the carbon emission associated with the electricity or heat supplier. Different suppliers have different carbon emission factors which may differ significantly from that of the considered region. 
    \item \textbf{SCOPE3 Emission} covers a variety of other \textit{indirect} emissions which are not covered by SCOPE2 Emission, e.g., the electricity transmission and distribution loss that are not covered in SCOPE2 Emission.   
\end{itemize}
\begin{figure}[t]
	\centering
	\includegraphics[width=.48\textwidth]{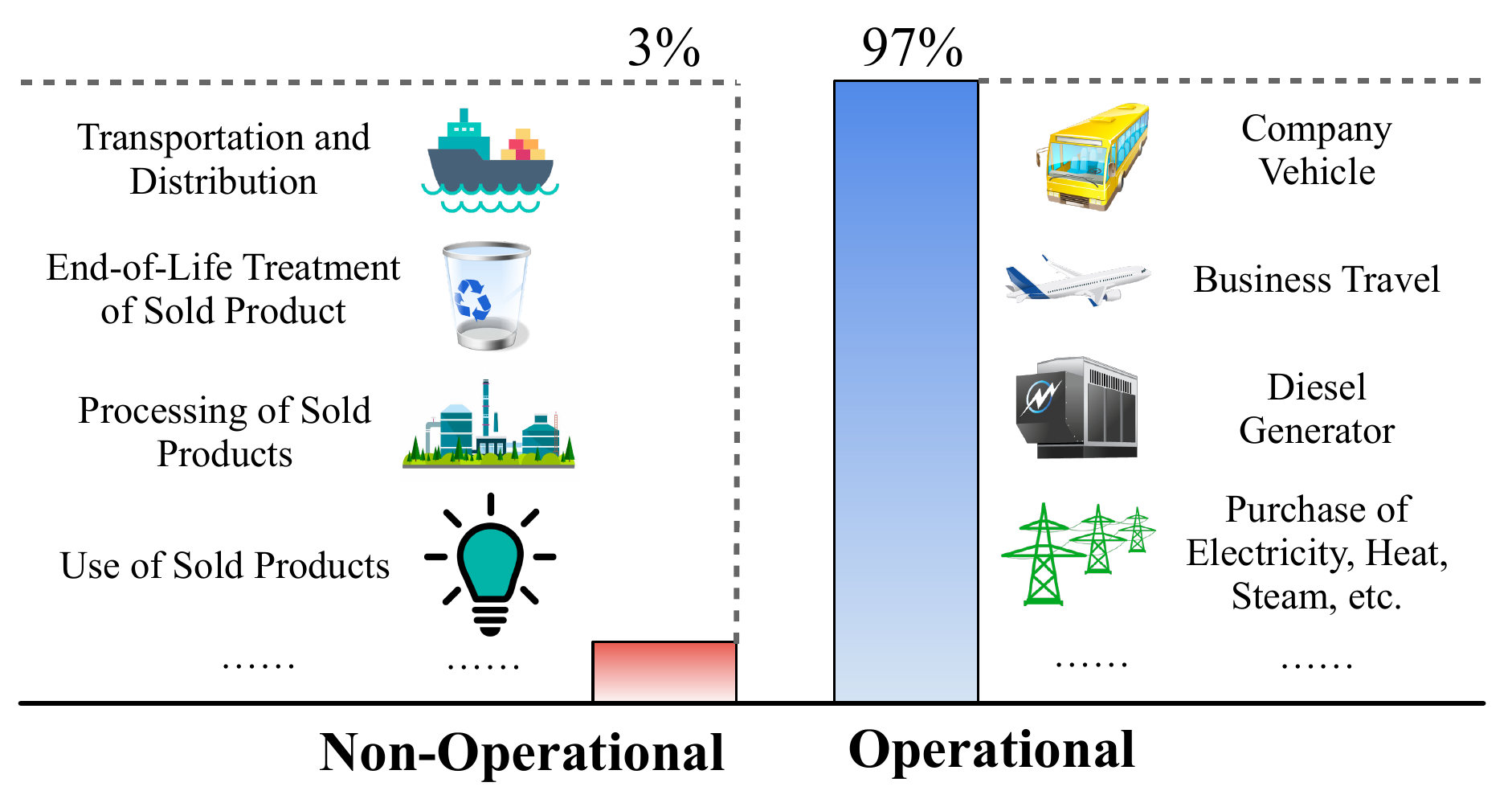}
	\caption{Share of operational (SCOPE1 + SCOPE2 Emission) and non-operational (SCOPE3 Emission) carbon emissions of a data center according to the LCA result from \cite{LCAWhitehead2015}. Carbon emissions in daily operations significantly exceed emboddied carbon emissions.} 
	\label{CarbonFootprintShare} 
\end{figure}
\begin{table*}
    \centering
    \begin{tabular}{c!{\vrule width 1pt}c}
         \noalign{\hrule height 1pt}
         Metric & Parameters \\
         \noalign{\hrule height 1pt}
         \text{CUE} & \makecell[l]{(1) $\beta$:  carbon emission factor of the electrical grid;\\(2) Power Usage Effectiveness (PUE): ratio between total energy consumption and the energy consumed by IT devices}\\
         \hline
         CFE Score (\%) &  \makecell[l]{(1) $CFE_{contracted}$: carbon-free energy from energy contracts (PPAs, RECs);\\(2) $CFE_{grid}$: carbon-free energy from the electrical grid;\\(3) $E_{total}$: data center total energy consumption;\\(4) $\gamma$: renewable energy penetration rate of the electrical grid;\\(5) $CFE_{total}$: total carbon-free energy yields of the contracted projects}\\
         \hline
         Avoid Emission ($e_{avoid}$) & \makecell[l]{(1) $CFE_{contracted}$: carbon-free energy from energy contracts (PPAs, RECs);\\(2) $\beta$:  carbon emission factor of the electrical grid}\\
         \noalign{\hrule height 1pt}
    \end{tabular}
    \caption{Metrics and key parameters for evaluating carbon efficiency of a data center.}
    \label{CarbonMetrics}
\end{table*}
In the context of data centers, carbon emissions can also be classified into these three scopes and their relative importance can be analysed. 
For SCOPE1 Emissions, diesel generators are often integrated in the data center microgrid, and corresponding carbon emissions due to diesel combustion are classified as SCOPE1 Emissions. As for SCOPE2 Emissions, the carbon emissions relating to the purchased electricity from the local grid are regarded as SCOPE2 Emissions. Moreover, data center operators may purchase heat from local heat utilities to provide domestic heating in the heating months, resulting in carbon emissions from heat production. In comparison to SCOPE1 and SCOPE2 emissions, SCOPE3 emissions are much more difficult to account for because they include carbon emissions from the entire supply chain, except for the electricity and heating supply. For example, carbon emissions embodied in the purchased servers and the construction of data centers should both be included in SCOPE3 Emissions. To accurately calculate such carbon emissions, sophisticated Life Cycle Assessment (LCA) of a data center must be adopted \cite{LCAWhitehead2015}, which is extremely time and cost-consuming. Due to the opacity of manufactural and operational information of a data center, only a few work has been conducted in the LCA of a data center \cite{LCAWhitehead2015,LCAShah2012sources,LCASweden,TotalGreen}. According to the reported results, the majority of the authors discovered that carbon emissions from daily operations of a data center accounts for the majority share of total carbon emissions, with the exception of a data center in Sweden where renewable penetration in the national electrical grid is high \cite{LCASweden}. Fig. \ref{CarbonFootprintShare} illustrates the proportion of different scopes of emission in a data center in the United Kingdom \cite{LCAWhitehead2015}. 

According to Fig. \ref{CarbonFootprintShare}, more efforts should be made in the data center industry to reduce and offset SCOPE2 emissions in order to achieve carbon neutrality more efficiently. 

\section{Carbon Neutrality of Data Centers}\label{CarbonNeutrailityPlans}
As we gain deeper understanding of the data center carbon footprint, a couple of natural questions raise: a) what is the definition of carbon-neutral data centers, b) how to evaluate the carbon neutrality status of a data center, c) what is the current industrial efforts towards it, and d) what are the approaches in general to achieve it? In this section, we answer these questions by first providing the definition of carbon neutrality. In the following, some metrics for evaluating data center carbon efficiency are presented. We then discuss the plans and efforts of global cloud providers towards carbon-neutral data centers to shed some light on the global progress on this area. Finally, general approaches for realizing carbon-neutral data centers are summarized based on some industrial consensus. 

\subsection{Carbon Neutrality Definition}
By definition, carbon neutrality means that carbon emissions from an entity throughout its lifecycle are \textit{completely offset} by the carbon emissions it removes from the atmosphere. From the technological stand point of view, carbon neutrality technologies can be categorized into two classes, i.e., carbon removal technologies and carbon-neutral technologies. Carbon removal technologies aim to remove existing carbon dioxide from the atmosphere. Typical carbon removal technologies include BioEnergy with Carbon Capture and Storage (BECCS) \cite{BECCS}, direct air capture \cite{keith2018process} and so on. Carbon-neutral technologies focus on zero carbon emissions by replacing carbon-intensive fossil fuels with carbon-free renewable energy, improving energy efficiency etc. 

In the context of data centers, we restrict the scope of carbon neutrality in this survey by only discussing the approaches to realize zero carbon emissions for a data center with carbon-neutral technologies because data centers are unlikely to participate in direct carbon capture. As a carbon emitter, a data center can reduce its carbon emissions by improving energy efficiency and replacing carbon-intensive energy with renewable energy. On the other hand, a data center can also participate in carbon offsetting programs by selling its waste heat or excess renewable energy to offset other entities' carbon emissions and obtain corresponding carbon credits to offset its own carbon emissions. Furthermore, carbon credits are available from the carbon markets to offset data center carbon emissions in a cost-efficient manner. The multi-facets property of the goal towards carbon neutrality calls for a multi-pronged solution which considers carbon emission reduction and carbon offsetting simultaneously. 

\subsection{Carbon Neutrality Metrics}
When it comes to benchmarking the carbon efficiency of a data center, the first step is to design a proper metric, reflecting the total carbon footprint of a data center. In the past decade, various metrics have been proposed to evaluate the carbon efficiency of a data center and they are summarized in Table \ref{CarbonMetrics}. 

\subsubsection{Carbon Usage Effectiveness}
To the best of our knowledge, the metric called Carbon Usage Effectiveness (CUE) proposed in the Green Grid White Paper \cite{CUE} is the first widely adopted metric in this area. CUE is defined in the following way:
\begin{align}\label{CUE}
\begin{split}
    \text{CUE} &= \beta\cdot\frac{E_{total}}{E_{IT}}\\
    & = \beta \cdot \text{PUE},
\end{split}
\end{align}
where $\beta$ is the carbon emission factor (kg$\text{CO}_2$eq/kWh) of the electrical grid, which represents the amount of carbon emissions when producing 1 kWh electricity and it is reported by either the government authorities like EPA or non-governmental organizations-based reports such as IEA reports \cite{IEAData}. $E_{total}$ and $E_{IT}$ are the energy consumption of a data center and the IT devices within it respectively, and their ratio is Power Usage Effectiveness (PUE), which is widely used in the data center industry to evaluate the energy efficiency of a data center. The minimum value of CUE is 0 and it is not upper bounded. It should also be noted that when on-site renewable energy is integrated in the power supply system of a data center, the electricity from renewable sources such as wind or solar should be subtracted from total electricity consumption in CUE calculation. From the standpoint of a data center operator, CUE shall be as small as possible to reduce its impacts on global warming.
\begin{table*}
    \centering
    \begin{tabular}{c!{\vrule width 1pt}c}
         \noalign{\hrule height 1pt}
         Company & Carbon Neutrality Plan \\
         \noalign{\hrule height 1pt}
         Google \cite{Google} & $24\times7$ carbon free operation in 2030 \\
         \hline
         Facebook \cite{Facebook} & Achieve carbon neutrality (SCOPE1 and SCOPE2)  in  2020  and  realize  SCOPE3  net  zero  emission in  2030\\
         \hline
         Microsoft \cite{Microsoft} & Become carbon negative in 2030 and remove its historical carbon footprint by 2050\\
         \hline
         Ant Group \cite{AntGroup} &  Achieve  zero  direct  and indirect  carbon  emissions  in  2020  and  canceling  out  carbon emissions  from  its  supply  chain  in  2030\\
         \hline
         OVHCloud \cite{OVHCloud2} & 100 \% renewable energy supply and 0\% waste to landfill by 2025 \\
         \noalign{\hrule height 1pt}
    \end{tabular}
    \caption{Carbon neutrality plans of selected high-tech icons. “Zero-carbon” and “carbon-free” can be used interchangeably. “Carbon neutral” is a relaxation of “zero-carbon” and “carbon-free” because it allows a data center to have a certain amount of carbon emissions which can be offset by purchased carbon credits or carbon emission permits.  “Carbon negative” is close to “carbon neutral” but it requires the data center to eliminate more carbon emissions than it emits. “100\% renewable energy supply” is close to “zero-carbon” and “carbon-free” since the major carbon emissions from a data center come from carbon-intensive electricity from the grid.}
    \label{CNPlan}
\end{table*}
\subsubsection{Carbon Free Energy Score}
Despite CUE captures the carbon emission incurred by electricity consumption, it is becoming less representative for the sustainability and greenness of a data center since more and more data center operators resort to off-site renewable energy like Power Purchase Agreement (PPA) or Renewable Energy Certificate (REC) for offsetting their overwhelmingly large amount of electricity consumption. Such off-site renewable energy is not included in CUE calculation, leading to unfair evaluation. 

Recently, in accompany with their ambitious goal of achieving round-the-clock operation with carbon free energy, Google has proposed the Carbon Free Energy (CFE) Score \cite{Google} for evaluating the degree to which the energy consumption of a data center is matched by the CFE on an hourly basis. The CFE Score is defined in the following way:
\begin{align}
    \text{CFE Score} \ (\%) = \frac{CFE_{contracted} + CFE_{grid}}{E_{total}},
\end{align}
where $CFE_{contracted}$ is the CFE from the contracted supplier via PPA or REC. $CFE_{grid}$ is the CFE from the electrical grid. $E_{total}$ is the total load of a data center. 

$CFE_{contracted}$ is the minimum between total load ($E_{total}$) and the actual delivered CFE from the contracted supplier ($CFE_{total}$):
\begin{align}
    CFE_{contracted} = \min\{E_{total}, \ CFE_{total}\}.
\end{align}

Because renewable energy is penetrating the electrical grid, a portion of the purchased electricity from the grid is produced by carbon free sources. Therefore, $CFE_{grid}$ should be calculated as follows:
\begin{align}
    CFE_{grid} = (E_{total} - CFE_{contracted})\cdot \gamma,
\end{align}
where $\gamma$ is the renewable energy penetration rate of the local electrical grid. 

Compared to CUE, CFE Score is a more comprehensive metric since it takes into account both the contracted CFE and the CFE from the electrical grid. It is also worth noting that CFE Score cannot exceed 100\% since the CFE that exceeds the total load is not counted. To achieve $24\times7$ carbon free operation, the CFE Score should be 100\% in every hour. 

\subsubsection{Avoided Emission}
Along with the CFE Score, another metric called avoided emission is proposed in Google's recent white paper\cite{Google} to prioritize different renewable energy projects. The reason for introducing this metric is that CFE Score may conceal significant variation in carbon intensity of the electrical grid as well as the source of grid emissions. For example, assuming that there are two data centers with the same total load of 1 MWh in two different locations, A and B, and both data centers have contracted 0.8 MWh CFE with local renewable energy projects. Supposing that the grid CFE ratio $\gamma$ for both sites is identical, say 30\%, the CFE Score for these two data centers will be identical (86\%). However, if the nonrenewable electricity is generated by coal in site A and natural gas in site B, the actual carbon emissions for the data center in site A will be much higher than those in site B since coal is more carbon-intensive than natural gas. Therefore, it is necessary to introduce an emission-related metric for comparing carbon neutrality status of different data centers. 

Avoided Emission ($e_{avoided}$) is the difference between the carbon emission without and with contracted CFE:
\begin{align}\label{AvoidedEmission}
\begin{split}
    e_{avoided} &= E_{total}\cdot\beta - (E_{total} - CFE_{contracted})\cdot\beta\\
    &= CFE_{contracted}\cdot\beta,
\end{split}
\end{align}
where $\beta$ represents the carbon intensity of the electrical grid. From Eq. (\ref{AvoidedEmission}), one insight can be gained: if the carbon intensity $\beta$ of a region is high, the data center operator should invest more in the renewable project in the region and sign CFE contract with the project in order to increase its Avoided Emission. Such investment will also provide a strong incentive for the decarbonization of the local electrical grid, which has been pointed out by Google as a key strategy for achieving $24\times7$ carbon neutrality in the future \cite{Google}.

\subsection{Industrial Efforts Towards Carbon-Neutral Data Centers}
As global warming is accelerating in recent years, many cloud providers as well as data center operators have announced ambitious plans for neutralizing their environmental impacts, even offsetting historical environmental impacts in the next several decades \cite{Microsoft}. Table \ref{CNPlan} summarizes the carbon neutrality plans of several celebrated cloud providers. The technological focus for achieving carbon-neutral data centers of different cloud providers will be discussed in the rest of this section. 

\subsubsection{Efforts from Cloud Providers in the United States}
The United States is home to a large number of high-tech companies such as Google, Facebook, Amazon, Microsoft, and Apple, which consume more than 1.8\% of total annual electricity in the United States \cite{USDCElectricity}. It is reported that data center operations consume a significant portion of ICT companies' total electricity consumption. Therefore, pressed by the rising electricity bills and an ever-increasing environmental crisis, some leading companies including Google and Facebook have published their plans and roadmap towards carbon-neutral data centers. 

As a global leader in green energy-powered operations, Google has achieved carbon neutrality as early as 2007 \cite{GoogleReport2020} and acted as a pioneer in sustainable development. Furthermore, the electricity consumption of Google has been matched via the purchased renewable energy since 2017. According to Google's opinion, the following aspects should be prioritized over the next decade to achieve round-the-clock carbon neutrality:
\begin{itemize}
    \item \textbf{Decarbonization of the entire electrical grid} is urgent and essential for achieving carbon neutrality. On the one hand, decarbonization of the electrical grid will increase the availability of renewable energy everywhere, relieving data center operators of the burden of investing in on-site renewable energy generators or renewable energy projects. On the other hand, it will also reduce carbon emissions from data center supply chains, as the manufacture of ICT equipment may consume a large amount of electricity, leading to huge embodied carbon emission for a data center. Owing to the decarbonization of the electrical grid, such embodied emissions will be significantly reduced.   
    \item \textbf{Energy portfolio optimization} is another key enabling approach. Due to the intermittent and unstable nature of popular renewable energy sources such as wind and solar, it is unlikely to rely on one kind of renewable energy to operate a data center. Therefore, efficient capacity planning and energy portfolio optimization shall be conducted in order to reduce the carbon footprint while ensuring the continuous operation of a data center.
    \item \textbf{Carbon-aware computing platform} will be a critical aid on the path towards carbon neutrality. The design philosophy of such a system is that non-critical jobs can be scheduled both spatially and temporally to improve the utilization of renewable energy. 
\end{itemize}

As another pioneer in sustainable development, Microsoft has announced a new goal of becoming carbon negative by 2030 and removing its historical carbon footprint by 2050 \cite{Microsoft}. By 2030, Microsoft and its value chain will have eliminated more than half of their carbon emissions. Achieving such a goal is nontrivial and Microsoft has planned to conduct some actions to fulfill its commitment. 
\begin{itemize}
    \item \textbf{Improve renewable energy penetration}, where the electricity used in daily operations will be fully matched with renewable energy in 2025.
    \item \textbf{Charge internal carbon fee} from both the daily operations activities and the entire value chain to fund the innovative technology in carbon removal and energy efficiency improvement.
    \item \textbf{Emphasize carbon removal} rather than carbon neutrality, where Microsoft has established a 1 billion dollars climate innovation fund for developing negative emission technology (NET) such as afforestation and reforestation, soil carbon sequestration, and other methods to remove more carbon than it emits, with the goal of completely eliminating its historical carbon footprint by 2050.
    \item \textbf{Strive for low-carbon solutions} for daily operations. For example, Microsoft is collaborating with Vattenfall to create a first-of-its-kind $24\times7$ renewable energy matching solution, allowing its customers to choose the best renewable energy source for them.  
\end{itemize}

Other high-tech icons, e.g., Facebook and Amazon, have also launched their carbon neutrality goal recently. Facebook claims to achieve carbon neutrality (SCOPE 1 and 2 Emission) in 2020 and realize net zero emissions (SCOPE1, 2 and 3 Emission) in 2030 \cite{Facebook}. To accomplish this, Facebook has constructed over 5,400 MW of new solar and wind power plants globally, equivalent to reducing its GHG emissions by more than 2.6 million metric tons. In addition, Facebook is supporting novel carbon removal solutions including nature-based emission avoidance projects like reforestation and regenerative agriculture to achieve net zero emissions. Moreover, another novel load balancing scheduler called Autoscale \cite{Autoscale} has been developed by Facebook to improve the energy efficiency of its data centers via increasing server utilization and shutting down idle servers. 

Amazon has also made significant efforts to reduce its carbon footprint in recent years. In December 2020, Amazon announced plans to add 26 utility-scale wind and solar energy projects, bringing its total renewable energy investment in 2020 to 35 projects and more than 4 GW of capacity \cite{Amazon}. To improve the energy efficiency of the AWS infrastructure, an intelligent direct evaporative cooling system is leveraged to fully utilize outside air to remove the heat generated by the servers in both cooler and hotter months, resulting in a considerable PUE improvement.    
\begin{figure*}[t]
	\centering
	\includegraphics[width=1.\textwidth]{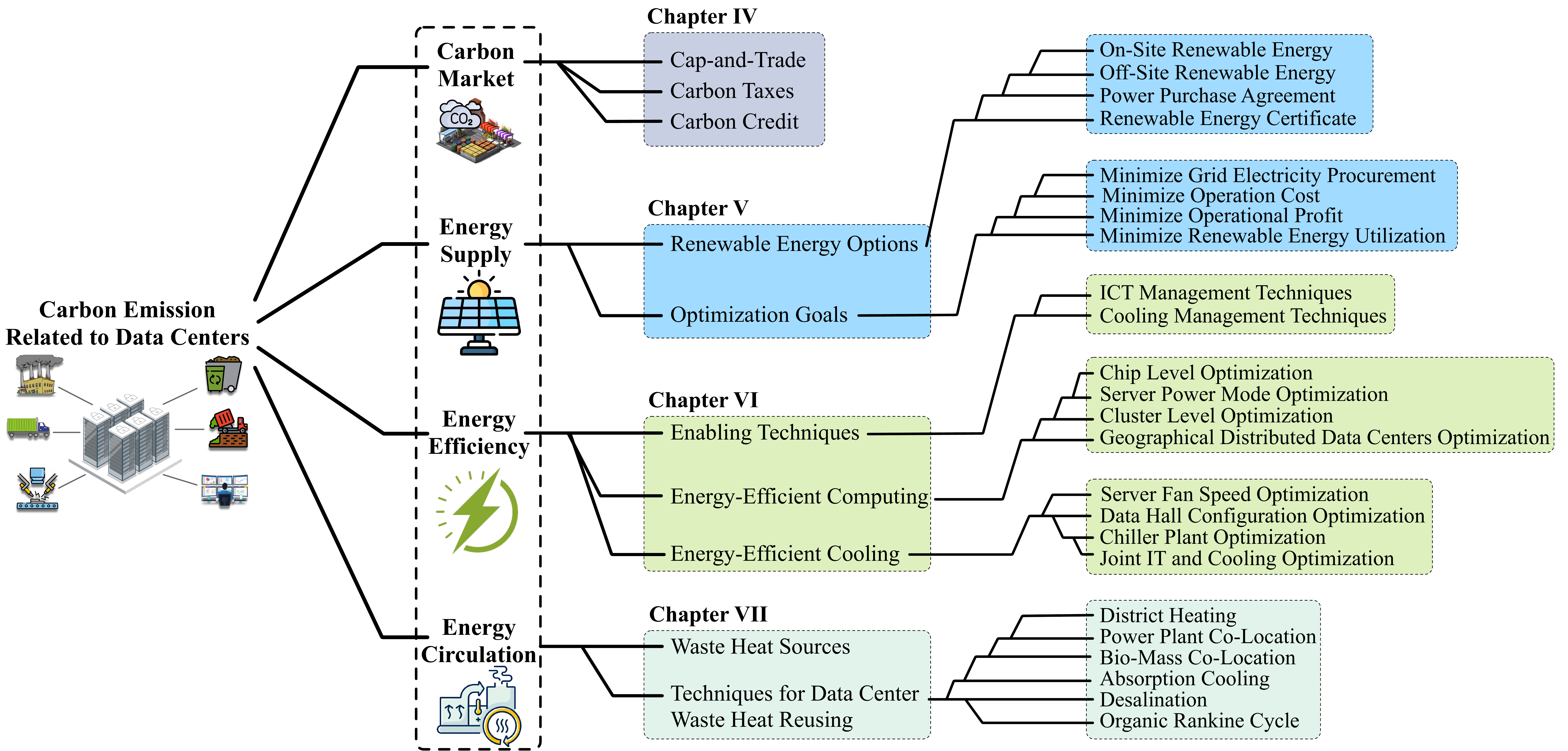}
	\caption{Approaches towards carbon-neutral data centers. The approaches can be categorized into policy instruments (carbon market mechanisms) and technological methods. From the technological standpoint, a comprehensive consideration of energy supply, energy utilization and energy circulation is critical to realize carbon-neutral data centers.}
	\label{Approches} 
\end{figure*}
\subsubsection{Efforts from Cloud Providers in Asia}
As another leading force in technology innovation, driven by both the policy and environmental pressure, many Asian ICT companies have also published their plans for reducing carbon footprint. For example, Ant Group, a celebrated internet finance company, has set the goal of  achieving zero direct and indirect carbon emissions by 2020 and canceling out carbon emissions from its supply chain by 2030 \cite{AntGroup}. In addition, Huawei has spent over a decade working to improve its sustainability. In its recent white paper \cite{HuaweiReport}, it envisioned the energy systems ``using bits to manage watts and using data flows to optimize energy flows" to achieve national-wide carbon neutrality. As a pioneer in sustainable ICT in China, Huawei has implemented numerous solutions to mitigate its carbon footprint. As an example, an intelligent automatic cooling system, iCooling \cite{HuaweiiCooling}, is deployed in the data center reducing total power consumption of refrigeration stations by 8\%-10\%.  It will also save 3.85 million kWh electricity, which is the same as planting 79,500 trees. Likewise, Huawei also participates actively in the transition to renewable energy. It uses renewable energy in its operations wherever possible and is expanding its photovoltaic (PV) plants across its campuses, which generated 12.6 million kWh of electricity in 2020.

Singapore, which aspires to be Southeast Asia's green data center hub, has developed a roadmap to use renewable energy and reduce energy consumption in data centers through the ingenious application of new technologies~\cite{DCD_singapore}. However, as a space-constrained country, Singapore's data center industry has an extremely high power density. Furthermore, Singapore is required to reduce its emission intensity by 36\% from 2005 levels until 2030. Such challenges call for a collaborative solution from both the government and the local data center industry. In this regard, the data center providers such as Keppel Data Centers (Keppel) and ST Telemedia Global Data Centers (STT) have also launched their plans. Keppel championed the harnessing of the cold capacity from LNG regasification to cool the near-shore floating data centers which alleviated land use and implemented seawater cooling \cite{Keppel}. In addition, STT also announced its Environmental, Social, and Governance (ESG) plan, which was a commitment to transitioning its data center to be carbon neutral by 2030~\cite{STT}. To achieve carbon neutrality, STT will purchase green energy and renewable energy certificates (REC) or power purchase agreements (PPA) as well as deploy new facilities and technologies to continuously elevate energy efficiency and improve PUE.

\subsubsection{Efforts from Cloud Providers in Europe}
Data center operators and trade associations in the EU are committed to the European Green Deal, a plan aiming to transform the EU into a modern, resource-efficient, and competitive economy \cite{EUGreenDeal}, in order to achieve Europe's climate neutrality goal by 2050\cite{EUPact}. Major European cloud and data center operators, including AWS, Google and OVHCloud, as well as smaller and national providers, have agreed to become climate neutral by 2030 \cite{EUDCGoal}. Under the Pact, all facilities must use 75 percent renewable or carbon-free energy by 2025, and be 100\% carbon-free by 2030 \cite{EUDCGoal}. All new data centers in cool climates will meet an annual PUE target of 1.3 by the start of 2025. Facilities in warm climates will only have to meet a PUE of 1.4, because they must use energy to cool their IT servers. Legacy data centers have higher PUEs, and they will have to meet these targets until 2030 \cite{EUDCGoal}. Similarly, targets for recycling and repairing servers are emerging: the Pact commits signatories to assessing all servers for reuse, repair or recycling, whereas the specific target percentage will not come to an agreement until 2025 \cite{EUDCGoal}. 

For example, OVHCloud, a leading European cloud provider with a global presence, has issued its data center carbon emission reduction roadmap towards carbon neutrality in 2020 \cite{OVHCloud1}. To further reduce its carbon footprint, OVHCloud intends to pay more attention in the following issues \cite{OVHCloud2}:
\begin{itemize}
    \item \textbf{Contract PPAs in Europe} for high quality renewable energy supply, with the goal of reaching 100\% renewable energy supply by 2025,
    \item \textbf{Optimize lifespan of hardware equipment} via encouraging circular economy among its partners and developing a new brand for products running on refurbished servers;
    \item \textbf{Improve industrial ecodesign} by following the philosophy of ``use smarter, use longer, use less". 45\% servers are made from repurposed components and 24/31 data centers are reused industrial buildings;
    \item \textbf{0\% waste to landfill} by 2025 via efficient waste management.
\end{itemize}

\subsection{Approaches Towards Carbon-neutral Data Centers}
Based on the previous introduction about industrial efforts toward carbon-neutral data centers, one can find that although different entities place diverse emphasis on the path towards carbon-neutral data centers, the following issues are widely acknowledged:
\begin{itemize}
    \item Leverage carbon market to offset data center carbon emissions in a cost-efficient way.  
    \item Increase the use of renewable energy in daily operations of a data center.
    \item Keep improving the energy efficiency of both the IT system and the infrastructure.
    \item Boost the circulation of both energy and material throughout the lifecycle of a data center.
\end{itemize}

Therefore, Section \ref{Carbon Market}, \ref{Energy Supply}, \ref{Energy Utilization}, and \ref{Energy Circulation} will summarize and discuss existing methods on the these issues raised above and the outline of the rest of this survey is illustrated in Fig. \ref{Approches}. 

\section{Introduction to Carbon Market} \label{Carbon Market}
As humans have become more aware of their huge negative environmental impacts since the first industrial  revolution, an increasing number of programs and regulations have been implemented to control carbon emissions. These emission control programs can be coarsely categorized into three classes: carbon taxes, cap-and-trade, and carbon credits. These emission control programs comprise the carbon market, in which each ton of carbon emissions is assigned a price based on a specific pricing mechanism. The impetus of establishing a carbon market is that people will be able to choose whether or not to emit carbon dioxide given the carbon price. According to the latest report by the World Bank \cite{WorldBank}, 10 carbon pricing mechanisms are issued in 2019, equaling the increase in the 3 years preceding 2019 and the global carbon market will fund more than 45 billion dollars in 2019 with more than half of the funds going to environmental projects. Furthermore, Chinese government announced its giant sustainable development goal, i.e., 30.60 plan (achieving peak carbon emission in 2030 and carbon neutrality in 2060) and the global carbon market will cover around 21.5\% carbon emissions with the opening of the carbon market in China. In this section, we introduce these three carbon emission control mechanisms and discuss their merits and drawbacks in order to provide some insights for data center operators in selecting proper carbon pricing programs to reduce their carbon footprint in a cost-efficient manner.

\begin{figure}[t]
	\centering
	\includegraphics[width=.48\textwidth]{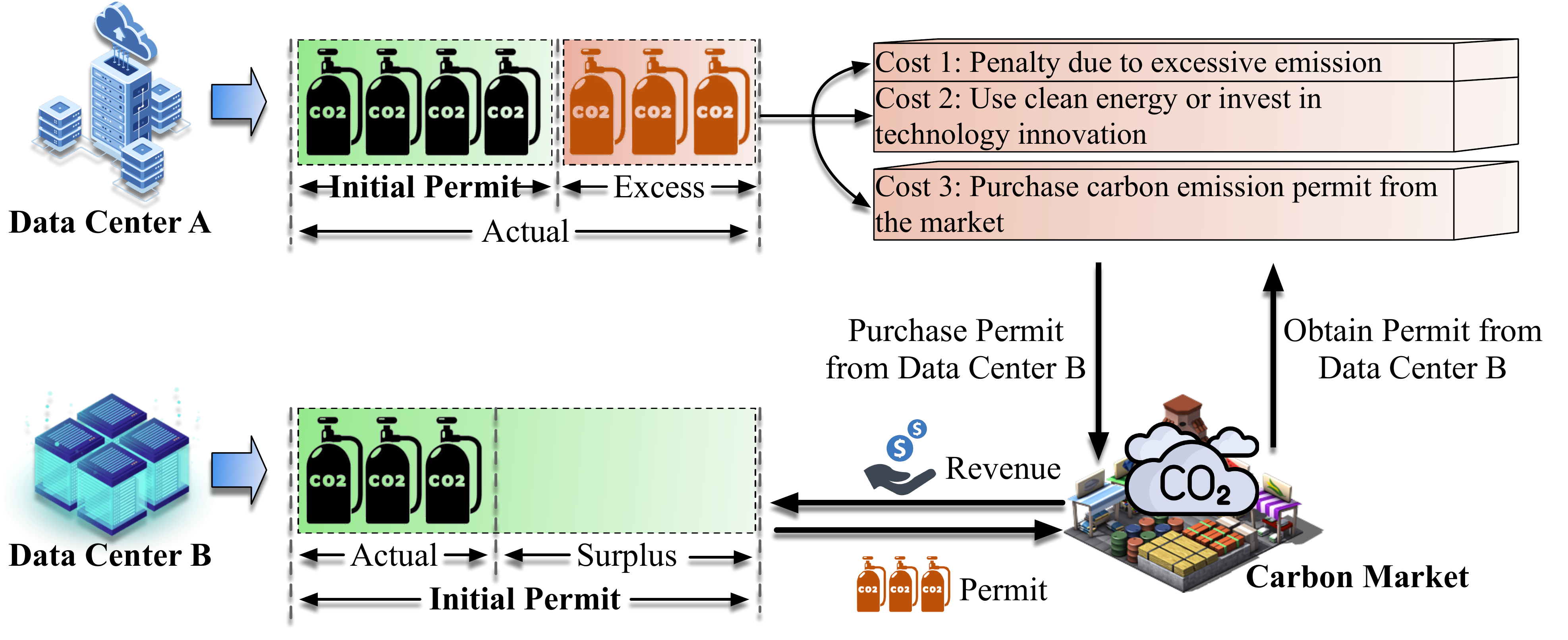}
	\caption{Illustration of two data centers involving in carbon emission permit trading. Once the cost of purchasing carbon emission permits is less than the cost of penalty and technological innovation, a data center operator prefers to purchase carbon credits to offset its carbon emissions.}
	\label{CapTrade} 
\end{figure}
\subsection{Cap-and-trade}
In the cap-and-trade system, the government establishes a ``cap" which decreases annually, on a data center's total carbon emissions and allocates permits for carbon emissions to data center operators for each compliance period (e.g., 1 year). Due to the ``cap", the total carbon emission of a data center is expected to decrease gradually. Each permit entitles the holder to one metric ton of carbon emissions. After one metric ton of carbon emissions, the data center should return the permit to the supervision department. The data center operators are able to leverage the allocated carbon emission permits to meet their carbon emission. In addition, they can save some permits by keeping their total carbon emissions below the ``cap" via technological innovation and then sell them back to the carbon market to obtain revenues. Certainly, if a data center continues to emit more carbon dioxide than its holding permits, it must purchase the permit from the carbon market to cover the excess.

In the cap-and-trade system, the ``cap" controls total carbon emissions and ``trading" makes the process more cost-effective. In this regard, we use two data centers with carbon emission permit trading to show how the cap-and-trade system makes emission reduction more cost-efficient, as shown in Fig. \ref{CapTrade}.

In Fig. \ref{CapTrade}, the data center A fails to control its carbon emission under the ``cap", whereas data center B even saves some permits via technology innovation. The operator of data center A finds that the cost of purchasing the carbon emission permit is much smaller than investing in carbon emission reduction technologies or paying the penalty for excessive carbon emissions, and therefore it decides to purchase carbon emission permits from data center B to meet its ``cap". In this process, data center A satisfies the ``cap" with minimum cost and data center B obtains cash revenues by selling its permits. It seems to be a win-win solution and that's why the cap-and-trade system is cost-effective in achieving the climate goal. However, due to the current relative low carbon price, some critics suspect the effectiveness of the cap-and-trade and they claim that the low carbon price will incentivize the emitter to procure carbon emission permits to offset their carbon emissions instead of investing in emission reduction technologies to reduce their carbon emissions \cite{gilbertson2009carbon}. 
\begin{figure}[t]
	\centering
	\includegraphics[width=.46\textwidth]{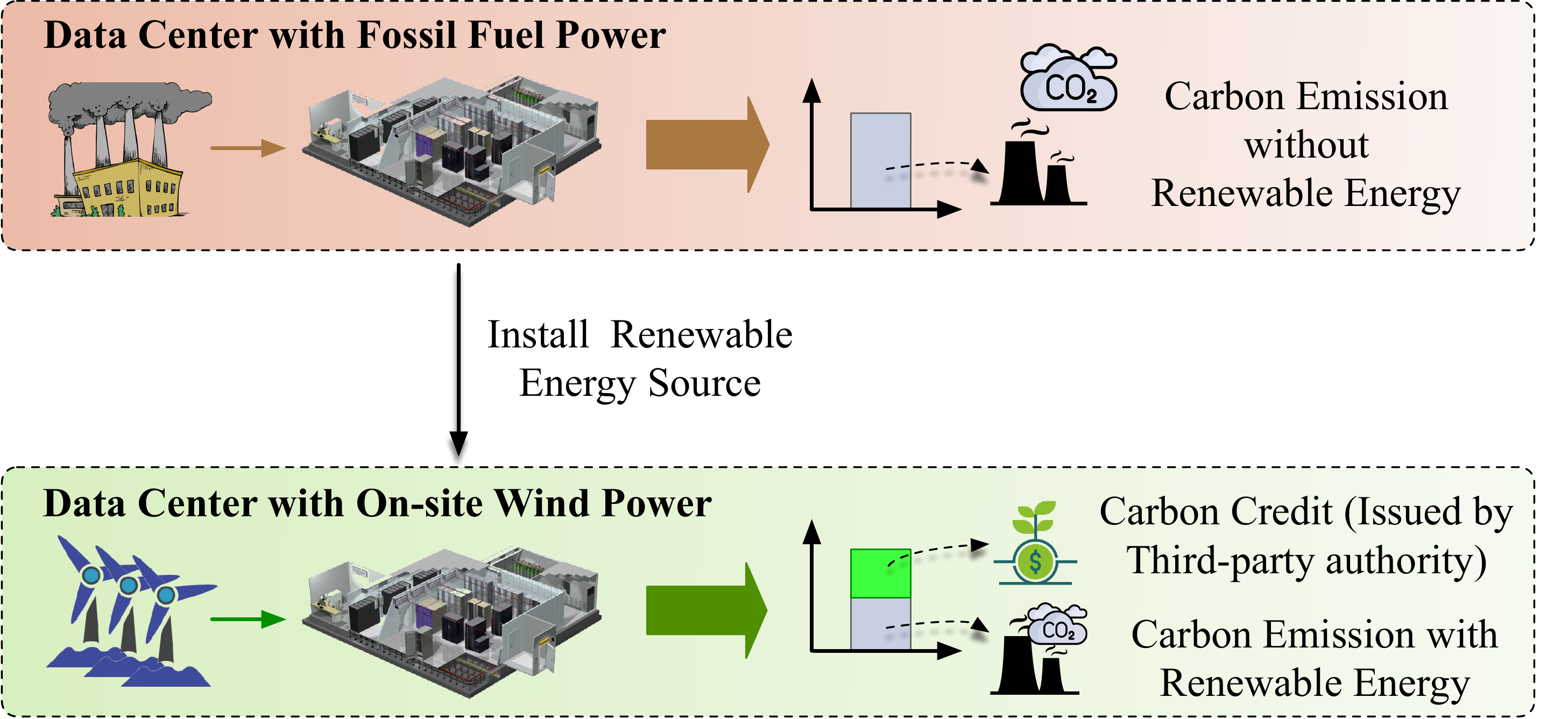}
	\caption{Illustration of the carbon credit mechanism. If a data center reduce its carbon emissions by technological innovation or changing its energy supply, the reduced carbon emissions can be issued carbon credits from a third-party authority, which can be sold in the carbon market to obtain cash revenues.}
	\label{CarbonCredit}
\end{figure}
\subsection{Carbon Taxes}
The carbon tax refers to the government-designated price for one metric ton equivalent carbon dioxide emission. The prices exist strong spatial variation and the global average prices are still far below the desired price (40-80 USD/t$\text{CO}_2$e) for meeting the climate goal set in Paris Agreement \cite{WorldBank}. The data center operator must pay for its carbon emissions after one year of operation.

The carbon tax is different from the cap-and-trade system in three aspects. First, the carbon tax charges the data center operator based on its annual amount of carbon emission, whereas in the cap-and-trade system, the data center operator must purchase carbon emission permits in advance based on the estimation of its future carbon emission. Secondly, the carbon tax does not support trading between emitters. Most importantly, unlike the cap-and-trade system, there is no ``cap" imposed on the data center, and therefore there is no guarantee on the reduction of total carbon emissions in the carbon tax system. In order to reduce total carbon emission, the government should conduct thorough investigation and design the carbon tax carefully to strike for a balance between economic development and the fulfillment of the climate goal.

\subsection{Carbon Credit}
Recent years have witnessed the rapid growth in the carbon credit mechanism. According to the report from World Bank \cite{WorldBank}, the registered carbon credit projects were 18,644 in 2020, bringing the carbon credits issued since 2002 to about 4.3 billion t$\text{CO}_2$e which accounts for 7.9\% annual carbon emissions or equals to the total $\text{CO}_2$ absorbed by 200 billion trees. Carbon credit mechanism provides another way of offsetting the carbon emission of a data center. Different from the Cap-and-Trade mechanism, carbon credit is voluntary, instead of compulsory. In the carbon credit mechanism, data centers can obtain carbon credit via adopting emission-reduction technology such as on-site renewable energy generators. The principle of the carbon credit mechanism is illustrated in Fig. \ref{CarbonCredit}.

In Fig. \ref{CarbonCredit}, the carbon emission of a data center declines due to the installation of the on-site wind driven generator, which generates corresponding carbon credits for the reduced carbon emission. It is important to note that the carbon credit is valid only if it is issued by an acknowledged third-party authority like Golden Standards. After obtaining the carbon credits, the data center operator can utilize them to offset its carbon emissions, or to sell them in the carbon market to obtain cash revenues. Noted that the carbon credit can coexist with the Cap-and-Trade mechanism and acts as a supplement for offsetting carbon emission. However, carbon credits are not universally accepted by all carbon pricing mechanisms yet. For example, the carbon credit issued by Joint implementation (JI) is accepted only in the EU ETS. Carbon credit mechanism offers more flexibility for data center operators to achieve their carbon neutrality goals.

\section{carbon-neutral data center: From Energy Supply Perspective}\label{Energy Supply}
Recently, to achieve their ambitious carbon neutrality goal, more and more cloud providers such as Google \cite{Google} and Microsoft \cite{Microsoft} have attempted to integrate carbon free energy, i.e., renewable energy into daily operations of their giant cloud data centers. Research opportunities as well as challenges occur when the intermittent and variable renewable energy has been introduced. In this section, we first introduce several renewable energy sources will be discussed in order to provide some insight for data center operators when a set of renewable energy is available for them to choose from. Subsequently, we will summarize the optimization objectives in the existing literature and discuss their mutual relationship. In what follows, we introduce existing works on this topic according to their optimization goals. 
\begin{figure}[t]
	\centering
	\includegraphics[width=.48\textwidth]{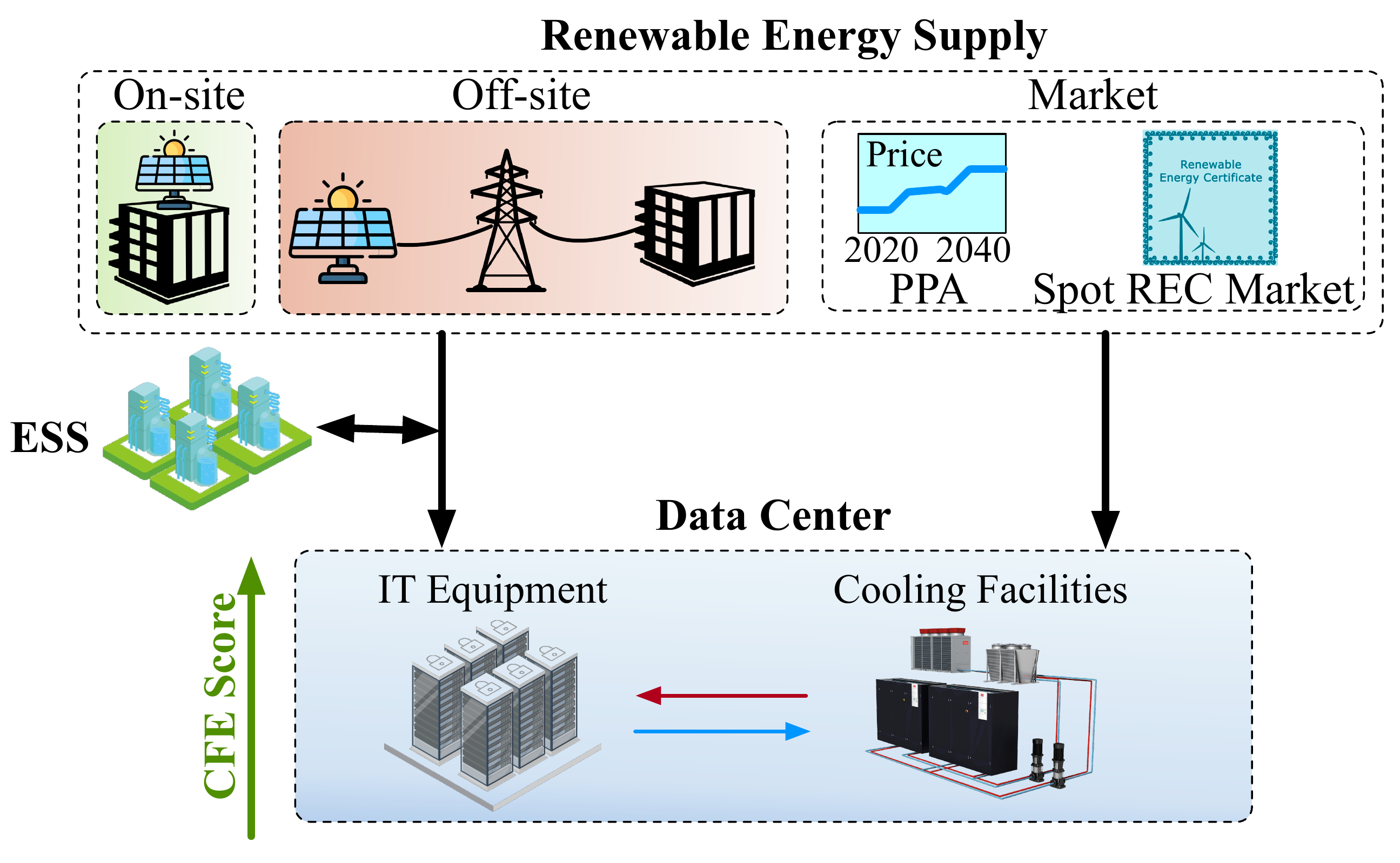}
	\caption{Overview of a carbon free energy integrated data center. The electricity comes from a mix of green and brown energy sources. The ESS is also installed to store intermittent and unstable renewable electricity. PPAs and RECs serve as two alternative renewable energy options.}
	\label{CFEDataCenter} 
\end{figure}
\subsection{Renewable Energy Options for Data Centers}
Generally speaking, there are four renewable energy options for a data center operator to choose from and the architecture of a renewable energy integrated data center is illustrated in Fig. \ref{CFEDataCenter}. In this system, an on-site renewable energy plant is installed to supply renewable electricity into the power supply system of the data center. Furthermore, in order to cope with the intermittent and unstable nature of renewable energy, an energy storage system (ESS) is often introduced into the data center microgrid so that the surplus energy can be stored for future usage and the energy shortage can also be remedied via the battery discharge. In addition, as many cloud providers have contracted PPAs with local renewable projects or purchased RECs from the market, the electrical grid acts as a wheel, transporting renewable energy to the data center. According to Kirchhoff's law, once the electricity flows into the grid, the customer is no longer able to distinguish its source. Therefore, PPAs or RECs enable data center operators to claim that a certain amount of their electricity comes from renewable energy.  It is also important to note that, as an increasing number of renewable energy plants have been established around the world, the electrical grid is becoming greener. Therefore, part of the grid electricity comes from renewable energy sources and it should be factored into the carbon emission factor of the electricity drawn from the grid. In the rest of this section, we briefly introduce these four options and discuss their advantages and disadvantages relating to data center management.
\subsubsection{On-site Renewable Energy}
Increasingly, on-site renewable energy becomes an alternative energy supply for green data centers, with wind and solar being the top two popular options. When the on-site renewable energy generator is integrated into the data center microgrid, the data center operator takes the responsibility for managing, maintaining, and operating the renewable energy plant. All or a portion of the renewable electricity can be utilized in its daily operations including powering the servers and infrastructure. The flexibility gives data center operators more opportunities and freedom in capacity planning as well as power management. Furthermore, a data center becomes an energy prosumer instead of only an energy consumer with on-site renewable energy generation. Data centers are able to sell the surplus renewable energy back to the electrical grid to offset their grid electricity consumption and obtain corresponding carbon credits via net metering in the United States \cite{NetMetering}.  However, since typical renewable energy sources such as solar and wind are intermittent and variable (because the sun doesn’t always shine and the wind doesn’t always blow), relying on on-site renewable energy purely may impede the operation of a data center because a number of critical tasks hosted in the data center requires round-the-clock timely response. To tackle this shortage, increasing number of data centers are installing the on-site Energy Storage System (ESS) to shift the renewable energy temporarily to match the workload demand. However, the on-site ESS has the following shortcomings \cite{huang2020review}:
\begin{itemize}
    \item Cloud data centers host thousands of high performance servers which draw a large amount of electricity. In order to meet this current demand, the demanded space for hosting the ESS may be prohibitive for most data centers;
    
    \item The turn-around efficiency of current battery systems is too low to agilely respond to workload variation due to internal resistance and self-discharging;
    
    \item The lifetime of the ESS will be significantly shortened due to frequent charge and discharge in the daily operation of a data center; 
    \item The disposal of the ESS is not environmentally-friendly since the ESS contains some harmful chemical instances. 
\end{itemize}
Moreover, the location of the data center may not be the best location for producing renewable energy. Therefore, for some data center locations, installing on-site renewable energy generation systems may not be an cost-efficient solution.

\subsubsection{Off-site Renewable Energy}
Off-site renewable energy generation offers the opportunity to both the data center operator and the electricity utility to do the work they are best at: data centers take charge of managing the IT facility as well as the infrastructure to improve the energy efficiency, while the utility is responsible for managing the generation of renewable energy and the electrical grid and transport the green electricity to the data center. In such a scenario, the grid acts as the wheel for delivering renewable electricity. Furthermore, with the separation of renewable energy generation and data center operations, renewable energy generators can be placed in the most suitable place to work in their most productive configuration. The challenge in leveraging such a solution lies in the establishment of a charge and accounting mechanism which records the contribution of the renewable energy projects and incorporates this into the utility bill of the data center.

\subsubsection{Power Purchase Agreement (PPA)}
PPA is a contract between a data center operator and a renewable energy producer which allows the data center operator to purchase part of all of the renewable energy generated by the producer at a negotiated and fixed price \cite{GooglePPA}. After signing the PPA, the data center operator is able to claim that part of the electricity drawn from the grid comes from a carbon-free energy source with bundled certification. With the PPA, the data center can control the future expenditure in renewable energy procurement and control their operational risks simultaneously. Furthermore, the data center operator does not take part in the daily management of the renewable energy plant, allowing them to focus on the work it is best at. On the other hand, PPA is also beneficial for the decarbonization of the local electrical grid since PPA involves the investment in the construction of additional renewable energy plants. Many celebrated cloud providers such as Google have signed PPA with local renewable energy providers. As an example, Google signed a PPA to add new wind and solar plants in Netherland so that over 70\% electricity consumption of the data center in Eemshaven comes from carbon-free energy now, whereas less than 20\% renewable energy penetration was available before signing the PPA \cite{Google}.    

\subsubsection{Renewable Energy Certificate (REC)}
The Renewable Energy Certificate (REC), also known as ``renewable energy credit", is a certification of one MWh electricity coming from a certain renewable energy source such as wind or solar. After the data center purchases a REC from the market, it is capable of claiming that one MWh electricity consumption is from the specific carbon free source certificated by the REC. It seems that the REC and the PPA are similar since the data center operator is able to announce its usage of carbon-free energy to reduce its carbon footprint, two stark differences distinguish the REC and the PPA. On the one hand, REC is a tradable commodity whereas the PPA is not. The tradable property makes the data center operator more flexible in terms of carbon footprint reduction: if the data center operator cannot meet its carbon capping target, it can purchase REC from the market to fulfill the target instead of investing into new renewable energy projects since the price of REC in the spot market is relatively low (\$0.8/MWh in the United States in 2018 \cite{RECMarket}). Furthermore, if the on-site and off-site renewable energy generator is integrated in the data center microgrid, the data center operator can also sell RECs to the market provided that its excess renewable energy is certificated via a third-party authority. On the other hand, the REC is decoupled with physical electricity, while the claimed usage of renewable energy is coupled with physical electricity in the PPA. Therefore, when a data center operator claims that the carbon emissions corresponding to one MWh electricity consumption is reduced due to the usage of the REC, it may just transfer its carbon emissions into another place without reducing the total carbon emissions to the atmosphere. However, because additional renewable energy plants are constructed with the PPA, the total carbon emission due to the renewable electricity procurement is literally reduced. Therefore, cloud providers increasingly choose to utilize PPA to cut down their carbon emissions, and demonstrate their leadership in sustainable development.

\subsection{Optimization Goals}
In the works covered in this survey, the optimization for a carbon-free energy integrated data center may target for one or more goals listed in the following:

\begin{itemize}
    \item \textbf{Minimize grid (brown) electricity procurement.} Research works with such an objective aim to reduce the brown electricity procurement from the electrical grid.
    \item \textbf{Minimize operational costs.} These research works target to reduce the (long-term) operation cost including the grid electricity procurement cost, energy storage system cost, the cost due to service level agreement (SLA) violation and so on.
    \item \textbf{Maximize operational profits.} These research works aim to maximize operational profits by improving the revenues coming from SLA fulfillment, energy trading and offering ancillary service in smart grid, and reducing operational costs simultaneously by joint IT, cooling and power distribution system control. 
    \item \textbf{Maximize utilization of renewable energy.} These works aim to better match the workload demand and the generation of renewable demand so that the waste of precious renewable energy is minimized.  
\end{itemize}

It is obvious that these optimization goals overlap to some extent. For example, maximizing the utilization of renewable energy may decrease the amount of electricity drawn from the grid, leading to less electricity procurement costs. However, since the presence of renewable energy does not always align with the peak of electricity price, it may be more economic to save the surplus renewable energy and utilize it when the electricity price is high. Therefore, there exists multiple trade-offs between these optimization goals and it is important to find proper equilibrium among optimization goals which are seemingly overlapping but also conflicting to some extents . 

\subsection{Grid Electricity Procurement Minimization}\label{Bills}
\begin{figure}[t]
	\centering
	\includegraphics[width=.48\textwidth]{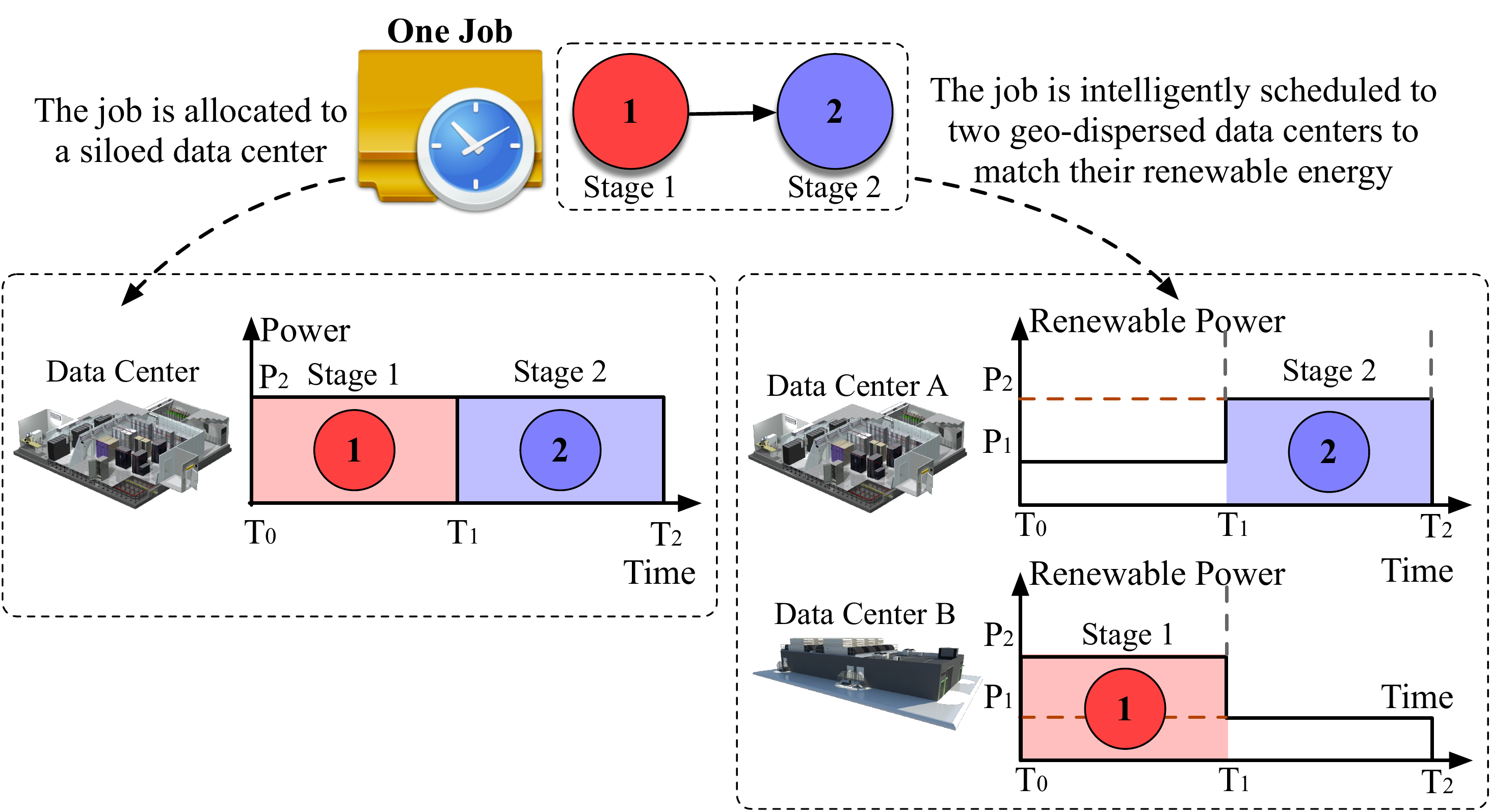}
	\caption{Illustration of the intuition of optimization over a geo-dispersed data center network. If the two jobs can be scheduled spatiotemporally, their energy demands can be perfectly matched with the renewable energy. Otherwise, additional brown energy should be used to meet their energy demands.}
	\label{GeoIntuition} 
\end{figure}

Intuitively, many researchers have attempted to reduce electricity bills for operating the data center with renewable energy integration because renewable energy is almost free (except for marginal maintenance fee) in the operation phase once the renewable plant has been established. Among these works, the majority of them focus on the optimization over a cluster of the geo-dispersed data centers because the spatial variation of renewable energy offers more opportunities for improving its utilization.  In Fig. \ref{GeoIntuition}, we provide a graphical illustration of such intuition.

In Fig. \ref{GeoIntuition}, there are two jobs, i.e., job 1 and job 2, to be executed and their power consumption is $P_2$. The renewable energy supply in the data center located in site A is $P_1$ between $t_0$ and $t_1$ and $P_2$ between $t_1$ and $t_2$. The renewable energy supply in the data center located in site B is $P_2$ between $t_0$ and $t_1$ and $P_1$ between $t_1$ and $t_2$. Without the geo-dispersed data centers, job 1 and job 2 have to be scheduled to either data center A or data center B with total power consumption be $P_2-P_1$. However, with the geo-dispersed data centers, we are able to schedule job 1 to data center B and job 2 to data center A. The power demand is  perfectly matched with renewable energy supply, leading to reduced grid electricity procurement. Furthermore, such an idea can be easily extended to leverage the spatial variation of electricity price as well as the carbon emission factor to further reduce the electricity procurement cost or the carbon footprint. Furthermore, an excellent survey on geographical load balancing based data center power management can be found in \cite{GLBSmartGrid}.

 When it comes to the optimization of electricity bills, some researchers focus on the instantaneous optimization, i.e., optimization on the current time slot without the knowledge of future system states, whereas many other researchers aim to minimize the electricity bill over a specific time horizon. In the following, we will present the related works on this topic in accord with this taxonomy and the literature summary is available in Table \ref{MinPRenewable}. 

\begin{table*}
    \centering
    \begin{tabular}{c!{\vrule width 1pt}c!{\vrule width 1pt}c!{\vrule width 1pt}c!{\vrule width 1pt}c!{\vrule width 1pt}c}
        \noalign{\hrule height 1pt}
        Reference  & Constraints & Control Knobs & Formulation & Algorithms & Achieved Results \\ 
        \noalign{\hrule height 1pt}
        \cite{AbbasCL2016} & \makecell[c]{Total energy\\ budget} & \makecell[c]{Brown energy consumption} & \makecell[c]{Information Graph} & \makecell[c]{Heuristic sorting} & \makecell[c]{Prove that\\ minimum brown \\electricity consumption\\ exists} \\
        \noalign{\hrule height .1pt}
    
        \hline
        \cite{Hogade2018Mini} & \makecell[c]{QoS constraint} & \makecell[c]{Workload distribution among\\ multiple data centers;\\Workload assignment within\\ a data center} & \makecell[c]{Integer Programming} & \makecell[c]{Force Direct\\ Load Distribution\\ (FDLD)} & \makecell[c]{61\% cost saving} \\
        \hline
        
        \cite{YuanTASC2019} & \makecell[c]{Bandwidth;\\Response time;\\Total energy consumption} & \makecell[c]{Workload distribution among\\ multiple data centers} & \makecell[c]{Nonlinear Programming} & \makecell[c]{Simulated Annealing} & \makecell[c]{around 30\% cost saving} \\
        \hline
        
        \cite{Gupta2019} & \makecell[c]{No specific constraints} & \makecell[c]{Solar power;\\Free cooling} & \makecell[c]{No specific formulation} & \makecell[c]{Greedy Heuristic} & \makecell[c]{60\% brown energy\\ reduction} \\
        \hline
        
        \cite{Gu2016ICC} & \makecell[c]{QoS constraint;\\Total Budgets} & \makecell[c]{Workload distribution among\\ multiple data centers} & \makecell[c]{Linear Integer \\ Programming} & \makecell[c]{CPLEX} & \makecell[c]{25\% electricity\\ reduction} \\
        \hline
        
        \cite{Le2010} & \makecell[c]{SLA constraints} & \makecell[c]{Workload distribution among\\ multiple data centers;\\Brown energy procurement} & \makecell[c]{Nonlinear Programming} & \makecell[c]{Simulated Annealing} & \makecell[c]{24\% brown energy\\ reduction} \\
        \hline
        
        \cite{Aksanli} & \makecell[c]{No specific constraints} & \makecell[c]{Batch job scheduling} & \makecell[c]{No specific formulation} & \makecell[c]{Heuristic algorithm} & \makecell[c]{Improved green energy\\ usage efficiency} \\
        \hline
        
        \cite{Chen} & \makecell[c]{No specific constraints} & \makecell[c]{Batch job scheduling} & \makecell[c]{No specific formulation} & \makecell[c]{Heuristic algorithm} & \makecell[c]{40\% brown energy\\ reduction} \\
        \hline
        
        \cite{DengMultiGreen} & \makecell[c]{Renewable capacity} & \makecell[c]{Long-term ahead\\ power purchase;\\Power drawn from\\ on-site renewables, batteries\\and real-time market} & \makecell[c]{Two-stage Lyapunov\\ Optimization} & \makecell[c]{Interior Point Method\\ for per-time slot\\ convex optimization} & \makecell[c]{Significant electricity\\ cost reduction} \\
        \hline
        
        \cite{Lee2015} & \makecell[c]{TES evolving equation;\\DC server capacity} & \makecell[c]{Workload distribution;\\TES discharge and charge } & \makecell[c]{Lyapunov Optimization} & \makecell[c]{LP solver for per time \\ slot optimization} & \makecell[c]{98\% cooling energy\\ saving in summer} \\
        \noalign{\hrule height 1pt}
    \end{tabular}
    \caption{Summary of existing works on minimizing brown electricity consumption in renewable energy-integrated data centers in terms of constraints, control knobs, problem formulation, solving algorithm, and achieved results.}
    \label{MinPRenewable}
\end{table*}

Some researchers aim to minimize electricity bills at the beginning of each time slot. In this formulation, given the current state of the data center, e.g., the incoming workload, and the yield of renewable energy,  data center operator should seek for the optimal decision variables such as the workload distribution vector so that current electricity procurement cost is minimized without violating some performance constraints. In \cite{AbbasCL2016}, Abbas \textit{et al.} studied the fundamental trade-off between the total and brown energy consumption of geo-dispersed data center clusters. A new service efficiency metric is proposed based on the M/GI/1 Processor Sharing (PS) queue model to evaluate the service efficiency of a data center. At the beginning of each time slot, an information flow graph is established for the geo-dispersed data centers and each data center is represented as a node in the graph and sorted according to its service efficiency in descending order. Closed-form workload distribution policy is derived based on the queue theory subsequently. The authors then proved that given the total power consumption, there existed a minimum value of brown energy consumption so that the SLA could be satisfied. The researchers also showed that increasing the total power consumption would increase the utilization of renewable energy and decrease the consumption of brown energy as well. Different from \cite{AbbasCL2016} which only considered a simplified data center system, Hogade \textit{et al.} proposed a holistic framework for the data center, taking into consideration the cooling power, on-site wind and solar power generation, net metering, peak electricity price, and the co-location inference when a multi-core machine was processing a memory-intensive task \cite{Hogade2018Mini}. The objective is to minimize the electricity bill while satisfying the QoS requirements of the geo-dispersed data center via intelligent workload distribution among multiple data centers as well as the workload assignment to a specific server within a single data center. Three solutions based on Force-Directed Load Distribution (FDLD) \cite{FDLD}, greedy heuristic and genetic algorithms are proposed to solve the optimization problem and 61\% cost saving is achievable. Further, Yuan \textit{et al}. incorporated the bandwidth cost of transmitting data between users and data centers into the electricity bills of geo-dispersed data centers \cite{YuanTASC2019BandWidth}. On-site solar and wind energy are also considered and the problem is formulated as nonlinear programming with the total bandwidth, average response time and total energy consumption as constraints and workload distribution vectors as optimization variables. The constrained optimization is first transformed into an unconstrained one with the penalty method \cite{penaltymethods} and solved with a  heuristic algorithm. Compared with two scheduling methods prioritizing the cheapest electricity price \cite{deng2014eco} and the utilization of renewable energy \cite{bi2016trs}, the proposed method takes electricity price, bandwidth price, and the presence of renewable energy into consideration, leading to 30.58\% and 30.82\% cost saving on average, respectively. Recently, Gupta \textit{et al}. investigated the potential of cost saving when the on-site solar energy power, outside free cooling and the ESS were available for a data center \cite{Gupta2019}. A greedy algorithm which prioritizes the utilization of solar power, outside free cooling is presented to reduce brown energy consumption. Simulation results show that 60\% brown energy consumption is achievable with only on-site solar power and the combination of on-site solar power and outside free cooling can reduce the brown energy procurement more significantly. 

Other researchers consider the electricity minimization problem over a specific time period. Intuitively, optimization over a time horizon may require the awareness of future system states and the value of the exogenous variables. A slice of research works assume the availability of perfect future information and leverage offline optimization techniques to derive the optimal policy. For example, Gu \textit{et al}. \cite{Gu2016ICC} studied the green scheduling of multiple data centers with ESS integration. Two constrained mixed integer linear programming (MILP) problems: a) minimizing brown energy consumption with QoS constraints, and b) minimizing carbon emissions with limited budgets are formulated and solved optimally in offline manner with MATLAB CPLEX solver. 

On a different track, some researchers leverage some prediction techniques to predict either the future workload or the presence of renewable energy so that a better policy considering the future variation of system input can be derived. In \cite{Le2010}, Le \textit{et al}. proposed a two time scale carbon management framework with the Cap-and-Trade mechanism. In the coarse time scale (a year), the framework decides the power mix of each data center based on the energy usage of the previous year. In the finer granularity, it distributes requests among multiple data centers so that the overall electricity cost is minimized. The problem is formulated as a nonlinear programming problem and AutoRegressive Integrated Moving Average (ARIMA) \cite{ARIMA} is utilized to predict the workload in the near future. A solver based on Simulated Annealing \cite{SA} is proposed to obtain the request distribution policy. Simulation results show that 24\% brown energy saving is achievable with 10\% increase in total cost, demonstrating the trade-off between carbon emission reduction and cost saving. Different from \cite{Le2010} where the future workload was predicted, quite a few researchers attempted to predict the production of renewable energy to perform workload distribution in order to improve the utilization of renewable energy. Aksanli \textit{et al}. proposed a batch job scheduler based on the prediction of the on-site solar and wind power in the future 30 minutes \cite{Aksanli}. After obtaining the prediction results, the scheduler can determine how many batch jobs can be supported with the predicted renewable energy. The researcher leveraged the well acknowledged Weather Conditioned Moving Average (WCMA) algorithm \cite{piorno2009prediction} for solar energy prediction and weighted nearest neighbors method for wind power prediction by considering the seasonal effect of wind power. These two prediction techniques achieve good prediction results with mean error of 9.6\% and 17.2\% respectively. Compared with the solution without renewable energy prediction, the future-aware scheduler is able to improve the renewable energy utilization by three times and reduce 4x batch job cancel rate. In addition, Chen \textit{et al}. developed a batch job scheduler for a cluster of high performance computing (HPC) data centers with the knowledge of predicted renewable energy production \cite{Chen}. The scheduler can be decomposed into the static part and the dynamic part. In the static part, an HPC task is decomposed into a Directed Acyclic Graph (DAG) and each part is pushed to a task queue. In the dynamic part, with the renewable energy prediction results, the scheduler migrates the running task with maximum remaining time into the site with enough renewable energy until there is no task to migrate or no renewable energy to support the task. In terms of starting a new job, the scheduler makes decisions according to the predicted renewable energy as well as the outside temperature at each site in order to realize the green-aware and free-cooling-aware scheduling. Simulation results demonstrate that the proposed scheduler achieves 40\% brown energy reduction compared with the baseline round robin scheduler.

Despite the success of many predictive schedulers \cite{Le2010}\cite{Aksanli}\cite{Chen}, prediction based optimization requires the highly accurate prediction algorithm for renewable energy, going against the intermittent and volatile property of renewable energy. To overcome the performance loss with inaccurate prediction results, a plethora of researchers strive to design online algorithms without the knowledge of future renewable energy production. Lyapunov Optimization \cite{neely2010} is the most popular choice for deriving such solutions. Lyapunov Optimization aims to solve the optimization problem with the time-averaged objective function and constraints. By transforming the time-averaged constrains into virtual queues and integrating virtual queue stability into the Drift-plus-Penalty function \cite{neely2010}, the optimization over a time horizon can be transformed into a set of new optimization problem in each time slot and the asymptotic optimal solution over the time horizon is provably achievable if the sub-problem in each time slot is solved optimally without any knowledge of future information, making it attractive in the presence of future uncertainty. Due to the strong theoretical guarantee, a considerable number of researchers have incorporated Lyapunov Optimization into their solution to realize optimal online schedulers.

In \cite{Urgaonkar}, Urgaonkar \textit{et al}. studied the optimal power management within a data center with ESS. The objective is to minimize the time averaged cost of purchasing grid electricity as well as that of battery charging and discharging. The framework proposed by Deng \textit{et al}. called MultiGreen extended the system in \cite{Urgaonkar} by incorporating on-site renewable energy generation \cite{DengMultiGreen}. MultiGreen is a two-stage online algorithm based on Lyapunov Optimization. In the coarse time granularity, the data center operator makes decisions on procuring grid electricity from the long-term electricity market where electricity price is lower than real time market. In the fine time granularity, grid electricity is purchased from the real time market to fulfill the operation of the data center. By managing the charge and discharge of the battery, and the procurement of grid electricity from both the long-term and real time market, the time averaged electricity procurement cost is minimized. Simulation results show that MultiGreen achieves near the same performance compared with the offline optimal algorithm and outperforms the single stage solution remarkably. In addition, simulation results also suggest that increasing the penetration of renewable energy and the capacity of the battery is beneficial in terms of cost saving. Different from previous work focusing on a single ESS integrated data center, Lee \textit{et al}. proposed a new model which considers the workload distribution among a cluster of data centers incorporating a novel cooling system with free air cooling and a thermal energy system (TES) \cite{Lee2015}. TES is a complementary cooling technology where thermal energy is stored in chilled water or chilled ice tanks, and this stored energy is used to cool data center servers when needed——e.g., when free outside air cooling is infeasible. Similar to \cite{Urgaonkar} and \cite{DengMultiGreen}, Lyapunov Optimization is utilized to derive optimal online workload distribution and TES management policy. The optimization at each time slot is formulated as a linear programming problem and can be solved optimally. Simulation results demonstrate that at least 64\% and 98\% cooling energy savings are achievable during summer and winter respectively.

\subsection{Operational Cost Minimization}\label{Cost}
Going beyond the minimization of electricity bills, a large number of researchers have investigated the potential of operational cost reduction with renewable energy penetration. Operating costs cover many aspects, including but not limited to electricity bills, renewable energy installation and usage costs, SLA violation cost, carbon taxes, etc. In this section, we will summarize existing works aiming to operational cost minimization according to different kinds of operation costs they considered, and they are summarized in Table \ref{MinOperationCostRenewable}. 

\begin{table*}
    \centering
    \begin{tabular}{c!{\vrule width 1pt}c!{\vrule width 1pt}c!{\vrule width 1pt}c!{\vrule width 1pt}c!{\vrule width 1pt}c}
        \noalign{\hrule height 1pt}
        Reference  & Constraints & Control Knobs & Formulation & Algorithms & Achieved Results \\ 
        \noalign{\hrule height 1pt}
        
        \cite{dou2017carbon} & \makecell[c]{Wind power capacity;\\Task queue stability} & \makecell[c]{Server scheduling\\for batch jobs;\\Wind and solar\\ power drawn} & \makecell[c]{Lyapunov Optimization} & \makecell[c]{LP solver for per time \\ slot optimization} & \makecell[c]{Reduced electricity cost\\and carbon taxes} \\
        \hline
        
        \cite{khosravi2017dynamic} & \makecell[c]{Server capacity;\\Local brown and\\ green energy capacity} & \makecell[c]{VM placement} & \makecell[c]{Linear Integer Programming} & \makecell[c]{Greedy Heuristic} & \makecell[c]{57.3\% cost saving} \\
        \hline
        
        \cite{ren2012carbon} & \makecell[c]{Grid power capacity;\\Renewable power capacity;\\Carbon capping} & \makecell[c]{Power drawn from\\different sources\\(grid, on/off-site\\ renewable energy)} & \makecell[c]{Linear Programming} & \makecell[c]{LP Solver} & \makecell[c]{Different carbon capping\\targets generate different\\energy portfolio} \\
        \hline
        
        \cite{mahmud2013online} & \makecell[c]{QoS constraint;\\Carbon neutrality} & \makecell[c]{Server provision for\\each type of job} & \makecell[c]{Lyapunov Optimization} & \makecell[c]{LP solver for per time \\ slot optimization} & \makecell[c]{20\% cost saving;\\Carbon-neutral operations}\\
        \hline
        
        \cite{ren2013coca} & \makecell[c]{QoS constraint;\\Carbon neutrality} & \makecell[c]{Server provision for\\each type of job;\\server speed scaling} & \makecell[c]{Lyapunov Optimization} & \makecell[c]{Decentralized solver\\ for per time \\slot optimization} & \makecell[c]{25\% cost saving;\\Carbon-neutral operations}\\
        \hline
        
        \cite{mahmud2013dynamic} & \makecell[c]{QoS constraint;\\Carbon neutrality} & \makecell[c]{Server provision for\\each type of job;} & \makecell[c]{Lyapunov Optimization} & \makecell[c]{LP solver for per time \\ slot optimization} & \makecell[c]{Over 20\% cost saving;\\Carbon-neutral operations;\\Further 2.5\% cost saving\\ with demand-responsive\\ electricity price}\\
        \hline
        
        \cite{liu2011greening} & \makecell[c]{Data center capacity} & \makecell[c]{Workload distribution\\ among multiple\\ data centers;\\Server provision} & \makecell[c]{Linear Programming} & \makecell[c]{Distributed solver} & \makecell[c]{``Follow-the-renewable"\\may be best;\\Energy storage system\\is critical;\\Wind power is preferred}\\
        \hline
        
        \cite{goudarzi2013geographical} & \makecell[c]{Server capacity} & \makecell[c]{VM placement} & \makecell[c]{Nonlinear Programming} & \makecell[c]{Force-Directed based\\ heuristic algorithm} & \makecell[c]{27-40\% cost saving}\\
        \noalign{\hrule height 1pt}
    \end{tabular}
    \caption{Summary of existing works on minimizing operational costs in renewable energy-integrated data centers in terms of constraints, control knobs, problem formulation, solving algorithm, and achieved results.}
    \label{MinOperationCostRenewable}
\end{table*}

As for the optimization related to carbon emissions, researchers either aim to minimize the carbon emissions explicitly or to treat carbon capping as a constraint in the optimization problem. In \cite{dou2017carbon}, Dou \textit{et al}. investigated the joint optimization of electricity bills and the carbon tax of a single data center with on-site wind power generation. In the system, the workload is categorized into delay-sensitive tasks which should be scheduled to execute immediately and delay-tolerant tasks which can be delayed within a certain period of time. The delay-tolerant task is pushed to a task queue once it arrives and the stability of the task queue is imposed as a constraint. As the optimization involves time-averaged constraints as well as the objective, Lyapunov Optimization is leveraged to derive optimal control law which decides the number of servers for executing  delay-tolerant tasks, the power drawn from the grid, and the on-site wind power generator at each time slot. Simulation results show that the electricity cost as well as the corresponding carbon tax will decrease with the increase in the capacity of on-site wind plants. In addition, carbon taxes will account for more than 30\% of total costs, demonstrating the need for improving renewable energy penetration. In contrast to \cite{dou2017carbon} which only considered a single data center, Khosravi \textit{et al}. targeted to the minimization of electricity bills and carbon footprint in geo-dispersed data centers with dynamic PUE via Virtual Machine (VM) placement optimization \cite{khosravi2017dynamic}. By exploiting the spatial variation of carbon intensity as well as carbon taxes, the broker decides to place a VM to a site with a greedy heuristic which prioritizes the data center with the minimum increase in total costs. The authors have shown through simulation that the solution considering renewable energy can reduce total costs by up to 57.3\% compared with the renewable energy-agnostic one. Except for joint optimization of carbon taxes and electricity bills, some researchers impose the carbon capping in the optimization framework. Ren \textit{et al}. investigated the power management problem for a single data center integrated with on-site and off-site renewable energy and the battery system \cite{ren2012carbon}. The problem is formulated as a linear programming problem with the objective of joint optimization of electricity bills, renewable energy procurement costs and battery maintenance costs. In this framework, the reduced carbon emissions must exceed a predefined threshold so that the carbon capping is realized. With the framework, the authors studied the trade-off between the carbon capping target and the best renewable energy source option. The results show that when the carbon capping target is below 30\%, on-site renewable energy is preferable, otherwise off-site renewable energy is more suitable. Different from \cite{ren2012carbon} which only considered carbon capping, some researchers pursued a more ambitious goal——achieving carbon neutrality (all electricity usage is offset by renewable energy) while reducing operation costs. 
\begin{table*}
    \centering
    \begin{tabular}{c!{\vrule width 1pt}c!{\vrule width 1pt}c!{\vrule width 1pt}c!{\vrule width 1pt}c!{\vrule width 1pt}c}
        \noalign{\hrule height 1pt}
        Reference  & Constraints & Control Knobs & Formulation & Algorithms & Achieved Results \\ 
        \noalign{\hrule height 1pt}
        
        \cite{kiani2016profit} & \makecell[c]{QoS constraint} & \makecell[c]{Workload distribution;\\server service rate} & \makecell[c]{Convex Programming} & \makecell[c]{CVX} & \makecell[c]{Profits will increase with more\\ wind power penetration}\\
        \hline
        
        \cite{EnergyPortofolio} & \makecell[c]{Minimum average profits;\\Server capacity} & \makecell[c]{Energy portfolio} & \makecell[c]{Mixed Integer Stochastic\\ Programming} & \makecell[c]{Convex\\ relaxation} & \makecell[c]{Profits will increase with more\\ wind power penetration}\\
        \hline
        
        \cite{chen2015cooling} & \makecell[c]{Time-averaged pending\\ rate for batch jobs} & \makecell[c]{Battery charge\&discharge;\\Diesel generator;\\Power distribution} & \makecell[c]{Nonlinear Programming\\over time slots} & \makecell[c]{Dual approach\\ with relaxation} & \makecell[c]{35\% cooling energy saving;\\
        60\% net cost saving}\\
        \noalign{\hrule height 1pt}
    \end{tabular}
    \caption{Summary of existing works on maximizing operational profits in renewable energy-integrated data centers in terms of constraints, control knobs, problem formulation, solving algorithm, and achieved results.}
    \label{MaxProfit}
\end{table*}
In \cite{mahmud2013dynamic}, Mahmud \textit{et al}. proposed to utilized Lyapunov Optimization to derive an online policy which controlled the number of servers for a specific type of service at each time slot. The objective is to jointly optimize the electricity cost and the delay cost which is derived with M/M/1 queue modeling. The carbon neutrality constraint is expressed as a time-averaged term——the time-averaged electricity consumption must be smaller than the time-averaged available renewable energy. Simulation results demonstrate that the proposed algorithm is able to achieve carbon neutrality while reducing 20\% operational cost compared with prediction based solutions. The work presented in \cite{ren2013coca} extended \cite{mahmud2013dynamic} by considering the server speed scaling. The per time slot optimization problem is formulated as a mixed integer linear programming and an optimizer called GSD (Gibbs Sampling-based Distributed optimization) based on a variation of Gibbs sampling is proposed to solve the optimization in a decentralized manner. Simulation results show that 25\% operation cost reduction is achievable. Since a data center is a huge electricity consumer, it may have non-negligible impacts on the electricity price. Therefore, demand response, a pricing mechanism which offers electricity price based on the electricity consumption of a consumer, is incorporated in the framework proposed in  \cite{mahmud2013online} to further reduce the operational cost \cite{mahmud2013dynamic}. By considering demand-response electricity prices, it can further decrease average costs by approximately 2.5\%.

Except for the previous work taking carbon taxes or carbon capping into consideration, a host of researchers incorporate SLA into optimization objectives. In \cite{liu2011greening}, Liu \textit{et al}. investigated whether geo-dispersed data centers were helpful to encourage the utilization of renewable energy and to reduce the carbon footprint of data centers. Three optimal distributed optimization algorithms are derived to achieve optimal renewable-energy-aware workload distribution and server provisioning. The target is to jointly optimize electricity bills and the delay cost which is a strictly increasing and convex function in terms of total delay. The authors suggested that geographical workload distribution was quite effective in terms of reducing brown energy consumption and ``following the renewable" workload distribution policy might be the best choice. Further, the battery system is introduced in the framework to study the role of storage in powering data centers with renewable energy. Simulation results suggest that a small scale battery system can overcome the spatiotemporal variation of renewable energy and provide further operational cost reduction. In addition, the authors also discussed the optimal renewable energy portfolio and found that wind power was more reliable than solar power because wind availability was not correlated with the geographical locations and the variation can be considerably smooth out when wind power from different locations was aggregated. 

In addition, in a virtualized data center, VM migration facilitates server consolidation, leading to energy consumption reduction. However, it will result in additional data transmission and request response time. In \cite{goudarzi2013geographical}, besides the grid electricity cost and SLA violation cost, VM migration cost is integrated into the operational cost, which is the penalty for service outage due to VM migration determined by the SLA. Geographical load balancing is studied in a virtualized data center and a force-directed online scheduler is proposed to achieve 27\%-40\% operational cost reduction.

\subsection{Operational Profit Maximization}
In Section \ref{Bills} and \ref{Cost}, it is shown that incorporating renewable energy into the power supply of a data center can effectively reduce grid electricity bills and operational costs. However, as the ultimate goal of operating a data center is earning profits for the stakeholder, some researchers study the profit maximization problem of a renewable energy integrated data center. The operational revenues come from the following three major sources:
\begin{itemize}
    \item \textbf{Fulfillment of SLA.} Typically, a SLA will regulate the maximum service delay for an Internet service. If the service time of a request exceeds the maximum delay, the data center administrator will be punished. Otherwise, a service fee will be paid to the data center operator. Besides revenues from serving Internet requests, a data center can earn its income from executing batch jobs such as training AI models.
    
    \item \textbf{Energy trading.} As more and more data centers are equipped with on-site renewable energy generators, they can sell excess renewable energy back to the grid via net metering \cite{NetMetering} and obtain corresponding revenues.
    
    \item \textbf{Offering ANcillary Service (ANS) to the grid} \cite{ghamkhari2012data}.  As data centers are massive electricity consumers, an increasing number of researchers have realized the importance of data centers as \textit{load resources} in keeping the power supply and demand balance of the grid. Equipped with the ESS, a data center is able to offer electricity to the grid in case of electricity shortage and reserve electricity in the storage in case of grid electricity surplus. By taking part in the operation of smart grid, a data center will obtain revenues from the grid. 
\end{itemize}
In this section, we will present exisiting works in profit maximization based on the source of revenues, and they are listed in Table \ref{MaxProfit}.

In \cite{kiani2016profit}, the profit maximization of geo-dispersed data centers with on-site wind power are studied. The revenues come from serving the request within the maximum delay and the probability of SLA violation is derived in closed-form via the G/D/1 queue modeling of the request arrival and service process. The problem is formulated as a convex programming with the workload distribution policy and the server service rate as decision variables. Trace-based simulation is conducted to validate the framework and it is found that the proposed method outperforms the baseline method without geographical load balancing in terms of profit maximization. In addition, profit will increase with the increasing penetration of wind power, showing the importance of renewable energy in profit maximization. 
\begin{table*}
    \centering
    \begin{tabular}{c!{\vrule width 1pt}c!{\vrule width 1pt}c!{\vrule width 1pt}c!{\vrule width 1pt}c!{\vrule width 1pt}c}
        \noalign{\hrule height 1pt}
        Reference  & Constraints & Control Knobs & Formulation & Algorithms & Achieved Results \\ 
        \noalign{\hrule height 1pt}
        
        \cite{zhang2011greenware} & \makecell[c]{QoS constraint;\\Renewable energy\\ capacity;\\Monetary cost budget} & \makecell[c]{Workload distribution to\\servers powered by\\different renewable energy} & \makecell[c]{Linear Fractional Programming\\(IFP)} & \makecell[c]{IFP solver} & \makecell[c]{42\% brown energy reduction} \\
        \hline
        
        \cite{goiri2011greenslot} & \makecell[c]{No specific constraint} & \makecell[c]{Batch job scheduling} & \makecell[c]{No specific formulation} & \makecell[c]{Least-slack\\-time-first\\ heuristic} & \makecell[c]{40\% improvement in\\ renewable energy\\ utilization;\\20\% cost saving} \\
        \hline
        
        \cite{goiri2012greenhadoop} & \makecell[c]{No specific constraint} & \makecell[c]{Hadoop job scheduling} & \makecell[c]{No specific formulation} & \makecell[c]{Least-slack\\-time-first\\ heuristic} & \makecell[c]{31\% improvement in\\ renewable energy\\ utilization;\\39\% electricity saving} \\
        \hline
        
        \cite{li2019thermal} & \makecell[c]{Inlet temperature redline;\\Job deadline;\\resource capacity} & \makecell[c]{Batch job scheduling;\\Pre-cooling} & \makecell[c]{Integer Programming} & \makecell[c]{Thermal\\-aware\\ heuristic} & \makecell[c]{Near 100\% solar\\ energy utilization;}\\
        \hline
        
        \cite{li2012iswitch} & \makecell[c]{No specific constraint} & \makecell[c]{VM placement} & \makecell[c]{Nonlinear Integer Programming} & \makecell[c]{Simulated\\ Annealing} & \makecell[c]{Over 80\% wind\\ power utilization} \\
        \hline
        
        \cite{zhao2021deep} & \makecell[c]{No specific constraint} & \makecell[c]{Task shifting} & \makecell[c]{Markov Decision Process} & \makecell[c]{Proximal Policy\\ Optimization} & \makecell[c]{26.87\% energy saving} \\
        \hline
        
        \cite{wang2021multi} & \makecell[c]{No specific constraint} & \makecell[c]{VM placement} & \makecell[c]{Multi-Agent\\ Reinforcement Learning} & \makecell[c]{MiniMax\\ Q-learning} & \makecell[c]{19\% energy saving;\\33\% carbon emission\\ reduction} \\
        \noalign{\hrule height 1pt}
    \end{tabular}
    \caption{Summary of existing works on maximizing renewable energy utilization in renewable energy-integrated data centers in terms of constraints, control knobs, problem formulation, solving algorithm, and achieved results.}
    \label{MaxRenewable}
\end{table*}
Further, some researchers consider the cooperation of smart grid to maximize the operational profits \cite{EnergyPortofolio}\cite{chen2015cooling}. Ghamkhari \textit{et al}. investigated the energy portfolio optimization of a data center in the presence of multiple energy sources \cite{EnergyPortofolio}. The operational revenues come from the fulfillment of the SLA and offering ANS to the grid. The problem is formulated as a mixed integer stochastic programming with the expected profit over a period of time as the objective. In addition, risk management is conducted by regulating the minimum profit above a predefined threshold. Simulation results show that the profit increases with more wind power turbines. In \cite{chen2015cooling}, a holistic framework was proposed, which considered the on-site renewable energy, on-site conventional generator, on-site energy storage, free air cooling and energy trading. Profits are maximized by the joint control of outside free air cooling, the storage and the conventional generator. The problem is formulated as stochastic programming over several time slots. Queue-based relaxation techniques \cite{marques2012optimal} are exploited to decouple the optimization variables at different time slots and a solver based on Lagrange Dual approach is proposed to solve the online optimization problem at each time slot. Simulation results demonstrate that the online algorithm can produce a near optimal solution compared with the optimal offline algorithm. In addition, cooling energy consumption is reduced by 35\% with combined cooling sources compared with the chiller-only cooling solution and the net-cost can be reduced by more than 60\% with on-site renewable energy. Therefore, renewable energy and free cooling should play an important role in profit maximization. 

\subsection{Renewable Energy Utilization Maximization}
Unlike the aforementioned works which improved the renewable energy utilization in an implicit way (from minimizing electricity bills or operational costs, or maximizing profits), some researchers have proposed solutions which aim to maximize the renewable energy utilization so that carbon emissions can be reduced. Related works are summarized in Table \ref{MaxRenewable}.

In \cite{zhang2011greenware}, Zhang \textit{et al}. proposed GreenWare, a renewable-energy-aware geographical workload scheduler and renewable energy manager, to maximize the renewable energy utilization with a limited budget. On-site wind and solar power generation are considered and the optimization problem is formulated as a linear fractional programming which can be solved via a standard solver \cite{lfp}. The authors found that increasing the allowable budget would improve the renewable energy utilization but with diminishing  marginal benefits. In addition, reducing brown energy consumption is not always cost-friendly since renewable energy is more expensive than brown energy. 

Different from \cite{zhang2011greenware} which considered the workload distribution for delay-sensitive tasks, the works in \cite{goiri2011greenslot} and \cite{li2019thermal} scheduled delay-tolerant jobs temporally so that the workload could be aligned with the presence of renewable energy to increase renewable energy utilization. In \cite{goiri2011greenslot}, a greedy scheduler based on renewable energy prediction was proposed. When preparing a new schedule, a task is popped out of the task queue according to the Least Slack Time First principle. The cost for executing a task at each time slot in the scheduling window (2 days) is estimated based on the price of electricity, the predicted renewable energy, and it is scheduled to the time slot with minimum cost to be executed. Future solar power is estimated as the product of the power generated in an ideal sunny day and a weather related attenuation factor (0$\sim$1). By the simulation results from a production scientific workload, GreenSlot can increase the renewable energy consumption by 117\% and reduce the total cost by 39\%. Furthermore, Goiri \textit{et al}. applied GreenSlot scheduler in Hadoop \cite{dean2008mapreduce}, resulting in a solar energy- and electricity cost-aware MapReduce framework called GreenHadoop \cite{goiri2012greenhadoop}. The Experiment results show that GreenHadoop is able to improve the utlization of solar power by 31\% and reduce electricity cost by 39\% compared with Hadoop.  In \cite{li2019thermal}, a thermal-aware batch job scheduler was proposed. A long short-term memory (LSTM) \cite{hochreiter1997long} model is used to accurately forecast the future solar power, facilitating the temporal job scheduling. By taking into account the heat recirculation model \cite{zhou2011holistic}, the inlet temperature of server racks can be estimated. After obtaining the solar power prediction, the workload and air supply temperature management are considered simultaneously. If excess solar power is available, air supply temperature is decreased to perform pre-cooling so that surplus renewable energy will not be wasted. In case of the  workload peak, if pre-cooling has been executed and the maximum inlet temperature is still below the redline, more batch jobs are allocated to provide additional computing power. Simulation results demonstrate that the proposed scheduler can effectively increase the utilization of renewable energy while decreasing the consumption of brown energy compared with the round robin or static scheduler. 

The aforementioned works only accounted for the power demand side optimization in order to improve the utilization of renewable energy. In \cite{li2012iswitch}, Li \textit{et al}. proposed iSwitch, a light-weight and simple-to-implement supply/demand cooperative VM scheduler. In order to perform power supply side optimization, the statistical characteristics of wind power generation are studied carefully and wind power generation is categorized into three regions: wind power outage region, high fluctuation region with low power production, and stable output region. On the supply side, a lazy tracking mechanism with fine time granularity was developed, which increases VMs allocation to servers connected to wind power only when wind power production was stable. On the demand side, switch assignment of each cluster is calculated by solving an optimization problem, which minimizes the standard deviation of server utilization over two consecutive time slots. Simulated annealing is used to solve the optimization and the simulation results show that it is capable of utilizing more than 80\% of wind power.

In recent years, as deep learning techniques become prevailing in data center energy management, some researchers have attempted to explore their feasibility in renewable energy utilization maximization. In \cite{zhao2021deep}, Zhao \textit{et al.} proposed using deep reinforcement learning (DRL) in workload shifting and cloud-bursting in a geographically dispersed hybrid multi-cloud environment which consists of multiple private and public clouds. The objective is to maximize the renewable energy utilization while avoiding the task deadline violation.  A model-free task scheduler based on Proximal Policy Optimization (PPO) \cite{schulman2017proximal} is proposed to either postponing the incoming task by several seconds or scheduling it into a private or public cloud which offers VMs to serve the task. Experiment results show that the task scheduler can save 26.87\% energy compared with the round robin scheduler. Unlike the centralized scheduler proposed in \cite{zhao2021deep}, Wang \textit{et al.} proposed a distributed scheme based on Multi-Agent  Reinforcement Learning (MARL) to solve the renewable energy demand-supply matching problem for a set of geo-distributed data centers in order to maximize the renewable energy utilization among multiple data centers while achieving monetary cost saving as well as carbon emission reduction. In order to implement long-term energy plan, a renewable energy predictor based on Seasonal ARIMA (SARIMA) \cite{chen2018time} is utilized to predict future renewable energy yield. Based on the prediction results and the energy demand of each data center, the Reinforcement Learning (RL) agent in each data center decides how much energy should be requested from each energy generator. Furthermore, a Deadline-Guaranteed Job Postponement (DGJP) method is also proposed to tackle the intermittent renewable energy supply by shifting the workload to the period with sufficient renewable energy according to the urgency of the workload. It is reported that the proposed scheme can reduce 19\%  monetary costs and 33\% carbon emissions.  

\textbf{Summary:} the renewable energy management problem is often formulated as a constrained nonlinear programming problem over time horizon. Lyapunov Optimization is leveraged to overcome the uncertainty of future information. Heuristic based algorithms such as the simulated annealing, the genetic algorithm, and the greedy algorithm, are frequently leveraged to solve the optimization problem. Recently, reinforcement learning is emerging in the maximizing renewable energy utilization for geographical-dispersed data centers and achieve significant progress in this area. 

\section{carbon-neutral data center: From Energy Utilization Perspective}\label{Energy Utilization}

As shown in Fig. \ref{CarbonFootprintShare}, carbon emissions from daily operations (energy utilization) accounts for 97\% of a data center's total carbon footprint. Hence, significant carbon emission reduction is possible by improving data center energy efficiency. The electricity drawn from the ICT system and the cooling system accounts for approximately 86\% of total energy consumption in a data center \cite{zhang2016towards}. Therefore, greater efforts should be made to improve energy efficiency of these two systems.

\begin{figure}[t]
	\centering
	\includegraphics[width=.47\textwidth]{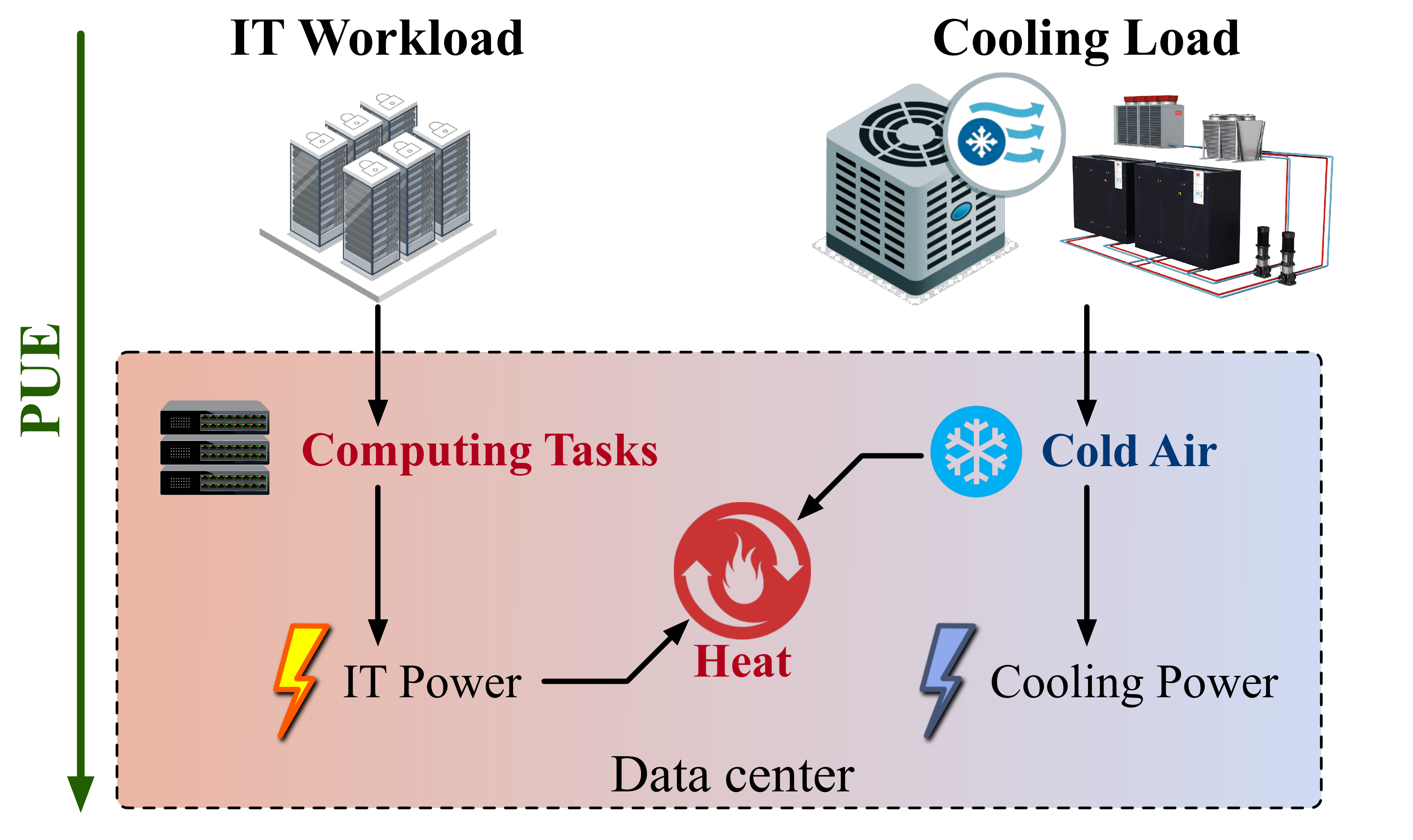}
	\caption{Illustration of the coupling of the IT and cooling system in a data center. The IT and cooling systems are coupled by the heat dissipation and rejection process within the data hall. A joint optimization of both systems is preferred to improve PUE.}
	\label{CoupleITCooling} 
\end{figure}

Many existing works optimize the energy efficiency of either the ICT system or the cooling system. However, as stated in \cite{zhang2016towards}, heat connects the IT and cooling systems, and the relationship between the IT and cooling systems is depicted in Fig. \ref{CoupleITCooling}. In this regard, some researchers also considered the joint optimization of the IT and the cooling systems to further improve energy efficiency.
In this section, we will first discuss the techniques that can be utilized in ICT and cooling system management. Subsequently, existing literature in energy-efficient computing and cooling, as well as joint optimization of the ICT and the cooling system, will be reviewed.  

\subsection{Enabling Techniques}
The management of the ICT system will determine electricity consumption as well as heat generation in a data center. The cooling system is utilized to reject heat in the data hall, making the environment suitable for IT devices. The cooling system management is responsible for ensuring the safe operation of IT devices while minimizing the electricity required to power them. In this section, we will divide the related techniques into two classes, i.e., IT management techniques and cooling management techniques. 
\begin{figure*}[t]
	\centering
	\includegraphics[width=1.\textwidth]{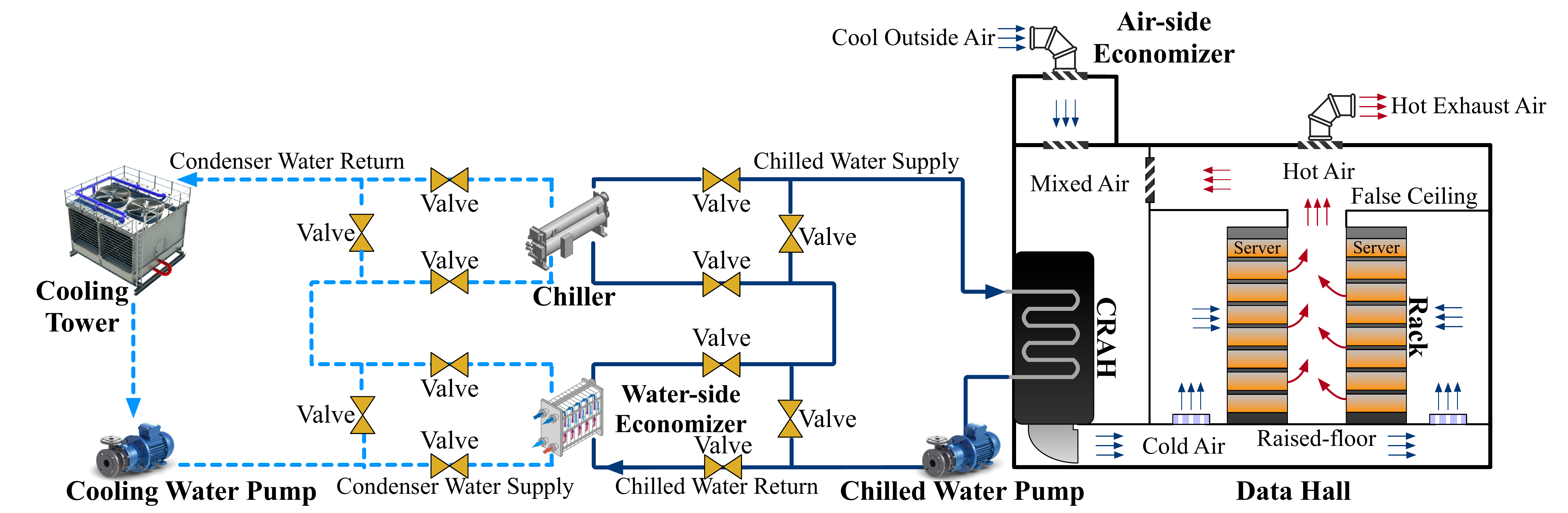}
	\caption{Illustration of a typical chilled-water air-cooled data center with a water-side economizer in the chiller plant and an air-side economizer in the data hall. The chiller is used for providing chilled water and the condenser water is cooled down by the cooling tower. The water-side economizer reduces cooling load of the chiller with outside cold water. The air-side economizer pumps outside cool and dry air into the data hall to provide free air cooling.}
	\label{DataHallChillerPlant} 
\end{figure*}
\subsubsection{ICT Management Techniques}
From the ICT system stand point of view, the techniques can be classified into three scales, i.e., server level, rack level and data center level. In this survey, we brief cover the basic concept of these energy-saving techniques and an detailed review on data center power modelling can be found in \cite{dayarathna2015data}. 

Dynamic Voltage Frequency Scaling (DVFS) on chips is commonly utilized at the server level. Chip energy consumption can be reduced significantly by dynamically adjusting the voltage and operating frequency of the chips. When a CPU operates at a frequency $f$, the energy consumption is proportional to $f^3$, whereas the task executing time is roughly proportional to $f$ for CPU-intensive tasks \cite{ge2007cpu}. Therefore, many researchers aim to significantly minimize the chip's energy consumption by slightly sacrificing task execution time. In addition, if the working frequency of I/O-intensive tasks decreases during memory access, a significant amount of energy can be saved as well.  

At the rack level, ICT energy consumption comes from both servers and networking devices such routers and switches. As reported in \cite{dayarathna2015data}, servers account for the majority of  total energy consumption. As a result, we will concentrate on techniques that improve a server's energy efficiency in this section, and the discussion on energy-efficient data center networking is out of the scope of this survey paper. Indeed, energy-aware communications and networking is also an active research field and judicious resource allocation \cite{Zhang2020Distributed} and load balancing \cite{LoadBalancingNetwork} can significantly improve the energy efficiency of the data center network. For example, Zhang \textit{et al.} present an excellent survey on the load balancing mechanisms in data center networks \cite{LoadBalancingNetwork}. They summarized the energy consumption of data center networks, and they also reviewed several load balancing schemes such as Data center Energy-efficient Network-aware Scheduling (DENS) \cite{kliazovich2013dens} which take data center network energy efficiency into consideration. We also refer to \cite{bilal2013green, DCArchitectureDesign} for other energy-saving techniques for networking devices such as adaptive link rate.

The method of distributing workload among servers and server power mode management are two major factors that influence the energy consumption of a server rack. In this regard, techniques such as dynamic server provisioning, workload dispatching, and VM migration can be exploited to reduce total energy consumption. 

Dynamic server provisioning means to consolidate the workload into a set of properly selected servers and other servers can be turned off to reduce total energy. It is important to note that turning on/off servers or changing the power mode will degrade the lifetime of a server and increase failure rate. Therefore, switching cost should be considered when dynamic server provision is applied \cite{guenter2011managing}.

Another important technique in rack level energy-efficient computing is workload dispatching, also known as task scheduling. On the one hand, workload dispatching can balance the workload distribution across multiple servers while avoiding local hotspots. On the other hand, workload can be purposefully assigned to servers with higher energy efficiency to reduce total energy consumption. The difference between dynamic server provisioning and workload dispatching is that the former turns off some unnecessary servers, whereas the latter just simply adjusts server utilization. Besides workload dispatching, workload can also be migrated from one server to another at a low cost using the live VM migration techniques \cite{le2020survey} in a virtualized environment. A running task can be migrated to another server using such techniques in the event of a thermal emergency or in the pursuit of total energy reduction. 

At the data center level, in addition to dispatching jobs within a single data center, jobs can be distributed among geo-dispersed data centers to reduce electricity costs because electricity prices vary spatiotemporally across multiple sites. 

\subsubsection{Cooling Management Techniques}
It's a common practice for data center operators to keep the server racks in an over-cooled ambient environment to avoid server overheating. However, it may waste a large amount of electricity to keep the data hall at an unnecessarily low temperature. In contrast, if the temperature in the data hall rises due to insufficient cooling power, it may cause local hot spots where server temperature exceeds the redline, leading to thermal emergency. Hence, managing the cooling system is also critical for energy-efficient data centers. A typical chilled-water air-cooled data center with the air-side and water-side economizer is illustrated in Fig. \ref{DataHallChillerPlant}, which includes a data hall with a air-side economizer and a chiller plant with a water-side economizer.

At server or rack level, fans attached to the server or the rack can be controlled to reject heat in an energy-efficient manner. The fans circulate air, carrying heat away from the server or rack. All heat generated by the server can be removed as the fan speed increases, and the server temperature will not rise \cite{piatek2015modeling}. As a result, the leakage power which monotonically increases with the server temperature will not increase, leading to less server power consumption. Otherwise, the server temperature will rise due to insufficient heat rejection by fans, and more cooling power will be required to remove the dissipated heat. It is worth noting that fan power consumption is proportional to the cubic of the fan speed, implying that there is a trade-off between power consumption of fans, servers and the cooling system. 

At the data hall level, several techniques can be applied to improve energy efficiency. These techniques can be coarsely categorized into two classes: (1) static optimization which optimizes the layout or configuration of the data hall as well as the server placement in the planning phase; (2) dynamic optimization which dynamically controls the Computer Room Air Conditioning (CRAC) inside a data hall or the outside chiller plant in the operation phase. 

As for the static optimization, data center operators are able to choose cooling systems with higher energy-efficiency to save cooling energy. Nowadays, legacy data centers usually adopt air-cooling system with raised floor layout, which is convenient for implementation and maintenance. However, the most significant drawback of such a system is heat recirculation, i.e., a portion of exhausted hot air fails to return to the hot aisle and mixes with the supply cold air in the rack inlet, which declines the cooling efficiency by a large margin \cite{HAR2018}. To tackle this problem, several other layouts have been proposed and the basic idea of these layout is to prevent dissipated hot air from intermingling with cold supply air via hot aisle containment \cite{HotColdAsile}, row-level air cooling \cite{lin2014row,wang2017airflow} and in-rack air cooling \cite{dunlap2012choosing}. As for the hot aisle containment, the hot aisle is sealed by installing enclosure between the rear of a server rack and the ceiling roof, as shown in Fig. \ref{DataHallChillerPlant}. Under such a layout, heat recirculation will be prevented to a large extent. For other data hall air flow management techniques with air cooling such as in-rack cooling, please refer to \cite{zhang2021survey} for detailed discussion. Moreover, data center operators can also choose cooling systems with higher energy efficiency such as liquid cooling systems \cite{li2015current} to achieve further cooling energy reduction. Since liquid has much higher specific heat capacity than air, liquid cooling has superior cooling efficiency and is suitable for future data centers with growing energy density. As for the liquid cooling system, it can be classified into direct liquid cooling and indirect liquid cooling \cite{zhang2021survey}. Direct liquid cooling means that dielectric liquid coolant contacts the electronic devices directly to absorb heat dissipated by them, e.g., immsersion cooling \cite{ImmersionCooling}. Since devices attach liquid coolant directly, there is no requirement for pipes or enclosure, which is highly convenient for data center operations. On the contrary, for indirect liquid cooling, hot exhausted air contacts the pipe or the cold plate with liquid coolant flowing inside and the heat is absorbed by the liquid coolant in an indirect way. Because the liquid coolant is not dielectric, indirect cooling may suffer from server shutdown due to liquid leakage. 

Besides choosing more energy-efficient cooling systems, nowadays increasing number of data centers are equipped with Thermal Energy Storage (TES) tanks to store cheap electricity or unstable renewable energy \cite{zheng2014exploiting, brannvall2020edge, ding2019integrated}. When discharging, the cold energy stored in the cold water or ice in the tank is leveraged to cool the warm return water, which reduces the cooling load of the chiller. In terms of recharging, the chiller utilizes the cheaper electricity or excess renewable energy to provide additional cold water or ice, which is charged into the tank. The merits of the TES compared to the ESS are three folds \cite{zheng2014exploiting}. Firstly, it is more environmental-friendly with less recycling costs. Second, it has a longer life time (typically 20-30 years). In addition, it is more suitable for long-lasting peak workloads since the batteries are designed to support the uninterrupted operation for about 3-5 minutes. 

Since the IT system and the cooling system are coupled,  server racks can be designed with high cooling efficiency to save cooling energy. For example, Low Power (LP) blade design that leverages low power mobile processors for handling I/O-bounded cloud workloads has been studied and the results showed promising electricity consumption saving without sacrificing the service quality \cite{lim2008understanding, hamilton2009cooperative, krioukov2010napsac}. In addition, some researchers designed dematerialized server racks to reduce energy consumption of server fans. The cooling for the entire rack is provided by a single array of shared fans along one side of the rack, which significantly reduce fan power consumption. Moreover, the servers are attached to a shared perforated spine for highly efficient air cooling \cite{chang2012totally}. Besides the server rack design, server placement can also be optimized by considering the heat recirculation and workload variation \cite{azimi2014thermal}.

In terms of dynamic optimization, there are several techniques that allow us to control the cooling system reactively or proactively. To save cooling power for an air-cooling system, operators are able to dynamically adjust the vent tile opening, CRAC blower rotational speed and air supply temperature. The opening of a vent tile affects server racks near a vent tile because it will affect air flow rate around the server racks. Such adjusting can be achieved via Adaptive Vent Tile (AVT) \cite{zhou2011holistic}. The rotational speed of the CRAC blower can be controlled by a Variable Frequency Drive (VFD) \cite{zhou2011holistic} which affects the air flow rate. By increasing the CRAC blower rotational speed, air flow rate will increase correspondingly, resulting in higher cooling capacity. Hence, operators can dynamically regulate the CRAC blower rotational speed to meet the cooling load. In addition, operators can also change the air supply temperature by controlling the supply chilled water temperature to keep data hall temperature  below the redline at all times. As CRAC power consumption is negatively correlated with air supply temperature \cite{pakbaznia2009minimizing}, overprovisioning will result in unnecessary CRAC power consumption. As a result, prudent air supply temperature control can achieve significant cooling energy savings, as demonstrated by \cite{zhou2011holistic,li2019transforming}. 

\begin{figure}[t]
	\centering
	\includegraphics[width=.4\textwidth]{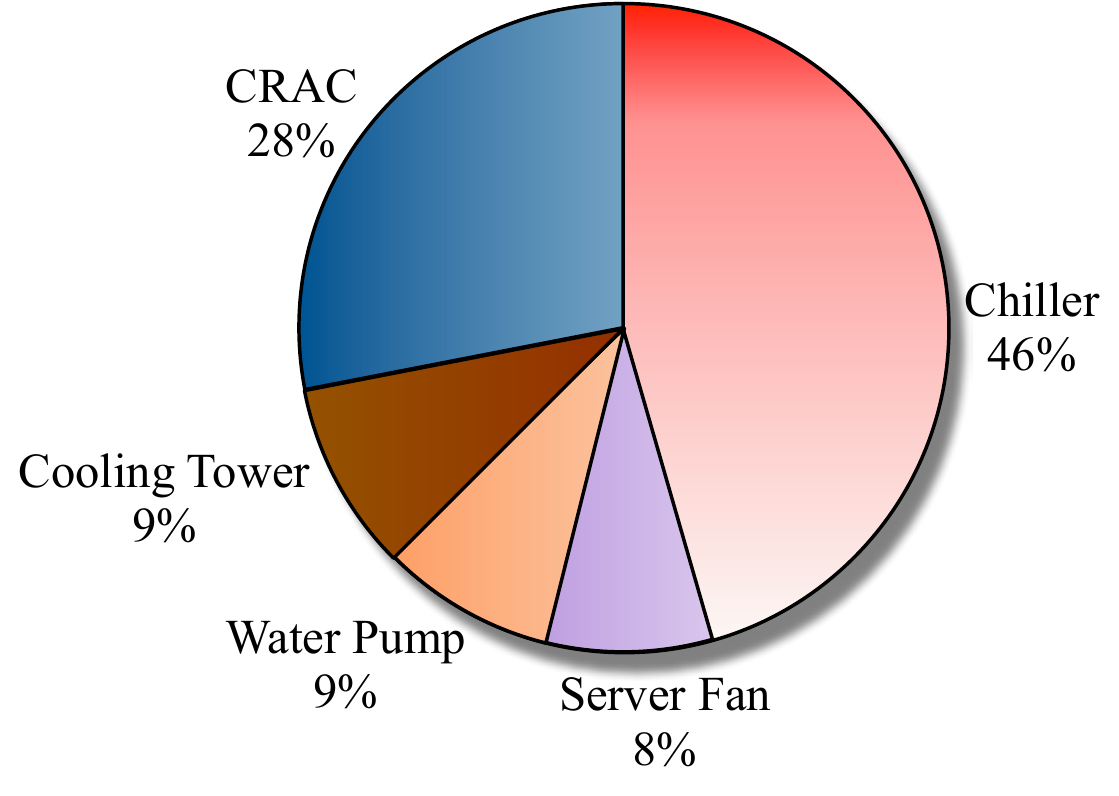}
	\caption{Power consumption of different components of a chilled-water air cooling system \cite{arghode2016air}. Power consumption of a chiller accounts for around 50\% of total cooling energy. Another major cooling energy consumer is the CRACs in the data hall.}
	\label{CoolingSystemBreakDown} 
\end{figure}

Aside from CRAC units, another important component in the data center cooling system is the chiller plant which includes chillers and cooling towers as shown in Fig. \ref{DataHallChillerPlant}. Typically, the chiller plant consumes about 50\% total cooling energy \cite{arghode2016air}, as shown in Fig. \ref{CoolingSystemBreakDown}.
For a chiller, its power consumption relates to the temperature of supply chilled water and supply condenser water, the flow rate of the chilled water and condenser water, as well as the partial cooling load \cite{yu2008optimization}. In terms of the cooling tower, its power consumption is related to the ambient weather (temperature, humidity, etc), temperature and flow rate of condenser return water, and the temperature set point of the condenser supply water \cite{yu2008optimization}. Typically, these knobs can be controlled via  model-based\cite{ahn2001optimal,yu2008optimization} and  data-driven\cite{vu2017data,zheng2018data,chen2019gnu} approaches.  

Besides controlling CRAC units and chiller plants, free cooling has gained increasing attention in recent years. Free cooling can be categorized into two classes, namely free air cooling (air-side economizer) \cite{liu2012renewable,xu2013temperature, li2014coordinating,ji2020energy} and free liquid cooling (water-side economizer) \cite{li2020model}. These two kinds of economizers are illustrated in Fig. \ref{DataHallChillerPlant}. For the free air cooling, outside cool air is pumped into the data hall directly and the requirement for a chiller plant to provide chilled water to cool the hot return air is partially or fully eliminated. In terms of free liquid cooling, when the cooling tower outlet water temperature is lower than the return chilled water temperature, the water-side economizer can be leveraged to precool the return chilled water so that the cooling load of the chiller will be reduced. Further, if the cooling tower outlet water temperature is even lower than the chilled water supply temperature, the chiller can be turned off and the full cooling load is satisfied with the water-side economizer, leading to significant cooling energy saving. Compared to traditional chilled-water cooling systems where the chiller consumes considerable energy, free cooling reduces the demand for the chiller and saves massive energy, as presented in \cite{liu2012renewable,xu2013temperature}. However, the major drawback of free cooling is that it is dependent on the outside weather and can only be utilized when the outside temperature and humidity are suitable for cooling data halls. 

\subsection{Energy-Efficient Computing}
From the ICT side, a plethora of works have been conducted to improve the energy efficiency of ICT devices on multiple scales. In this section, we present existing works on energy-efficient computing based on the control knobs that they employ, which are summarized in Table \ref{EnergyEfficientComputing}.

\begin{table*}
    \centering
    \begin{tabular}{c!{\vrule width 1pt}c!{\vrule width 1pt}c!{\vrule width 1pt}c!{\vrule width 1pt}c!{\vrule width 1pt}c}
        \noalign{\hrule height 1pt}
        Reference  & Constraints & Control Knobs & Formulation & Algorithms & Achieved Results \\ 
        \noalign{\hrule height 1pt}
        
        \cite{ge2007cpu} & \makecell[c]{Performance loss} & \makecell[c]{CPU DVFS} & \makecell[c]{No specific formulation} & \makecell[c]{Closed-form solution} & \makecell[c]{20\% energy saving}\\
        \hline
        
        \cite{dhiman2007dynamic} & \makecell[c]{Service delay} & \makecell[c]{CPU DVFS} & \makecell[c]{Online learning} & \makecell[c]{Boosting} & \makecell[c]{up to 49\%\\ energy saving}\\
        \hline
        
        \cite{ghasemazar2010minimizing} & \makecell[c]{Throughput} & \makecell[c]{CPU DVFS;\\CPU core on/off;\\Task assignment} & \makecell[c]{Mixed Integer \\ Linear Programming} & \makecell[c]{Multi-tier PI control} & \makecell[c]{17\% energy saving}\\
        \hline
        
        \cite{wang2008power} & \makecell[c]{Response time } & \makecell[c]{VM placement;\\CPU DVFS} & \makecell[c]{Feedback control} & \makecell[c]{LQR controller;\\PI controller} & \makecell[c]{Higher energy efficiency;\\
        Stabilized response time}\\
        \hline
        
        \cite{wang2010coordinating} & \makecell[c]{CPU frequency bound;\\Total power consumption} & \makecell[c]{CPU time allocation;\\CPU DVFS} & \makecell[c]{Feedback control} & \makecell[c]{MPC Controller;\\PID controller} & \makecell[c]{stabilized respond time}\\
        \hline
        
        \cite{meisner2009powernap} & \makecell[c]{No specific constraints} & \makecell[c]{Server power mode} & \makecell[c]{No specific formulation} & \makecell[c]{Greedy Heuristic} & \makecell[c]{Significant energy\\ saving gain compared\\ to optimal DVFS}\\
        \hline
        
        \cite{wang2011powersleep} & \makecell[c]{Response time} & \makecell[c]{Server power mode;\\CPU DVFS} & \makecell[c]{Stochastic Programming} & \makecell[c]{closed-form solution} & \makecell[c]{Significantly higher\\ energy efficiency\\ compared to \cite{meisner2009powernap}}\\
        \hline
        
        \cite{krioukov2010napsac} & \makecell[c]{No specific constraints} & \makecell[c]{Load balancing;\\Server Provision} & \makecell[c]{No specific formulation} & \makecell[c]{Greedy knapsack for\\ dynamic provision;\\Weighted round robin\\ for load balancing} & \makecell[c]{63\% energy saving}\\
        \hline
        
        \cite{guenter2011managing} & \makecell[c]{Server conservation} & \makecell[c]{Server power mode;\\Server provision} & \makecell[c]{Integer Linear Programming} & \makecell[c]{ILP Solver} & \makecell[c]{considerable energy\\ saving with\\ server provision}\\
        \hline
        
        \cite{kusic2009power} & \makecell[c]{Server capacity;\\Minimum VM resource} & \makecell[c]{Server provision;\\VM placement;\\Workload distribution} & \makecell[c]{Limited Lookahead\\ Control (LLC)} & \makecell[c]{LLC Solver} & \makecell[c]{26\% energy saving}\\
        \hline
        
         \cite{rao2010minimizing} & \makecell[c]{SLA constraint} & \makecell[c]{Server provision;\\Geographical workload\\ distribution} & \makecell[c]{Minimum cost flow} & \makecell[c]{Fast polynomial\\ time solver} & \makecell[c]{30\% energy cost saving}\\
        \hline
        
        \cite{abbasi2011dynamic} & \makecell[c]{QoS constraint;\\idle power;\\VM capacity} & \makecell[c]{Online user association} & \makecell[c]{Mixed Integer Programming} & \makecell[c]{Greedy Heuristic} & \makecell[c]{14\% energy cost saving}\\
        \hline
        
        \cite{Zhou2013} & \makecell[c]{Carbon capping;\\Service delay} & \makecell[c]{Geographical workload\\ distribution;\\Server provision;\\Server speed scaling} & \makecell[c]{Lyapunov Optimization} & \makecell[c]{Generalized Bender\\ Decomposition} & \makecell[c]{5\% carbon emission\\ reduction with 3\%\\cost rising}\\
        \hline
        
        \cite{doyle2013stratus} & \makecell[c]{No specific constraints} & \makecell[c]{Geographical workload\\ distribution} & \makecell[c]{Graph Partition} & \makecell[c]{Voronoi Partition} & \makecell[c]{26\% carbon emission\\ reduction;\\30\% electricity\\ cost saving}\\
        \hline
        
        \cite{liu2017hierarchical} & \makecell[c]{No specific constraints} & \makecell[c]{VM mapping;\\ Server timeout} & \makecell[c]{Continuous time event-driven\\ semi Markov Decision\\ Process (SMDP)} & \makecell[c]{DQN with LSTM \\workload predictor} & \makecell[c]{26\% carbon emission\\ reduction;\\53.97\%  energy saving}\\
        \hline
        
        \cite{yi2019toward} & \makecell[c]{No specific constraints} & \makecell[c]{Job placement} & \makecell[c]{Markov Decision Process} & \makecell[c]{DQN with LSTM\\ state predictor} & \makecell[c]{Significant cooling\\ energy saving}\\
        \noalign{\hrule height 1pt}
    \end{tabular}
    \caption{Summary of existing works on energy-efficient computing in terms of constraints, control knobs, problem formulation, solving algorithm, and achieved results.}
    \label{EnergyEfficientComputing}
\end{table*}

\subsubsection{Chip Level Optimization}
At the server level, a slice of researchers leverage DVFS to reduce CPU energy consumption. In \cite{ge2007cpu}, Ge \textit{et al}. proposed CPU MISER, a real-time DVFS scheduler, to minimize energy drawn by the CPU while maintaining service performance. It is well known that a task's lifecycle can be either CPU-intensive or memory-intensive. The intuition is to reduce CPU working frequency when the task is not in the CPU-intensive phase. The CPU intensity is forecasted using a time series analysis model named RELAX, which predicts future workload via both historical data and run-time profiling. Given the predicted CPU intensity and maximum performance loss, the minimum CPU working frequency is obtained with a closed-form solution. Experiment results show that the proposed scheduler is able to save more than 20\% energy. \cite{dhiman2007dynamic} investigated a similar problem and proposed a solution based on online learning. In this framework, CPU intensity is evaluated by Cycles Per Instruction (CPI) which can be obtained from the Performance Monitoring Unit of a CPU. Different $v-f$ pairs are regarded as experts with bundled working frequency. After obtaining the CPU intensity, the performance loss and the energy loss of each $v-f$ pair is evaluated and its weight is calculated in closed form. The $v-f$ pair with highest weight is selected to be implemented in the current tick. Furthermore, the authors also showed that provable convergence was achievable with finite scheduler ticks. Experiment results show that the scheduler can reduce a CPU's energy consumption by up to 49\%. Beyond modulating the frequency and voltage of a single CPU, Ghasemazar \textit{et al}. proposed a hierarchical scheduler for a multi-core system that controlled CPU on/off, DVFS for each working CPU and task assignment to each working CPU \cite{ghasemazar2010minimizing}. The objective is to minimize the energy consumption while maintaining the throughput above a certain threshold. These three types of decisions are made at different time granularities: a) decision time granularity of CPU switch is the largest, the b) decision time granularity of DVFS is the second, and c) decision time granularity of task assignment is the smallest. The decision to turn on or off CPU cores are made based on a local greedy search, which compares the energy consumption of turning on one more core versus turning off a running core given current DVFS state and task assignment decision. DFVS controllers for running cores are feedback-controlled Proportional-Integral (PI) controllers \cite{richard2008modern}. The task assignment problem is formulated a Knapsack problem and solved via dynamic programming. In comparison to the baseline policy which always turns on all cores, the proposed scheduler saves 17\% of energy. 

As virtualization technology becomes more prevalent in data center industry, some researchers also investigate the potential of applying DVFS in virtualized servers. In \cite{wang2008power}, Wang \textit{et al}. proposed a two-layer control hierarchy for a virtualized server, with a the primary control loop maintaining load balancing among all VMs and a secondary control loop minimizing energy consumption by scaling CPU frequency. The primary loop is built on a well-established multiple-input-multiple-output (MIMO) controller. To design the MIMO controller, a parametric model with respective to the control error and the resource allocation weight for each VM is first established via system identification, and Linear Quadratic Regulator (LQR) \cite{LQR} is adopted subsequently. Because a simplified linear model is adopted in the secondary loop, PI control is utilized for its robustness against system modeling error. Experiment results show that the proposed closed-loop controller consumes significantly less energy than the open-looped controller. Similar to \cite{wang2008power}, a control-theoretical approach was also leveraged to stabilize the response time of each VM and the power consumption of each server around the set point for a virtualized server clusters \cite{wang2010coordinating}. In this work, the response time controller is a Proportional-Integral-Derivative (PID) controller \cite{franklin1998digital} and the power control is based on Model Predictive Control (MPC) \cite{maciejowski2002predictive}. It is reported that the proposed hierarchical controller is able to effectively stabilize both response time and power consumption. 

\subsubsection{Server Power Mode Optimization}
In addition to controlling CPU frequency as a energy-saving approach, other researchers propose to manage the server power state to improve energy efficiency. Due to the lengthy setup time, data center administrators hesitate to implement server power management technology, fearing that it will severely degrade service quality \cite{gandhi2011case}. In \cite{gandhi2011case}, the authors conducted an empirical study on the relationship between server setup time and the server sleep power, and they found that setup time decreases monotonically with increasing server sleep mode power. Inspired by these findings, the researchers investigated the feasible regime of sleep power if server power state management was applied in the data center. Two algorithms, AlwaysOn and Reactive \cite{krioukov2010napsac} are evaluated. The AlwaysOn algorithm ensures that servers remain in operation regardless of workload fluctuations. The number of active servers in the Reactive algorithm is dynamically adjusted to meet varying workloads. It is demonstrated  that for some server sleep mode power, the Reactive algorithm outperforms the AlwaysOn algorithm by a large margin. Furthermore, the authors asserted that the feasibility of server power management relied on the workload statistics and power management was preferable for slowly varying workloads. In \cite{meisner2009powernap}, a simple energy-saving solution called PowerNap was proposed, in which the system rapidly switched between the low-power idle state and high-performance active state. The basic idea is to idle the server when there are no incoming requests and the task queue is empty, and then activate the server back into a high-performance mode when a request arrives. PowerNap is modeled using an M/G/1 queue model, and average power consumption is calculated in closed form. Furthermore, as a baseline for comparing the energy-saving capability of server provision and DVFS, an ideal DVFS controller with maximum energy-saving potential is introduced. Trace-based simulation results show that PowerNap outperforms DVFS significantly in terms of energy savings because turning the server into idle mode reduces energy consumption of all components in the server, whereas DVFS only reduces CPU power consumption. However, PowerNap does not account for the performance loss caused by frequent power mode transitions. In \cite{wang2011powersleep}, Wang \textit{et al}. combined server power mode management with DVFS to further reduce server energy consumption and propose a controller named PowerSleep. Unlike \cite{meisner2009powernap} where the server entered the high-performance mode as soon as a new request arrived and returned to the idle mode as soon as the request queue was empty, PowerSleep introduced three temporal parameters called idle period threshold, sleep period threshold, and procrastination period threshold. The server will not enter the idle mode until the request queue is empty for a period of time (idle period threshold). Similarly, if a new request arrives while it is in sleep mode, it will wait for a certain amount of time (procrastination period threshold) before switching to the active mode. In addition, the maximum sleep time (sleep period threshold) is optimized so that the server can response to incoming requests quickly. Furthermore, PowerSleep also takes DVFS into account and introduces a server speed scaling parameter to further reduce energy consumption. A M/G/1/PS with Starter queue model \cite{levy1986queue} is utilized to relate the average energy consumption with these four parameters and the optimal parameters are given in a closed form solution. PowerSleep outperforms PowerNap in terms of energy savings because it reduces unnecessary server power mode switches and integrates with DVFS for additional energy savings. 

\subsubsection{Cluster Level Optimization}
A host of researchers also attempt to perform provisioning for a cluster of servers in addition to a single server. In \cite{krioukov2010napsac}, Krioukov \textit{et al}. proposed a cluster manager named NapSAC for a heterogeneous server cluster that included servers, mobile devices and embedded devices. The system designer aims to minimize energy consumption by adjusting the power mode of each device. The basic idea behind NapSAC is to handle the majority of workload with high performance hardware while cooperating with edge devices to satisfy burst workloads by adjusting the power mode of each device. The cluster manager is divided into two parts: one for dynamic provisioning and another for load balancing. Dynamic provisioning is modeled as a Knapsack problem and solved using the greedy heuristic. Load balancing is accomplished through weighted round robin algorithm, with each device's weight proportional to its computing power. Compared to the baseline method without dynamic provisioning, NapSAC can achieve 63\% energy savings. By jointly considering the energy saving, transition costs and reliability costs, Guenter \textit{et al}. proposed an automated server provisioning system named ACES with the aim of minimizing energy consumption while meeting workload demands within a period of time \cite{guenter2011managing}. ACES is formulated as a Markov model that can be unrolled over time, resulting in a machine state transition graph with the node representing the number of servers in a specific power state (on/off) at a given time slot and the edge representing the costs of state transition. The costs include latency costs associated with server power state transitions as well as maintenance costs equaling to the server procurement costs divided by its duty cycles. Because of state transition, the optimization is coupled temporally, and thus future workload should be known in order for the problem to be tractable. To predict the future workload, a seasonal linear regression model is established. Due to the special structure of the problem,  integer linear programming can be relaxed to linear programming and solved with a well-established solver. Trace-based simulation results show that the proposed predictive provisioning method saves considerably more energy than the one without workload prediction. Similar to \cite{guenter2011managing}, Kusic \textit{et al}. also adopted predictive provisioning and proposed a two-layer VM scheduler to minimize energy consumption while fulfilling QoS constraints \cite{kusic2009power}. At a coarse time scale, the scheduler should decide on the active server and VM, while workload distribution and CPU share of each VM are determined at a fine time scale. The Limited Lookahead Controller (LLC) \cite{abdelwahed2004online} is adopted to perform predictive provision, and the workload is forecasted via Kalman Filter \cite{harvey1990forecasting}. Experiment results show that the proposed scheme can save up to 26\% of energy. 

Due to the complicated workload pattern in the real world, such prediction-based approaches seem to be unrealistic in practice and thus, it is preferable to design an online adaptive automatic decision-maker. In this regard, some researchers resorted to Reinforcement Learning (RL) \cite{sutton2018reinforcement}, a model-free online decision making algorithm to improve energy efficiency of a cloud computing system. In \cite{liu2017hierarchical}, a hierarchical framework for resource allocation and server power management was proposed in a virtualized data center based on deep reinforcement learning (DRL) \cite{mnih2015human}. The framework includes a global tier for allocating VM resources to servers and a local tier for distributed power management. The global tier VM resource allocator employs a continuous-time and event-driven decision making framework, with the control acting only when a VM request arrives. The continuous-time Q-learning for Semi Markov Decision Process (SMDP) \cite{bradtke1995reinforcement} is chosen as the underlying DRL technique. When a new VM request arrives, the global controller's state space is the union of all server clusters' states. The action decides which server will serve the new request. The reward is a linear combination of metrics for power consumption, VM latency and reliability. In order to cope with the high dimensionality of the state space, an autoencoder is utilized to compress the state vector of each server cluster and the weights of each autoencoder are shared to alleviate the training burden. In the case of local server power management, a similar event-driven DRL technique is leveraged and a LSTM predictor is trained for each server to perform request inter-arrival time estimation. The action is the optimal waiting time for a server when its task queue is empty. According to the simulation results, the proposed hierarchical DRL framework achieves up to 50\% energy saving compared to the round robin scheduler. In addition, the local dynamic server power manager achieves a better trade-off between power consumption and service latency compared to the fixed waiting time policy. Similarly, Yi \textit{et al}. investigated the computation-intensive job allocation problem via DRL technique \cite{yi2019toward}. In this work, server temperature is also included in the state space so that the DRL agent can be trained to avoid thermal emergencies. Typically, the temperature distribution after executing a job allocation decision is obtained via computational fluid dynamics (CFD) simulation \cite{chen2012high}, which introduces significant computational overhead. To address this problem, the authors design a cascaded temperature and utilization prediction model based on the LSTM model to estimate the server utilization and temperature given current states and the actions. The offline trained cascaded LSTM model mimics the thermal and ICT dynamics in a data center and enables the training of the DRL agent. Trace-based simulation results
demonstrate that the DRL agent significantly outperforms the control-theoretical solution and baseline round robin scheduler in terms of energy savings. 
\begin{table*}
    \centering
    \begin{tabular}{c!{\vrule width 1pt}c!{\vrule width 1pt}c!{\vrule width 1pt}c!{\vrule width 1pt}c!{\vrule width 1pt}c}
        \noalign{\hrule height 1pt}
        Reference  & Constraints & Control Knobs & Formulation & Algorithms & Achieved Results \\ 
        \noalign{\hrule height 1pt}
        
        \cite{wang2009optimal} & \makecell[c]{CPU temperature redline} & \makecell[c]{Fan speed} & \makecell[c]{Convex Programming} & \makecell[c]{CVX} & \makecell[c]{20\% fan energy reduction}\\
        \hline
        
        \cite{piatek2015modeling} & \makecell[c]{No specific constraints} & \makecell[c]{Fan speed} & \makecell[c]{No specific formulation} & \makecell[c]{Lookup Tables} & \makecell[c]{About 10\% total\\ energy saving}\\
        \hline
        
        \cite{piatek2017intelligent} & \makecell[c]{CPU temperature redline} & \makecell[c]{Fan speed} & \makecell[c]{Integer Programming} & \makecell[c]{Brute force search} & \makecell[c]{10\% total energy saving;\\80\% cooling\\ energy saving}\\
        \noalign{\hrule height 1pt}
    \end{tabular}
    \caption{Summary of existing works on server fan control in terms of constraints, control knobs, problem formulation, solving algorithm, and achieved results.}
    \label{FanSpeedControl}
\end{table*}
\subsubsection{Geographical Distributed Data Centers Optimization}
Cloud data centers are typically diverse in terms of energy efficiency, electricity prices, and carbon emission factors, and many researchers propose leveraging spatiotemporal diversity among a cluster of data centers to minimize energy-related metrics such as energy consumption, energy costs and carbon emissions while satisfying QoS constraints. In \cite{rao2010minimizing}, Li \textit{et al}. attempted to utilize spatiotemporal variation in electricity prices to minimize energy procurement costs of a cluster of data centers by judiciously deciding the number of active servers in each data center and the workload distributed to each data center. For request delay, M/M/n queue modeling is utilized, and the problem is formulated as a mixed integer linear programming. A two-layer heuristic solver is proposed to solve the problem efficiently. The solver first solves the workload distribution with a standard linear programming solver, and then the minimum number of active servers is derived based on workload distribution results with M/M/n queue modeling. The proposed framework can reduce electricity costs by up to 30.15\%, according to simulation results. In \cite{abbasi2011dynamic},  Dynamic Application Hosting Management (DAHM) was considered in a virtualized cloud where the manager should decided the online user association to each data center, and the number of VMs hosting each application in each data center. The goal is to minimize the total cost including energy consumption costs,  SLA violation costs, and VM migration costs. VM migration costs are proportional to the number of VMs migrated in a time slot. The cost of SLA violation is the sum of punishment values for each user whose delay requirement is not satisfied. Since the migration cost is non-zero, the optimal DAHM solution is only possible if the manager is aware of all future information. Hence, a greedy heuristic algorithm which always assigns users in an area to a data center with the lowest cost is proposed to solve DAHM in an online manner. Trace-based simulation results show that the proposed greedy solver can save 20\% of the cost compared to the cost-oblivious solution which assigns users to the data center with the least amount of delay. 

Furthermore, some researchers consider carbon emission issues as part of the global workload management problem. In \cite{zhou2013carbon}, Zhou \textit{et al}. considered the electricity cost minimization problem under the carbon emission capping constraint by jointly controlling the global workload distribution across multiple data centers, the number of active servers in each data center, and the server speed scaling in each data center. Carbon emission constraint is formulated as a long-term time-averaged carbon emission budget. The problem is expressed as a Lyapunov Optimization and the asymptotic optimal policy is obtained by solving the linear programming problem in each time slot. Simulation results show that there exists a clear trade-off between electricity cost savings and carbon emission reduction: electricity costs decrease monotonically as carbon emission budgets increase. Different from \cite{zhou2013carbon} which considered cost optimization over multiple time intervals, Doyle \textit{et al}. considered the instantaneous total cost minimization problem for cloud data centers \cite{doyle2013stratus}. The geographical workload balancing is formulated as a graph partitioning problem. The graph has two kinds of nodes: the source node represents requests from a specific area and the destination node represents a certain data center. The weight between a source node and a destination node is the cost considering electricity costs and carbon emissions. Graph partitioning is the process of assigning each source node to a destination node in order to minimize the weighted sum, and it is accomplished using Voronoi Partitions \cite{aurenhammer1991voronoi}. Simulation results reveal that the joint optimization solution can reduce total carbon emissions by 21\% compared to the baseline round robin scheduler. 
\begin{table*}
    \centering
    \begin{tabular}{c!{\vrule width 1pt}c!{\vrule width 1pt}c!{\vrule width 1pt}c!{\vrule width 1pt}c!{\vrule width 1pt}c}
        \noalign{\hrule height 1pt}
        Reference  & Constraints & Control Knobs & Formulation & Algorithms & Achieved Results \\ 
        \noalign{\hrule height 1pt}
        
        \cite{azimi2014thermal} & \makecell[c]{No specific constraints} & \makecell[c]{Server Layout} & \makecell[c]{Integer Linear Programming} & \makecell[c]{LP Relaxation} & \makecell[c]{Over 40\% cooling\\ energy saving}\\
        \hline
        
        \cite{zhou2011holistic} & \makecell[c]{No specific constraints} & \makecell[c]{Vent tile opening;\\CRAC blower\\rotational speed;\\Air supply\\ temperature} & \makecell[c]{Model predictive control} & \makecell[c]{MPC controller} & \makecell[c]{36\% cooling energy saving}\\
        \hline
        
        \cite{zhou2011modeling} & \makecell[c]{No specific constraints} & \makecell[c]{CRAC blower\\rotational speed;\\Air supply\\ temperature} & \makecell[c]{Model predictive control} & \makecell[c]{Decentralized\\ MPC controller} & \makecell[c]{Significant cooling\\ energy saving}\\
        \hline
        
        \cite{zhou2012data} & \makecell[c]{No specific constraints} & \makecell[c]{CRAC blower\\rotational speed;\\Vent tile opening} & \makecell[c]{Model predictive control} & \makecell[c]{Decentralized \\ MPC controller\\with coordination} & \makecell[c]{Improved CRAC\\load balancing}\\
        \hline
        
        \cite{LazicMPCCooling} & \makecell[c]{Safe operation constraint} & \makecell[c]{CRAC blower\\rotational speed;\\Vent tile opening} & \makecell[c]{Model predictive control} & \makecell[c]{MPC controller\\ with random walk\\ exploration} & \makecell[c]{Significant cooling energy\\ saving compared to\\PID controller}\\
        \hline
        
        \cite{wan2021intelligent} & \makecell[c]{No specific constraints} & \makecell[c]{Vent tile opening} & \makecell[c]{Markov Decision Process} & \makecell[c]{DQN controller\\with reward sharing\\and fingerprint} & \makecell[c]{Smaller rack inlet\\ temperature variance}\\
        \hline
        
        \noalign{\hrule height 1pt}
    \end{tabular}
    \caption{Summary of existing works on data hall configuration optimization in terms of constraints, control knobs, problem formulation, solving algorithm, and achieved results.}
    \label{DataHallOptimization}
\end{table*}
\subsection{Energy-Efficient Cooling}
On the cooling side, a slice of researchers focus on improving the energy efficiency of the cooling system by controlling the key components of the cooling system, e.g., the chiller plant and CRACs in a data hall. Other researchers state that the ICT system is inextricably linked with the cooling system and they propose joint ICT and cooling control in order to save even more energy. Although some vendors attempted to implement advanced cooling techniques with higher energy efficiency in their data centers, e.g., evaporative cooling \cite{Amazon}, direct-to-chip liquid cooling \cite{li2014coordinating} and immersion cooling \cite{ImmersionCooling}, most legacy data centers still utilize air-cooling techniques to reject heat generated by ICT devices. Therefore, in this section, we concentrate on the works towards energy-efficient air-cooled data centers.

\subsubsection{Server Fan Speed Optimization}
At the rack level, fans are installed in the server rack containing multiple servers to circulate air and remove heat. In addition to server energy consumption, fans also consumes a considerable portion of energy, sometimes up to 51\% in some server configuration \cite{tian2020spinsmart}. Hence, quit a few researchers investigated the fan management problem so that either the energy consumption of fans or the total server rack can be minimized \cite{wang2009optimal,piatek2015modeling,piatek2017intelligent}, and the approaches for optimizing server fan speed are summarized in Table \ref{FanSpeedControl}. In \cite{wang2009optimal}, Wang \textit{et al}. investigated optimal fan speed control for a blade server. In their framework, the power consumption of a fan is modeled as a cubic function with respect to fan speed. A simplified parametric CPU thermal model is established based on thermodynamic principals for each server. The CPU heating process is modeled by a Resistance-Capacitance (RC) circuit in which the thermal resistance, and thermal capacity are related to the air flow rate which is related to the fan speed, and the source is the heat dissipated from server CPUs. To determine model parameters, System identification is implemented. Based on the thermal model, transient CPU temperature in the near future can be predicted accurately. By solving a constrained optimization with the predicted CPU temperature, proactive fan speed control is implemented. Experiment results show that 20\% fan energy savings are achievable compared to the reactive control scheme. However, \cite{wang2009optimal} only looked at the fan energy savings without taking into account the trade-off between fan energy savings and potential total energy consumption increases. In \cite{piatek2015modeling} and \cite{piatek2017intelligent}, the authors pointed out that the provisioning of fan speed may increase the energy consumption of ICT devices and cooling systems because the leakage power of ICT devices will increase as the temperature rises, resulting in an increase in cooling system energy consumption. In their works, Piatek \textit{et al}. studied this trade-off systematically for micro servers and the impacts of server fan speed control on the energy consumption of a data center. Both \cite{piatek2015modeling} and \cite{piatek2017intelligent} used a similar CPU thermal model based on an RC circuit to enable proactive control. To minimize the energy consumption of both ICT devices and fans, various fan speed control schemes are proposed, including constant low fan speed policy and dynamic fan speed policy. The results from a proof-of-concept testbed demonstrate that fan speed control can significantly reduce fan energy consumption. Furthermore, despite the increase in energy consumption of servers and cooling systems, the total energy consumption is reduced in comparison to the fan speed-agnostic scheme. 

\subsubsection{Data Hall Configuration Optimization}
On the data hall side, researchers studied either the static server placement optimization problem or the dynamic cooling system optimization problem in order to minimize energy consumption of the cooling system and the related works are summarized in Table \ref{DataHallOptimization}.

In terms of static optimization, Azimi \textit{et al}. studied the thermal-aware server layout planning problem in a heterogeneous data center \cite{azimi2014thermal}. The goal of server layout planning is to properly place different servers with varying power specifications in different locations throughout the data hall in order to minimize total heat recirculation. Three algorithms based on greedy heuristic, integer linear programming and stochastic programming are proposed and simulation results show that around 40\% cooling energy saving was possible with proper server placement. 

Although strategic layout planning can significantly reduce cooling power, static optimization is incapable of handling varying workloads as well as periodic server upgrades throughout a data center's lifecycle. Therefore, growing emphasis is placed on dynamically controlling cooling system parameters such as CRAC blower rotational speed and air supply temperature \cite{zhou2011modeling}\cite{zhou2012data}\cite{zhou2011holistic}\cite{LazicMPCCooling}, water flow rate of the water pump \cite{vu2017data} and etc.  In \cite{zhou2011modeling}, the basic energy and mass conservation law was used to derive a simplified thermal dynamic model for describing the complex mass and energy flow in the data center. This model establishes the linear relationship between the future inlet temperature and the current inlet temperature, air supply temperature, CRAC blower rotation speed, and vent tile opening. System identification experiments are conducted to determine the parameters in the model. With this model, a MPC controller for minimizing the cooling energy in response to dynamic ICT workloads is proposed, which determines the air supply temperature, CRAC blower rotation speed and vent tile opening simultaneously. Experiment results show that the MPC controller could save 36\% cooling energy compared to adjusting the CRAC units' blower speed and vent tile opening separately. Since the number of servers and CRACs in modern data centers is usually very large, the centralized MPC controller proposed in \cite{zhou2011holistic} suffers from the curve of dimensionality. To overcome the computational challenge incurred by the centralized control scheme, based upon the thermal model established in \cite{zhou2011holistic}, a decentralized MPC controller for multiple CRACs is proposed based on the observation that each CRAC in a data center only has significant effects on the nearby racks  \cite{zhou2011modeling}. The cooling energy is minimized while the rack inlet temperature maintains below a threshold by controlling the CRAC blower rotational speed and air supply temperature. It was reported that the distributed control scheme was valid and effective in providing energy-efficient cooling for a large-scale data center. In \cite{zhou2012data}, the authors further extended the distributed control scheme developed in \cite{zhou2011modeling}, allowing for coordination between multiple controllers. The motivation for developing cooperative controllers stemmed from the fact that the coupling between neighboring thermal zones was ignored in \cite{zhou2011modeling}, resulting in CRAC load imbalance. Such imbalance would reduce energy efficiency and even shorten the CRAC's lifespan. In the cooperative control scheme, each controller responsible for a thermal zone was also a MPC controller and local controllers exchanged information with those in the nearby thermal zones so that the CRACs in the same group of thermal zones had the same air supply temperature to prevent load imbalance. The authors discovered that by coordinating multiple CRACs, they were able to achieve significantly better CRAC load balancing. In addition, a data-driven, model-based MPC controller was proposed for dynamic control of the CRAC blower rotational speed and the valve opening for chilled water in \cite{LazicMPCCooling}. The complicated thermal dynamic of a data center is linearized in terms of the state vector, control variables and exogenous variables. To achieve system identification, a randomized exploration over controls approach is utilized to collect sufficient data with a wide dynamic range to learn the system dynamics more robustly. This exploration strategy uses a simple, range-limited uniform random walk over each control variable. In terms of the control process, optimal policy is obtained in each step by optimizing a linear combination of the quadratic error between current state variables and set the points, as well as that between control variables and their corresponding minimum allowable values, with the goal of minimizing cooling energy while maintaining the data center's safety. Experiment results demonstrate that the MPC controller could effectively and safely operate a data center after only a few hours of exploration with little prior knowledge of system dynamics. Furthermore, energy efficiency of the cooling system is significantly improved compared to that with traditional PID controllers. Different the works based on model predictive control, Wan \textit{et al.} proposed using AVT to improve the local cooling delivery via MARL techniques \cite{wan2021intelligent}. They formulate the AVT control problem as a Markove Decision Process and control objective is to strike a balance beween rack inlet temperature distribution and the power consumption. An agent is trained offline for each AVT with Deep Q-Learning (DQN) \cite{mnih2015human}. Furthermore, the idea of reward sharing and fingerprint \cite{foerster2017stabilising} is adopt so that the individual agent can be trained in an cooperative manner. In addition, this approach has been evaluated in a prototype data center and the evaluation results show that the variance of rack inlet temperature is reduced significantly, leading to better cooling delivery and potential system energy saving. 
\begin{table*}
    \centering
    \begin{tabular}{c!{\vrule width 1pt}c!{\vrule width 1pt}c!{\vrule width 1pt}c!{\vrule width 1pt}c!{\vrule width 1pt}c}
        \noalign{\hrule height 1pt}
        Reference  & Constraints & Control Knobs & Formulation & Algorithms & Achieved Results \\ 
        \hline
        \cite{ahn2001optimal} & \makecell[c]{Safe operation constraints} & \makecell[c]{Chiller water temperature;\\Condenser water temperature;\\Air supply temperature} & \makecell[c]{Quadratic Programming} & \makecell[c]{Closed form solution} & \makecell[c]{Significant chiller\\ plant energy saving}\\
        \hline
        
        \cite{yu2008optimization} & \makecell[c]{No specific constraints} & \makecell[c]{Cooling tower fan;\\Condenser water pump} & \makecell[c]{No specific formulation} & \makecell[c]{Closed form solution} & \makecell[c]{About 5.3\% chiller\\ plant energy saving}\\
        \hline
        
        \cite{tyagi2006extremum} & \makecell[c]{Safe operation constraints} & \makecell[c]{Condenser water temperature} & \makecell[c]{No specific formulation} & \makecell[c]{ESC} & \makecell[c]{Significant chiller\\ plant energy saving}\\
        \hline
        
        \cite{mu2016optimization} & \makecell[c]{Safe operation constraints} & \makecell[c]{Cooling tower fan;\\Condenser water pump;\\Chilled water pump;\\Condenser water temperature} & \makecell[c]{No specific formulation} & \makecell[c]{Multivariate\\ESC} & \makecell[c]{Significant chiller\\ plant energy saving}\\
        \hline
        
        \cite{vu2017data} & \makecell[c]{Safe operation constraints} & \makecell[c]{Cooling tower fan;\\Condenser water pump;\\Chilled water pump} & \makecell[c]{Nonlinear Programming} & \makecell[c]{COBYLA solver} & \makecell[c]{5-10\% chiller plant\\ energy saving}\\
        \hline
        
        \cite{zheng2018data} & \makecell[c]{No specific constraints} & \makecell[c]{Chiller cooling\\load allocation} & \makecell[c]{No specific formulation} & \makecell[c]{Predictive planning} & \makecell[c]{About 30\% cooling\\ energy saving}\\
        \hline
        
        \cite{chen2019gnu} & \makecell[c]{Safe operation constraints} & \makecell[c]{Supply water temperature} & \makecell[c]{Markov Decision Process} & \makecell[c]{Differentiable MPC} & \makecell[c]{16.7\% chiller plant\\energy saving}\\
        \noalign{\hrule height 1pt}
    \end{tabular}
    \caption{Summary of existing works on chiller plant optimization in terms of constraints, control knobs, problem formulation, solving algorithm, and achieved results.}
    \label{ChillerPlantOptimization}
\end{table*}
\subsubsection{Chiller Plant Optimization}
The aforementioned solutions only focus on the data hall optimization, ignoring the optimization of the chiller plant which is another critical component in the DC cooling system, and many researchers have developed approaches to improve energy efficiency of a chiller plant. Related literature is summarized in Table \ref{ChillerPlantOptimization}.

Some researchers adopted model-based control approaches to optimize the energy efficiency of a chiller plant. The power modeling of a chiller plant is based on the underlying physical law of each component. After the power model for each component is established, an optimization problem is formulated to minimize cooling energy consumption. For example, a quadratic power model of a chiller plant with regard to the chiller water temperature, condenser water temperature and air supply temperature was established in \cite{ahn2001optimal}. Multiple simulations with varying external variables (temperature, humidity and etc) and three control variables were conducted to obtain massive data points and linear regression is utilized to determine the model parameters. Optimal linear control laws for these three control knobs were derived and the results show that the quadratic power model is a good fit to actual power consumption and significant chiller plant energy savings are achieved with the optimal control policy. In addition, Yu \textit{et al}. developed a load-based speed control scheme for the cooling tower fan and the condenser water pump \cite{yu2008optimization}. Thermodynamic models for each component of a chiller as well as the cooling tower are developed. Based on these models, the authors proposed adjusting the cooling tower fan speed and the condenser water flow rate linearly with regard to the Partial Load Ratio (PLR). In addition, optimal cooling tower water leaving temperature is derived as a quadratic function of the PLR. Simulation results show that 5.3\% annual electricity consumption saving was available when the condenser water flow rate and the cooling tower fan were controlled optimally. 

However, model-based methods require accurate modeling for the chiller plant, which may be infeasible in the real world because the power consumption of a chiller plant is related to the dynamic weather condition, aging of the chiller and etc. Therefore, some researchers adopted model-free approaches to regulate the chiller plant. Among them, extremum seeking control (ESC) has attracted widespread attention from the mechanical community due to its model-free and real-time optimization properties \cite{vu2017data}. In \cite{tyagi2006extremum}, a real-time control approach based on extremum seeking control (ESC), a model-free control scheme \cite{ariyur2003real}, was developed to optimize the condenser water temperature set point. By iteratively interacting with the oracle function and narrowing down the interval that contains the optimal set point, it is guaranteed to return the optimal set point for a convex oracle function within finite number of iterations. Since the power consumption of the chiller plant is a convex function of the condenser water temperature set point for a fixed PLR, the algorithm is guaranteed to find the optimal set point. In contrast to \cite{tyagi2006extremum} where only condenser water temperature set point was considered, the cooling tower fan, the condenser water pump, the chilled water pump and the condenser water set temperature were jointly controlled via multivariate ESC \cite{mu2016optimization}. To handle the constraints imposed on the inputs, a quadratic penalty function is added into the objective function that is the total power consumption of the cooling tower, the chillers and the water pumps. Simulation results based on Modelica\footnote{https://modelica.org/} demonstrate its significant energy saving potential. Despite the success of ESC-based approach, it was pointed by \cite{vu2017data} that ESC might not always find the optimum in the search space due to the sophisticated dynamics of a chiller plant.

With the proliferation of sensor big data, quite a few researchers attempt to incorporate domain knowledge with data-driven modeling to realize accurate energy modeling of a chiller plant, and then perform optimization based on such data-driven energy models. For example, a data-driven chiller plant modeling and optimization method were proposed to improve the energy efficiency of a chiller plant \cite{vu2017data}. Unlike the traditional modeling approach relying on domain knowledge of mechanical cooling, electrical and thermal management of a data center, the author decomposed the chiller plant into seven modules with domain knowledge and established parametric models for them. Cubic functions are used to model the power consumption of the water pumps and the cooling tower in terms of the variable speed drive (VSD) speed of the condenser water pump, the chilled water pump, and the cooling tower fan. The nonlinear relationship between VSD speed and water flow rate is modelled by a neural network. An optimization problem is formulated to minimize the total energy consumption of the chiller, the water pump and the cool tower under the safe operation constraint of each module. The simulation results show the superiority of the parametric model with domain knowledge over the black-box modeling, and it is capable of saving 5-10\% of cooling energy. A similar domain knowledge-integrated concept was used to assign the cooling load to each chiller in the chiller plant \cite{zheng2018data}. This issue is also known as Chiller Sequencing (CS). To accomplish this, the Coefficient of Performance (CoP) of each chiller under different cooling loads is predicted using historical data by machine learning. The prediction features are based on domain knowledge of the chiller plant, which includes temporal, meteorological, and mechanical information. 

After that, the controller is able to obtain the chiller optimal cooling load distribution with minimum electricity consumption while satisfying the cooling demand based on the predicted CoP and the cooling demand. The data-driven approach saved over 30\% of chiller plant energy, according to simulation results. In contrast to \cite{vu2017data,zheng2018data} in which domain knowledge was incorporated into the energy modelling of a chiller plant, Chen \textit{et al}. proposed to control the supply water temperature via reinforcement learning (RL) and integrated the knowledge of control laws into the training process of the RL agent through imitation learning \cite{silver2010learning} in order to accelerate training \cite{chen2019gnu}. The RL agent is first trained offline to mimic the behavior of a traditional PID controller. Once it is put into the environment, it consistently improves its policy in an end-to-end manner via Proximal Policy Optimization (PPO). In addition, the complex system dynamics are also incorporated into the RL agent with Differentiable Model Predictive Control \cite{amos2018differentiable} to improve sample efficiency. Experiment results showed a 16.7\% cooling energy reduction compared to existing PID controller. 
\begin{table*}
    \centering
    \begin{tabular}{c!{\vrule width 1pt}c!{\vrule width 1pt}c!{\vrule width 1pt}c!{\vrule width 1pt}c!{\vrule width 1pt}c}
        \noalign{\hrule height 1pt}
        Reference  & Constraints & Control Knobs & Formulation & Algorithms & Achieved Results \\ 
        \noalign{\hrule height 1pt}
        \cite{pakbaznia2009minimizing} & \makecell[c]{temperature redline;\\server capacity} & \makecell[c]{Air supply temperature;\\ Server provision;\\Server speed scaling} & \makecell[c]{Integer Linear\\ Programming} & \makecell[c]{ILP solver} & \makecell[c]{13\% energy saving}\\
        \hline
        
        \cite{pakbaznia2010temperature} & \makecell[c]{temperature redline;\\server capacity} & \makecell[c]{Air supply temperature;\\ Server provision;\\Classis on/off} & \makecell[c]{Integer Linear\\ Programming} & \makecell[c]{ILP solver} & \makecell[c]{Significantly outperform\\ greedy algorithm}\\
        \hline
        
         \cite{sansottera2011cooling} & \makecell[c]{temperature redline;\\service delay} & \makecell[c]{Air supply temperature;\\Job placement} & \makecell[c]{Mixed Integer\\ Non-linear Programming\\(MINLP)} & \makecell[c]{Greedy Least Power\\(GLP)} & \makecell[c]{8\% total energy saving}\\
        \hline
        
        \cite{li2011temperature} & \makecell[c]{No specific constraint} & \makecell[c]{Power distribution;\\Task scheduling} & \makecell[c]{No specific formulation} & \makecell[c]{TEL/TLE\\heuristic} & \makecell[c]{Significant computing\\ energy reduction}\\
        \hline
        
        \cite{abbasi2012tacoma} & \makecell[c]{Response time;\\Load balancing} & \makecell[c]{Server provision;\\Workload distribution} & \makecell[c]{Nonlinear Mixed \\Integer Programming} & \makecell[c]{TASP-LRH;\\TASP-CPLRH;\\TAWD-LRH;\\TAWD-CPLRH} & \makecell[c]{Up to 40\% energy saving}\\
        \hline
        
        \cite{mukherjee2009spatio} & \makecell[c]{Server capacity;\\VM resource requirement;\\VM execution time;\\VM deadline} & \makecell[c]{VM placement;\\VM migration} & \makecell[c]{Nonlinear Integer\\ Programming } & \makecell[c]{GA-based algorithm \\(SCINT);\\LRH heuristic} & \makecell[c]{Up to 60\% energy saving}\\
        \hline
        
        \cite{xu2013temperature} & \makecell[c]{QoS constraint} & \makecell[c]{Geographical \\workload distribution;\\Batch job\\scheduling} & \makecell[c]{Convex Programming} & \makecell[c]{Distributed\\ ADMM solver} & \makecell[c]{5\%-20\% cost saving}\\
        \hline
        
        \cite{liu2012renewable} & \makecell[c]{Server capacity;\\Battery evolving\\ equation;\\Batch job\\parallelism} & \makecell[c]{Free cooling;\\Battery charge\\and discharge;\\Batch job\\scheduling} & \makecell[c]{Convex Programming} & \makecell[c]{CVX} & \makecell[c]{Free cooling\\ is critical\\}\\
        \hline
        
        \cite{li2014coordinating} & \makecell[c]{Temperature redline} & \makecell[c]{Geographical\\ workload distribution;\\Workload placement\\within a data center} & \makecell[c]{Mixed Integer\\ Linear Programming} & \makecell[c]{LINGO solver} & \makecell[c]{Up to 38\% energy saving}\\
        \hline
        
        \cite{ji2020energy} & \makecell[c]{Server capacity} & \makecell[c]{Free cooling;\\Job scheduling} & \makecell[c]{Integer Programming} & \makecell[c]{Greedy heuristic\\with task grouping} & \makecell[c]{Over 50\% energy saving}\\\hline
        
        \cite{yi2019toward} & \makecell[c]{Safe operation of\\both IT and\\cooling system} & \makecell[c]{Job scheduling;\\Air flow rate} & \makecell[c]{Markov Decision Process} & \makecell[c]{PADQN} & \makecell[c]{15\% energy saving}\\
        \hline
        
        \cite{chi2020jointly} & \makecell[c]{Safe operation of\\both IT and\\cooling system} & \makecell[c]{Job scheduling;\\Air flow rate} & \makecell[c]{Markov Decision Process} & \makecell[c]{MARL} & \makecell[c]{Outperform \cite{yi2019toward} in\\terms of energy \\efficiency}\\
        \hline
        
        \cite{zhou2021joint} & \makecell[c]{No specific constraints} & \makecell[c]{Job scheduling;\\Air flow rate} & \makecell[c]{Markov Decision Process} & \makecell[c]{DQN for job\\ scheduling;\\DDPG for air flow\\ rate control} & \makecell[c]{15\% overall\\ energy saving}\\
        \noalign{\hrule height 1pt}
    \end{tabular}
    \caption{Summary of existing works on joint IT and cooling optimization in terms of constraints, control knobs, problem formulation, solving algorithm, and achieved results.}
    \label{JointITCoolingOptimization}
\end{table*}
\subsubsection{Joint IT and Cooling Optimization}
The previously mentioned works only attempt to save cooling energy by only controlling the cooling system while keeping the data center operating within the thermal safety region. However, a joint IT and cooling optimization is preferable because the heat generation and rejection process connects the IT system to the cooling system. Hence, a number of studies have been conducted to investigate the benefits of jointly optimizing IT and cooling systems and they are summarized in Table \ref{JointITCoolingOptimization}.

The fact that IT and cooling systems are nonlinearly correlated is a major challenge for joint control. For example, to determine the effects of a specific IT workload scheduling decision on the ambient temperature, sophisticated nonlinear CFD simulation should be conducted to obtain the temperature field. Therefore, if we want to obtain a temperature field using CFD simulation, it is nearly impossible to perform near real-time dynamic IT and cooling control. To overcome it, a simplified linear thermal model was widely adopted in the literature \cite{pakbaznia2009minimizing, tolia2009unified, li2011temperature, pakbaznia2010temperature, abbasi2012tacoma, mukherjee2009spatio, sansottera2011cooling}. In this model, each rack inlet temperature is modeled as a linear combination of the air supply temperature and the weighted sum of each rack's power consumption:
\begin{equation}\label{HRF}
    \mathbf{T}_{\text{in}} = T_s\cdot\mathbf{1} + \mathbf{Dp},
\end{equation}
where $\mathbf{T}_{\text{in}}$ is the vector form of the rack inlet temperature, $T_s$ is the air supply temperature, and $\mathbf{1}$ is the all-one vector with the dimension equaling to the number of server racks. $\mathbf{p}$ is the vector form of the power consumption of each server rack and $\mathbf{D}$ is the coefficient matrix obtained through multiple offline CFD simulations.

Some researchers explored the thermal-aware server provisioning and server speed scaling. In \cite{pakbaznia2009minimizing}, Pakbaznia \textit{et al}. proposed to minimize total energy consumption of a data center by jointly controlling the air supply temperature, server provisioning, and server speed scaling. For a fixed air supply temperature, the problem was formulated as an Integer Linear Programming (ILP). The ILP problem is solved for multiple air supply temperature candidates and the one with minimum total energy consumption is chosen as the final solution. The experimental results demonstrate that an average of 13\% energy saving is available. In \cite{pakbaznia2010temperature}, a similar problem was studied and a Temperature-Aware Dynamic Resource Provisioning (TA-DRP) framework was proposed to improve energy efficiency of a data center. To perform server provisioning, a workload prediction technique based on seasonal autoregression is utilized and the authors validate its effectiveness using real-world workload trace data. Following the forecasting of the workload, the required number of servers is determined based on the average number of servers required by a request. After the desired number of servers is calculated, the controller selects the locations of servers that will be employed in the next epoch with the goal of maximizing the air supply temperature to save cooling energy. The simulation results show that TA-DRP can save significantly more energy than the temperature-agnostic server consolidation scheme which prefers to turn on servers in an already-running racks. 

Unlike \cite{pakbaznia2009minimizing} and \cite{pakbaznia2010temperature} which focused on thermal-aware server provisioning, other researchers pay attention to reducing total energy consumption via workload scheduling. In \cite{sansottera2011cooling}, Sansottera \textit{et al}. proposed to minimize the energy consumption of a data center via judicious workload placement. The linear thermal model (\ref{HRF}) is utilized and the problem is formulated as a Mixed Integer Non-Linear Programming (MINLP). To find the approximate solution of the complicated MINLP, a greedy heuristic algorithm called Greedy Least Power (GLP) is proposed, which attempts to place the job in the server with the least increase in total energy consumption. It is reported that the cooling-aware workload scheduling scheme can save 8\% total energy compared to a performance-only strategy. In \cite{li2011temperature}, the trade-off between minimizing total energy consumption, reducing task latency, and lowering down peak inlet temperature is investigated. In this work, the highest priority is to minimize peak inlet temperature (T), which is accomplished by proportionally disitributing power to each server based on the Heat Recirculation Factor (HRF) \cite{moore2005making} that defines a server's contribution to total heat recirculation in a data center. Following the determination of power distribution, two heuristic algorithms called TEL/TLE are developed based on different priorities on latency (L) and computing energy (E). The TEL algorithm is able to minimize the peak inlet temperature and significantly reduce the computing energy by slightly sacrificing task latency, according to simulation results. In \cite{abbasi2012tacoma}, Abbasi \textit{et al}. took server provisioning and workload scheduling problems as a whole and proposed a thermal-aware scheme named TOMACA that included thermal-aware server provisioning (TASP) and thermal-aware workload distribution (TAWD). The goal of the TASP is to minimize total energy consumption while making the peak inlet temperature equal to the redline temperature by controlling the on/off of each server. Load balancing is also considered as a constraint in TASP formulation. Following a decision on server provisioning, TAWD is implemented at a fine time granularity (5-10 seconds) to minimize the same objective function as TASP with proper workload placement. A heuristic solutions is proposed for TASP and TAWD. The basic idea of the heuristic algorithms is to turn on the servers that contribute the least to total heat recirculation, and to place more workload on these servers in order to save cooling energy. Besides minimizing total energy consumption of a data center, Mukherjee \textit{et al}. considered a multi-objective optimization problem in a virtualized data center where the sum of total energy and VM migration cost was minimized by controlling VM placement and migration \cite{mukherjee2009spatio}. The reason for introducing VM migration is to more flexibly migrate workloads into cooling-efficient servers to save cooling energy. To solve nonlinear integer programming with perfect knowledge of future system states, one offline solution named SCINT based on genetic algorithm (GA) \cite{michalewicz2013genetic} is proposed. However, the high time complexity of SCINT, as well as the requirement of future information, makes it impractical for implementation. Therefore, two online greedy heuristic algorithms are developed. Similar to \cite{li2011temperature}, one heuristic algorithm prefers servers with less heat recirculation. Another heuristic algorithm is based on the idea of Early-Deadline-First (EDF) by relocating VMs approaching their deadline to thermal-efficient servers. It is  shown that the offline SCINT algorithm can achieve up to 60\% energy saving, while heuristic algorithms can also save up to 25\% of the energy. 

The preceding solution only considered the joint operation of the IT system and the CRAC unit in a data center (by changing air supply temperature). Nowadays, as novel energy-efficient cooling techniques such as free air cooling and direct liquid cooling gain popularity, some researchers have attempted to integrate such cooling techniques into data centers and investigate the energy-saving potential of joint IT and cooling optimization. In \cite{xu2013temperature} and \cite{liu2012renewable}, free air cooling was utilized to reduce cooling energy consumption. Because the availability of free cooling depends on local weather, which exists spatiotemporal variation, researchers consider scheduling the workload either spatially \cite{xu2013temperature} or temporally \cite{liu2012renewable}. By utilizing spatial variation of free cooling, Xu \textit{et al}. proposed a temperature-aware geographical load scheduling system which minimized the sum of the total energy cost of data centers and the request routing cost due to service delay. The problem is formulated as a convex programming and a distributed Alternating Direction Method of Multipliers (ADMM) \cite{boyd2011distributed} solver is developed to solve the large scale optimization efficiently. Simulation results show that free air cooling could save 5-20\% of total energy. In \cite{liu2012renewable}, a data center with free cooling, on-site renewable energy generation, and batteries was optimized with the objective of maximizing profits. The workloads include web requests with strict service delay constraints and batch jobs that can be deferred temporally to align with the presence of free air cooling or renewable energy. The revenues come from the execution of batch jobs and the costs are the total energy costs. With the knowledge of future workloads and renewable energy production, the problem is formulated as convex programming. To make the optimization tractable, a K-Nearest Neighbor (KNN) predictor is exploited to perform a renewable energy forecast and seasonal autoregressive model is adopted for workload prediction. Trace-based simulation results reveal that free air cooling is critical in terms of energy savings because it can meet the majority of cooling demand. Furthermore, the authors claimed that the best time of a day to execute batch jobs was at night, when free air cooling capacity was at its peak and the demand for serving web requests was at its valley. 

In addition, Liquid cooling was introduced in \cite{li2014coordinating} and \cite{ji2020energy} to  improve the energy efficiency of the cooling system even further. In \cite{li2014coordinating}, a data center with a hybrid cooling system including traditional chilled-water air cooling, direct liquid cooling and free air cooling was considered. As for the direct liquid cooling, the CPU in a server is directly connected with a cold plate containing the coolant to absorb heat, while the other components are still cooled by chilled air flow or free air cooling.  A hierarchical control scheme is proposed, which contains a global tier for distributing workloads among multiple geo-dispersed data centers and a local tier for thermal-aware workload placement within a data center. The authors claimed that the hybrid cooling system was able to reduce up to 38\% cooling energy and achieve near-optimal PUE ($\sim 1$) when the outside temperature was around 26 $^{\circ}$C. In \cite{ji2020energy}, Ji \textit{et al}. investigated the workload scheduling in a data center integrated with direct liquid cooling and free air cooling. The authors found that the optimal cooling system configuration is a piecewise function. When the amount of workloads is less than a certain threshold, the best option is to rely solely on free air cooling. Direct liquid cooling is preferable once the amount of workloads exceeds the threshold. Based on this, integer programming is developed to minimize the total energy consumption of the data center and a heuristic algorithm is proposed. The basic idea of the solver is to serve all requests with a minimum number of servers by task grouping so that server utilization can be improved and new servers can always be turned on with the lowest marginal cost. Simulation on different real-world trace data validates the superiority of the proposed solution in terms of energy savings. 

Most existing works on joint IT and cooling system optimization are based on model-based approaches. Due to the inherent complexity of the underlying physical processes, simplified models such as Eq. (\ref{HRF}) are often adopted. However, such models are sometimes either insufficient or inaccurate in capturing the complex dynamics of various interacting processes within a data center \cite{ran2019deepee}. Hence, researchers have recently begun to investigate joint IT and cooling control with data-driven model-free approaches such as DRL \cite{ran2019deepee, chi2020jointly}. In \cite{ran2019deepee}, a PArameterized action space based Deep Q-Network (PADQN) algorithm was proposed to jointly control IT task scheduling and air flow rate of CRACs in a data hall with highly dynamic and time-variant environments. The action space of the IT system is discrete and that of the cooling system is continuous, and thus existing DRL techniques cannot handle such hybrid action space well. Therefore, the authors incorporated the parameterized action space MDP (PAMDP) technique \cite{masson2016reinforcement} into the framework. Specifically, given the current system state, the controller uses a deterministic policy neural network to generate continuous actions for air flow rate adjustment. Subsequently, the action along with current state are sent to the Q-Network to generate the discrete action with highest Q-value for job scheduling. Furthermore, a two-time-scale control framework is developed to capture the different response time for IT system control (a few microseconds) and cooling system control (a few minutes). Trace-based simulation results show that PADQN outperforms the existing cooling system controller based on MPC by 15\%. Unlike \cite{ran2019deepee} that solved the hybrid action space issue with PAMDP technique, Chi \textit{et al}. proposed a more direct solution using Multi-Agent Reinforcement Learning (MARL) \cite{chi2020jointly}. Two DRL agents are trained offline: one for producing discrete actions for controlling the IT system and another for producing continuous actions for controlling CRACs in a data hall. In order to boost cooperation between these two agents, their states and actions are shared with each other. In this way, the cooling agent will learn to adjust the air flow rate based on the task scheduling decision. Similarly, an IT agent will also learn to adjust its task scheduling policy in response to a specific CRAC configuration. In terms of energy savings, the simulation results demonstrate that the MARL solution slightly outperformed PADQN \cite{ran2019deepee}. Furthermore, a unified framework for joint IT and cooling optimization based on DRL techniques is proposed by Zhou \cite{zhou2021joint}, where three control problems (load-aware target cooling, thermal-aware task scheduling, and iterative IT-facility optimization) can be addressed simultaneously. The IT task scheduling is achieved with a DRL agent based on DQN and the CRAC control is fulfilled with a Deep Deterministic Policy Gradient (DDPG) \cite{lillicrap2015continuous} controller, which can handle the continuous action space of the cooling system. In addition, such a framework is flexible because it can be adopted by either the IT manager in a data center who does not have access to the cooling system control or the facility manager who cannot control the IT task scheduling due to decoupling of the two controllers. Evaluations on the real-world trace data demonstrate that this approach can save up to 15\% total energy consumption of a hyper-scale data center.

\textbf{Summary:} most existing solutions for improving energy efficiency of a data center aim to optimize either the IT system or the cooling system. In addition, some efforts have been made in joint IT and cooling system optimization to improve energy efficiency even further. However, existing works often formulated the problem as a constrained optimization with the objective of minimizing energy consumption under QoS and thermal constraints. The accuracy of the modeling for both the IT and the cooling system is critical for these solutions' success in practice. However, high fidelity models of the IT and cooling system tend to be highly complicated, which impedes their application in optimization-based approaches because the optimization problems will become intractable with these models. In this regard, a slice of researchers investigate model-free controllers with emerging DRL techniques and obtain satisfactory results that outperform traditional optimization-based solutions, showing the potential of DRL in data center energy efficiency enhancement. 

\section{carbon-neutral data center: From Energy Circulation Perspective}\label{Energy Circulation}

As data centers have become a major energy consumer due to the rising demand for cloud-based connectivity and computing,  researchers have started to explore proper methods for reusing energy in a data center to reduce their energy consumption as well as carbon emissions. The energy conservation law states that energy can neither be created nor destroyed; rather, it can only be transformed or transferred from one form to another \cite{HASELI20201}. In a data center, most of the electricity drawn by the ICT system and cooling system is converted into waste heat. Such waste heat can be utilized as a renewable energy source \cite{sharma2010design}\cite{araya2021study} or providing complementary cooling for a data center \cite{haywood2010sustainable}\cite{haywood2012thermodynamic}, or being sold to other customers such as district heating networks \cite{oro2019DomesticHeat}\cite{he2018analysis}, nearby power plants \cite{marcinichen2012chip}, and seawater desalination plants \cite{sondur2018data} to reduce carbon emissions of these waste heat customers, resulting in carbon credits that can be leveraged to offset carbon emissions. Therefore, developing highly efficient energy recycling systems is also critical on the path towards carbon neutrality. The waste heat sources within a data center will be first presented in this section. Subsequently, several existing techniques for reusing waste heat in a data center will be reviewed and their merits and drawbacks will be discussed as well. Finally, several applications for data center waste heat will be introduced. 
\begin{table}
    \centering
    \begin{tabular}{c!{\vrule width 1pt}c!{\vrule width 1pt}c}
        \noalign{\hrule height 1pt}
        Cooling System & Locations & Temperature \\
        \noalign{\hrule height 1pt}
        Air cooling & \makecell[c]{Cold aisle (CRAC supply) \\ Hot aisle (CRAC return) \\ Chilled water supply \\ Chilled water return} & \makecell[c]{10-32 $^\circ$C \\ 50-60 $^\circ$C \\ 7-10 $^\circ$C\\ 35 $^\circ$C} \\
        \hline
        Liquid cooling & \makecell[c]{Liquid supply to servers \\ Liquid exits servers} & \makecell[c]{70-75 $^\circ$C\\ 72-80 $^\circ$C} \\
        \noalign{\hrule height 1pt}
    \end{tabular}
    \caption{Summary of thermal properties for the air cooling and liquid cooling system \cite{WasteHeatReview2014}.}
    \label{ThermalProperty}
\end{table}
\subsection{Waste Heat Sources Within a Data Center}
When it comes to waste heat recovery, it is critical to understand the best locations to recycle waste heat as well as the thermal properties of these locations because they will affect the available techniques for reusing waste heat. As excellent reviews for data center cooling techniques exist \cite{WasteHeatReview2014}\cite{huang2020review}, we briefly introduce these cooling systems and present the best locations and corresponding thermal properties for waste heat recovery for each system in this section. 
\begin{itemize}
     \item Air-cooled Data Center: In an air-cooled data center, waste heat from server racks is dissipated to the outside environment with CRACs in a data hall and the chiller plant. The best location for recycling waste heat is the CRAC return (hot aisle) with temperature of 50-60 $^\circ$C. 
     \item Liquid-cooled Data Center: In a liquid-cooled data center, servers are directly cooled by a cold plate through which liquid coolant flows. The liquid inlet temperature can reach 60 $^\circ$C and the outlet temperature can even reach 85 $^\circ$C, making it suitable for waste heat harvesting. The best location for waste heat recovery is the liquid exit of a server rack. 
\end{itemize}
In addition, a two-phase cooled system where coolant directly contacts server components such as CPUs has also been studied to cool data centers in a highly efficient manner \cite{huang2020review}. However, only air-cooled and liquid-cooled data centers are discussed due to practical difficulties in recycling waste heat for the two-phase cooling system. Thermal properties of the air cooling and liquid cooling are summarized in Table. \ref{ThermalProperty}.

\subsection{Existing Techniques for Data Center Waste Heat Reusing}
Recently, researchers have focused on the potential for reusing data center waste heat, and several solutions have been proposed. This section will go over the most commonly used applications for recycling data center waste heat. 
\begin{figure}[t]
	\centering
	\includegraphics[width=.48\textwidth]{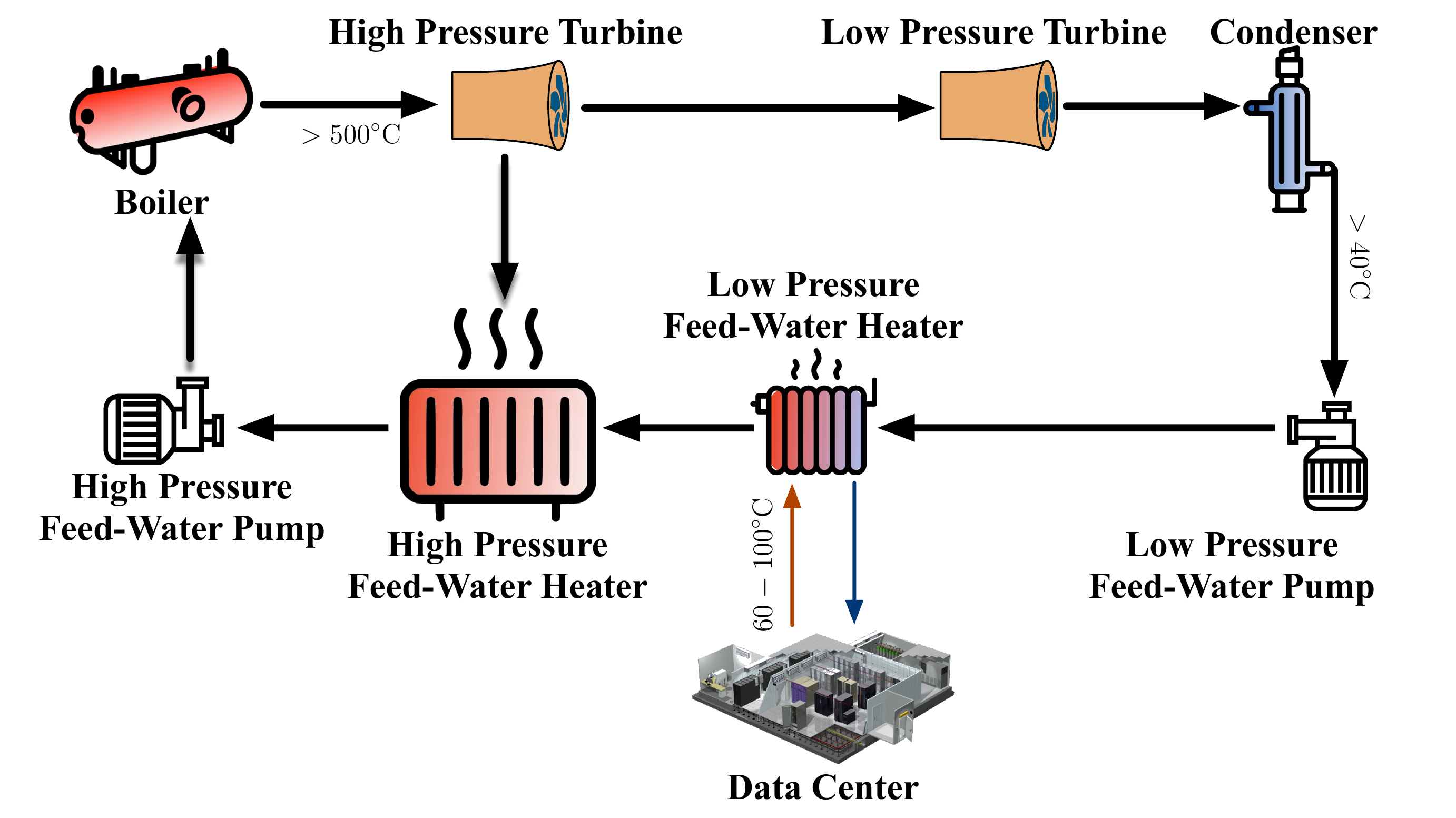}
	\caption{Illustration of how data center waste heat is utilized in a coal power plant Rankine cycle. Data center waste heat is utilized to preheat the feed-water to reduce the energy consumption of the boiler in the coal power plant.}
	\label{PowerPlantColation} 
\end{figure}
\subsubsection{District Heating}
District Heating is perhaps the most common application for recycling low-grade waste heat from a data center and its economical viability as well as implementation feasibility are validated by many research works \cite{oro2019DomesticHeat, deymi2019thermoeconomic, wahlroos2017utilizing, he2018analysis,davies2016using}. 

As the supply and return water temperatures of present district heating network are approximately 90 $^\circ$C and 70 $^\circ$C respectively \cite{skagestad2002district}, a heat pump is often utilized to upgrade the data center waste heat to meet the temperature requirements of the district heating network. In terms of the heat sources, some researchers choose to recycle the waste heat at the hot aisle \cite{deymi2019thermoeconomic}\cite{davies2016using} while others reuse it at the condenser of a chiller \cite{he2018analysis}. In \cite{oro2019DomesticHeat}, Or\'{o} \textit{et al}. analyzed the economical viability of recycling waste heat at the hot aisle and the condenser, and they found that implementing an air-to-water heat exchanger in the hot aisle was more cost-efficient with shorter payback period. In terms of the cost saving potential and ecological benefits, Davies \textit{et al}. found that installing a heat pump in the hot aisle of a 3.5 MW data center in London could save over 4000 metric tons of carbon emissions and a million pounds in costs.  In addition, He \textit{et al}. implemented a waste heat recycle system for a distributed cooling data center in Hohhot to reuse the waste heat in the return chilled water, claiming that over 18,000 metric tons of standard coal could be saved and the electricity savings for the district heating network were around 10\% per year. 

Due to the integration of the heat pump, additional electricity is required to power it, leading to extra energy consumption and carbon emissions. Meanwhile, the cooling load of the chiller can be reduced since the hot air is pre-cooled by the heat pump. Hence, various trade-offs regarding how to utilize the heat pump to recycle data center waste heat for district heating should be considered. In addition, whether the carbon emission reduction brought by waste heat recovery outweighs additional carbon emissions incurred by the heat pump still requires further exploration. Furthermore, the demand for district heating is high only in cold weather regions such as Nordic countries, it is not necessary for a plethora of warm and tropical countries, e.g., Singapore. As a result, the ubiquity of such a solution is another important issue because a large number of legacy data centers are located in warm regions. 

\subsubsection{Power Plant Co-location}
Data centers involving in the thermal cycle of a coal fired power plant are investigated in \cite{marcinichen2012chip} and the proposed system model is shown in Fig. \ref{PowerPlantColation}. 

The waste heat from the condenser is utilized to preheat the water before the feed-water heater. The cooling technique adopted is the on-chip two-phase cooling system with a standard waste heat temperature of 60 $^\circ$C. However, to implement such a waste heat recycling system in a traditional air-cooled data center, a heat pump is required. It was reported in \cite{marcinichen2012chip} that 2.2\% improvement in the power plant energy efficiency and up to \$45,000,000 annual cost savings are available for a 32.5 MW data center. In addition, \$1,000,000 cost savings are also achieved for the co-located power plant, leading to maximum annual cost savings of \$46,000,000 for both the data center and the power plant. 

Despite significant cost savings due to power plant co-location, the challenges are two fold. First, heat pumps must be integrated so that this technique can be applied in standard air-cooled data centers. Second, as the quality of waste heat degrades with distance, the data center should be located near the power plant, which is sometimes infeasible. 
\begin{figure}[t]
	\centering
	\includegraphics[width=.45\textwidth]{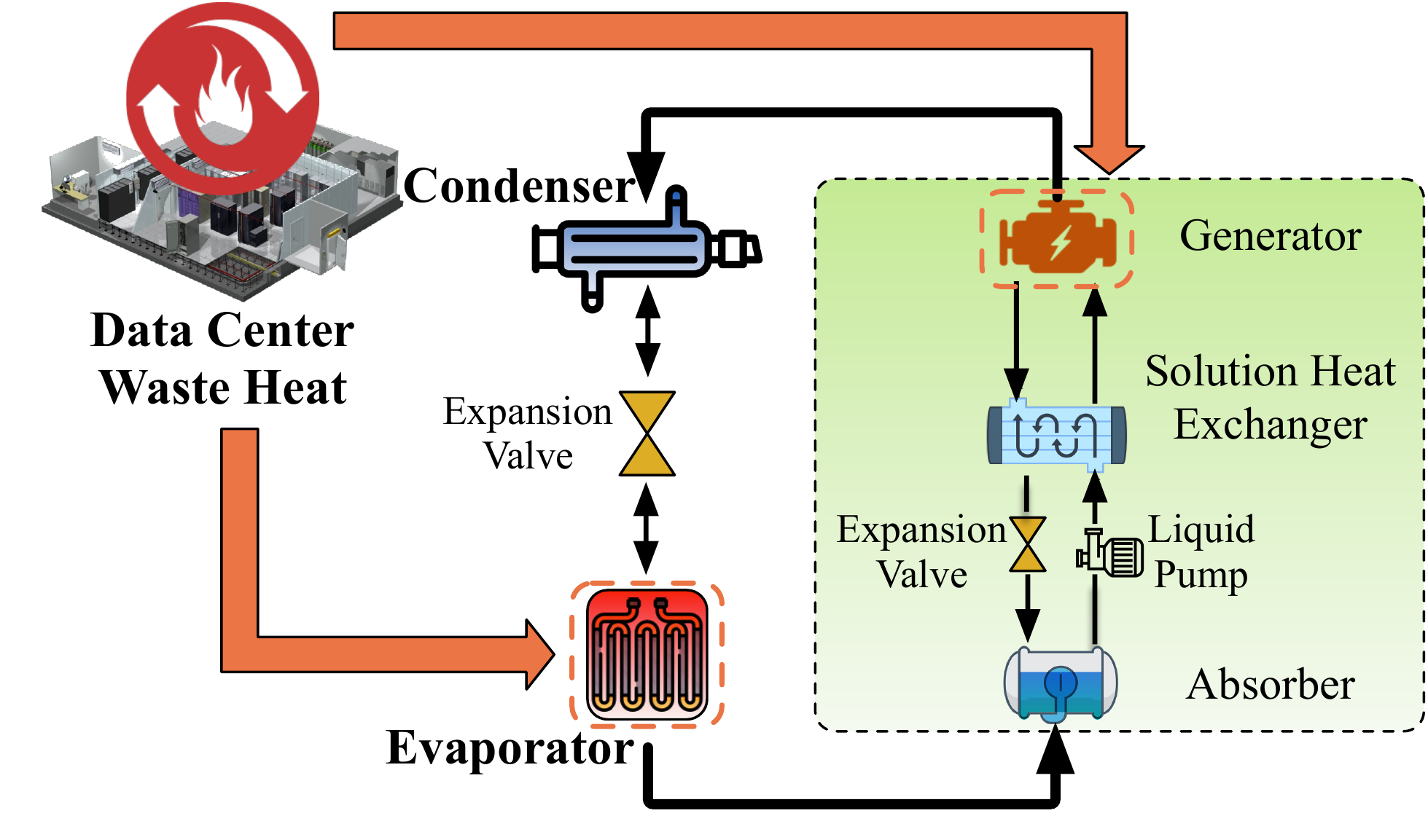}
	\caption{Illustration of an absorption cooling system driven by data center waste heat. The mechanical compressor in a chiller is replaced with a chemical compressor driven by data center waste heat. Additional cooling is also provided to absorb waste heat, which reduces loads for the cooling system at the same time.}
	\label{AbsorptionCooling} 
\end{figure}
\subsubsection{Bio-mass Co-location}
Although the marriage of data center waste heat to biomass fuel processing appears to be counterintuitive, it is exploited in a livestock farm to generate on-site biomass renewable energy \cite{sharma2010BiomassWH}. There are two ways of reusing waste heat in biomass processing. The first is to use the waste heat as a heat source for drying biomass materials. Another option is to keep the reactor of anaerobic digestion of biomass materials warm while reducing moisture content. Methane produced from anaerobic digestion can be used for power generation. However, to make the above two methods work efficiently, data center waste heat above 60 $^\circ$C will be required. As a result, unless a heap pump is installed, this technique is infeasible for widely adopted air-cooled data centers. Moreover, the data center should not be located too far away from the biomass processing plant because the quality of waster heat degrades rapidly as the heat transfer distance increases, which severely limits its application on a large scale.
\begin{figure}[t]
	\centering
	\includegraphics[width=.48\textwidth]{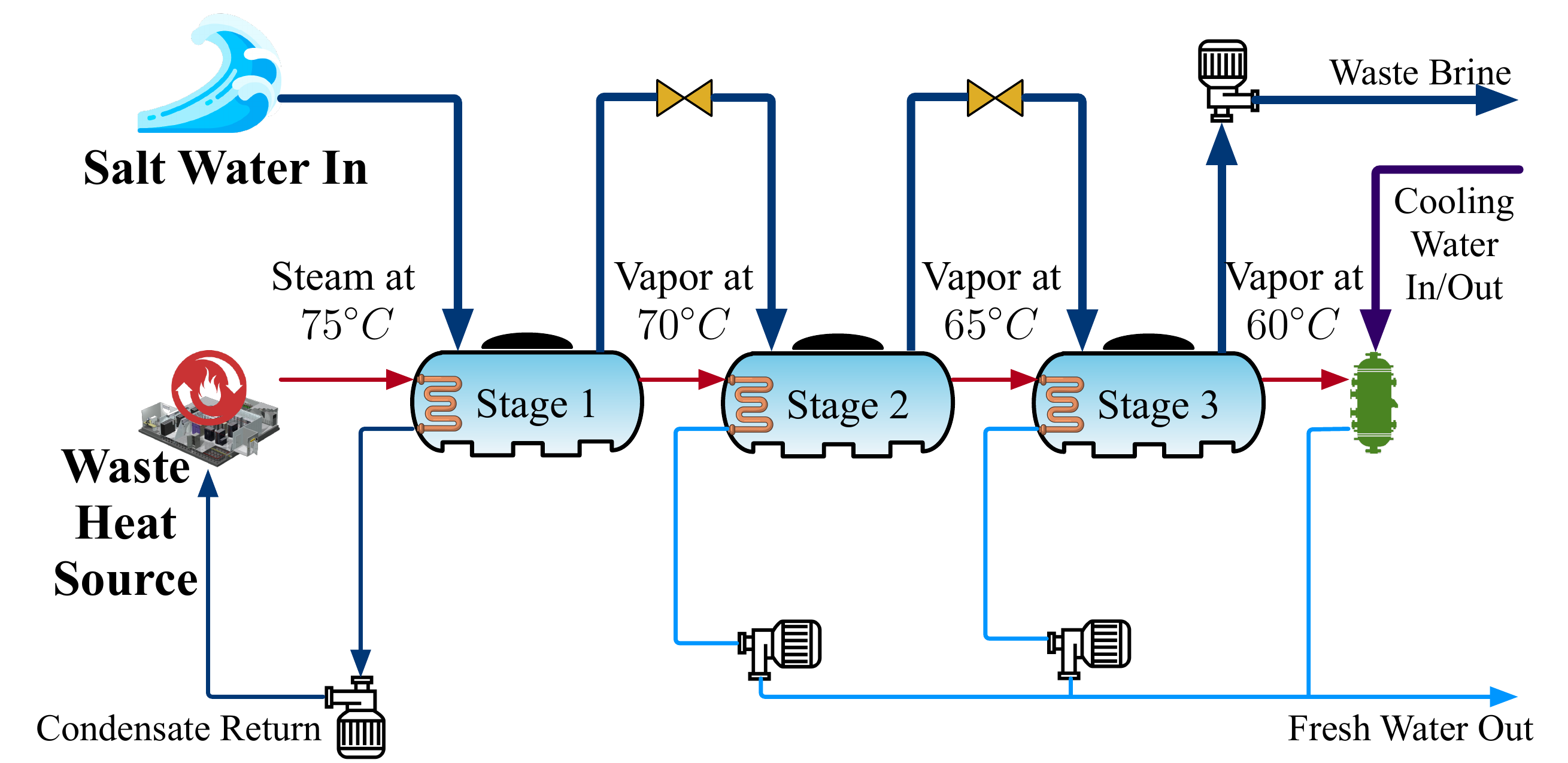}
	\caption{Illustration of Multiple Effect Distillation (MED) system driven by data center waste heat. Data center waste heat is leveraged to vaporized salty water at each stage. It is also possible to reduce the need for the chiller plant due to the heat extraction process during desalination.}
	\label{MED}
\end{figure}
\subsubsection{Absorption Cooling}
As data centers require substantial cooling, several researchers have studied the feasibility of applying waste heat-driven absorption chillers to provide additional cooling while reducing energy consumption of chiller plants \cite{haywood2010sustainable}\cite{haywood2012thermodynamic}. A typical absorption cooling cycle is depicted in Fig. \ref{AbsorptionCooling}.
After absorbing waste heat, low pressure refrigerant vapour enters the absorber and is absorbed by the absorbent, e.g., Lithium Bromide (Li-Br), through an exothermic process. The strong solution of the absorbent and the refrigerant is pumped to a solution heat exchanger via a liquid pump. The solution is preheated as it passes through the solution heat exchanger before entering the generator. A high pressure solution is heated in the generator by the data center waste heat, and the refrigerant and the absorbent are separated. Subsequently, the absorbent returns to the absorber, while the refrigerant continues to the condenser. The remaining components of the system are identical to those in a traditional vapour compression cooling system. Hence, in an absorption cooling loop, the compressor in a conventional vapour compression chiller is replaced by a chemical processor comprised of the loop of absorber, liquid pump, solution heat exchanger, generator, and expansion valve. 

The major merit of absorption cooling is saving cooling energy because it reduces the cooling load for the chiller plant and recovers part of the waste heat to drive the absorption cooling cycle. In \cite{haywood2010sustainable,haywood2012thermodynamic}, Haywood \textit{et al}. investigated the thermal feasibility of harvesting waste heat from a data center with a hybrid of direct liquid cooling and air cooling to drive a 10-ton single-effect lithium bromide-water absorption refrigeration system. The cold plate attached to the CPUs harvests the high quality waste heat, and the remaining waste heat is cooled by the absorption cooling system. In addition, a solar thermal storage system is introduced to stabilize the temperature of the input heat to the absorption chiller to keep it operating at the optimal setpoint. Their experiments reveal that the absorption cooling system achieves an optimal CoP of 0.86 (86\% cooling load can be met by the absorption cooling system) with an input heat temperature of 80 $^\circ$C and a near optimal PUE of 1.16 is possible. Furthermore, the authors also found that the system was able to work with slightly lower efficiency (CoP $\approx$ 0.6) at a waste heat temperature of 70 $^\circ$C. However, such waste heat temperature is still infeasible for traditional air-cooled data center without a heat pump, which precludes its practical implementation.  
\begin{table*}
    \centering
    \begin{tabular}{c!{\vrule width 1pt}c!{\vrule width 1pt}c!{\vrule width 1pt}c!{\vrule width 1pt}c}
        \noalign{\hrule height 1pt}
        Technique  & \makecell[c]{Heat Temperature Required} & Merits & Disadvantages & Suitable Data Center Type\\ 
        \noalign{\hrule height 1pt}
        
        District  Heating & 70 $^\circ$C & \makecell[c]{Reduce carbon emissions\\ of district heating network} & \makecell[c]{Additional heat pump;\\Infeasible for warm regions} & \makecell[c]{Liquid-cooled data center}\\\hline
        
        Power Plant Co-location & 60-100 $^\circ$C & \makecell[c]{Reduce carbon emissions\\ of power plants} & \makecell[c]{Need to be near power plants;\\Additional heat pump} & \makecell[c]{Liquid-cooled data center}\\\hline
        
        Biomass Co-location & $>$60 $^\circ$C & \makecell[c]{Alternative renewable\\ energy source} & \makecell[c]{Need to be near biomass plants;\\Additional heat pump} & \makecell[c]{Liquid-cooled data center}\\\hline
        
        Absorption Cooling & $>$70 $^\circ$C & \makecell[c]{Alternative renewable\\ cooling source;\\Less thermal energy loss} & \makecell[c]{Heat pump is required;\\Huge investments incurred} & \makecell[c]{Liquid-cooled data center}\\\hline
        
        Desalination & 40-75 $^\circ$C & \makecell[c]{Reduce carbon emissions\\ of desalination plants;\\Provide clean water} & \makecell[c]{Additional heat pump;\\Huge investments incurred;\\Infeasible for inland regions} & \makecell[c]{Liquid-cooled data center;\\Air-cooled data center}\\\hline
        
        Organic Rankine Cycle & 40-85 $^\circ$C & \makecell[c]{Alternative renewable\\ energy source;\\Less thermal energy loss} & \makecell[c]{Additional heat pump;\\Low efficiency} & \makecell[c]{Liquid-cooled data center;\\Air-cooled data center}\\
        \noalign{\hrule height 1pt}
    \end{tabular}
    \caption{Summary of existing data center waste heat recovery techniques.}
    \label{WasteHeat}
\end{table*}

\begin{figure}[t]
	\centering
	\includegraphics[width=.5\textwidth]{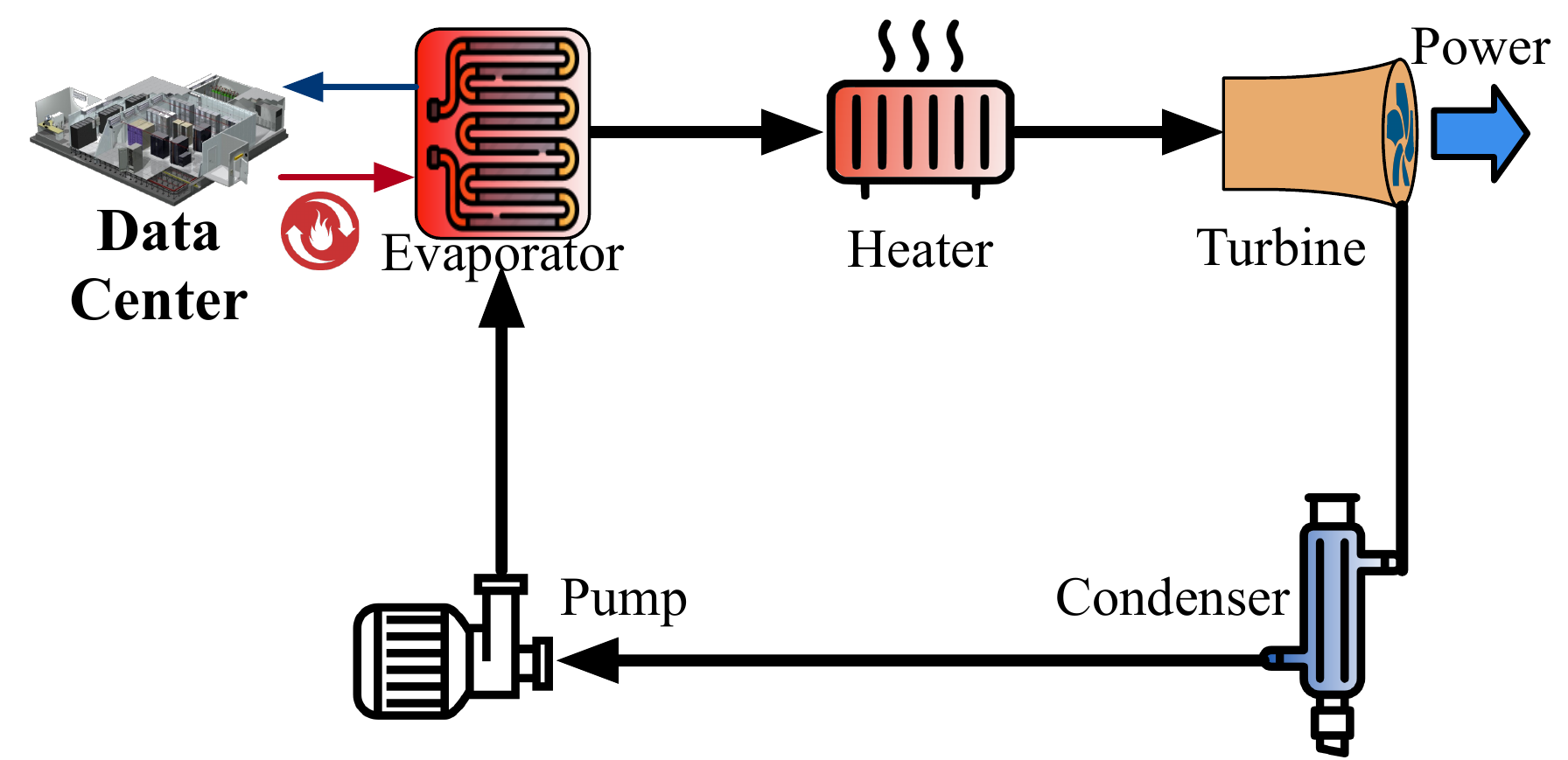}
	\caption{Illustration of Organic Rankine Cycle driven by data center waste heat. Organic working fluid is vaporized by data center waste heat at the evaporator and then proceeds to drive the turbine to generate electricity.}
	\label{ORC} 
\end{figure}
\subsubsection{Desalination}
Seawater desalination is another potential application of data center waste heat for producing clean water \cite{WasteHeatReview2014, sondur2018data} while reducing carbon emissions of a desalination plant. A typical way of desalination is using Multiple Effect Distillation (MED) as illustrated in Fig. \ref{MED}. 

Waste heat from a data center at 75 $^\circ$C is utilized to boil the water and generates steam. The salty seawater is then boiled and vaporized by the water steam. The vapour leaving the first stage is subsequently served as the heating medium for the salty water entering the second stage. The salty water at the second stage is boiled again and vapor at a lower temperature (65 $^\circ$C) is generated. This process is repeated until the quality of the vapor is too low to be recovered. It is important to note that clean water will be produced from the second stage, and the vapor temperature will decrease accordingly. 

However, the MED system requires waste heat at 75 $^\circ$C, which is impractical for liquid-cooled data centers, let alone common air-cooled data centers. In \cite{sondur2018data}, a system that reuses waste heat from data centers in the coastal region with the liquid-cooling system to produce drinkable water via low-pressure desalination process was deployed. The low-pressure desalination is based on the fact that water evaporates at lower temperature at low pressure \cite{gude2009desalination}. A vacuum is created at the top of an evaporation chamber in the low-pressure desalination, and water is able to vaporize to produce clean water at about 40-50 $^\circ$C with high efficiency \cite{sondur2018data}. Such requirements are appropriate for the liquid-cooled data centers as shown in \cite{sondur2018data}, whereas air-cooled data centers still do not fall within the feasible region to apply such technique. 

In conclusion, reusing data center waste heat for desalination has several benefits including the production of clean water, and the potential elimination of the need for chiller plants due to the heat extraction process during desalination. Whereas, several challenges remain unsolved. Firstly, similar to other techniques, it cannot be directly applied in prevailing air-cooled data centers without a heat pump. Secondly, a large number of inland regions are infeasible for clean water production from seawater. Furthermore, as the quality of data center waste heat degrades rapidly as heat transfer distance increases, data centers should be located in coastal regions, which limits their practical deployment.  

\subsubsection{Organic Rankine Cycle}
As an alternative approach to producing renewable electricity, data center waste heat has been adopted to drive an Organic Rankine Cycle (ORC). ORC, as depicted in Fig. \ref{ORC}, is similar to traditional steam Rankine cycle, except that the steam being replaced with organic fluid with lower boiling point. 

According to Fig. \ref{ORC}, organic working fluid is heated and vaporized with data center waste heat in the evaporator. It is then pumped to the turbine for electricity production. Although ORC requires a waste heat temperature of 65 $^\circ$C or higher, it can work at a waste heat temperature as low as 32 $^\circ$C with reduced efficiency, which is suitable for existing air-cooled data centers. Furthermore, Araya \textit{et al}. investigated the feasibility of reusing ultralow grade waste heat (40-85 $^\circ$C) from a rack server to drive an ORC \cite{araya2021study}. To accomplish this, low boiling point organic working fluids such as R-134a, Benzene, Toluene, and Propane are exploited. A lab-scale testbed is established and an estimation of 4\%-8\% power saving is available when ORC electricity is used to power the server rack at a waste heat temperature of 90$^\circ$C.

The benefits of ORC as a waste heat recycling approach are two folds. On the one hand, data center waste heat can be reused on-site, avoiding heat transportation which incurs considerable heat loss. On the other hand, renewable electricity can be generated as a supplementary energy source. Furthermore, because data centers continuously generate waste heat, this type of renewable energy is more stable than solar or wind energy. However, current ORC system has low thermal efficiency (1.9\%-4.6\%) \cite{araya2021study} and the economic viability for applying such systems should be investigated in the future. 

\subsection{Summary}
Table \ref{WasteHeat} summarizes existing data center waste heat recovery techniques in terms of their operation conditions, merits, and drawbacks. Multiple waste heat reusing techniques for data centers have been developed over the last decade. Some applications reduce carbon emissions of a data center by generating renewable energy or cooling, while some others reduces carbon emissions of customers from other economic sectors. Both of them are advantageous in terms of achieving carbon neutrality.

However, most of them aim to recycle high temperature waste heat in novel liquid-cooled or two-phase cooled data centers rather than low-grade waste heat in prevailing air-cooled data centers. When considering air-cooled data centers, heat pumps are required for the majority of waste heat recycling applications. In addition, waste heat recovery is geographically constrained. For example, district heating is only suitable for cold regions and only coastal regions have demands on seawater desalination. Moreover, the low heat recycling efficiency issue also precludes large-scale application of data center waste heat recovery systems, as stated in \cite{araya2021study}. Hence, data center waste heat recycling technology is still in its infancy, and a number of practical challenges, such as low efficiency and high cost, must be addressed before it can be implemented in practice.

\section{Digital Twin-Assist Industrial AI Framework for Carbon-Neutral Data Centers}\label{DT}
In this section, we propose a digital twin-based approach for data center carbon emission modeling and optimization, which combines all of the elements in Section \ref{Energy Supply}, \ref{Energy Utilization}, and \ref{Energy Circulation} with emerging digital twin and AI technologies. With the proposed framework, we then discuss future trends towards carbon-neutral data centers. 

\subsection{Multi-pronged Solution for Carbon Neutrality}
\begin{figure}[t]
	\centering
	\includegraphics[width=.45\textwidth]{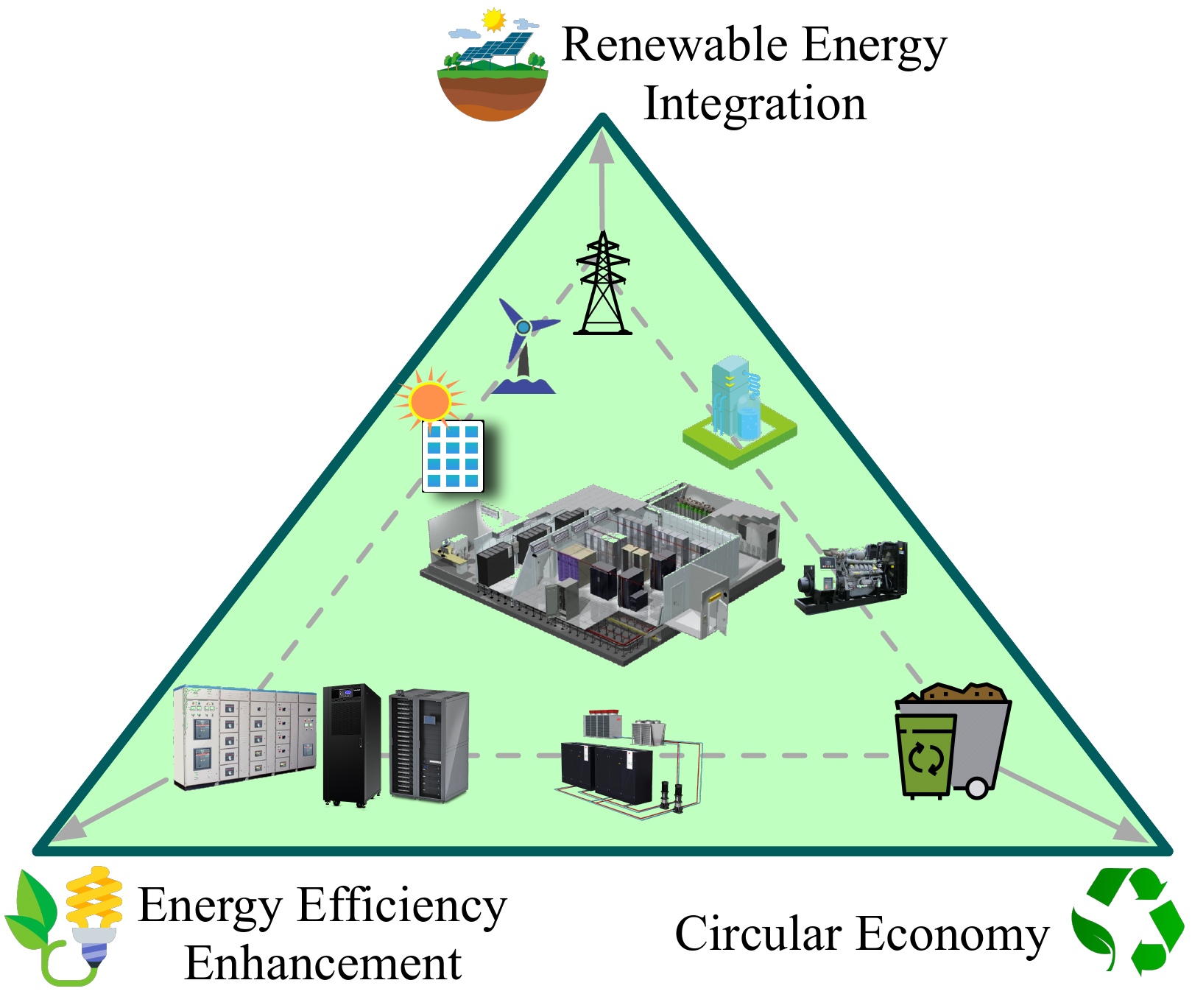}
	\caption{Proposed multi-pronged solution for carbon-neutral data centers. By simultaneously improving renewable energy penetration and energy efficiency, carbon emissions from the energy supply side and the utilization phase will decline significantly. Aided with enhanced energy and material circulation, it is envisioned to realize carbon-neutral data centers.}
	\label{Multi-pronged} 
\end{figure}

To achieve carbon neutrality in the data center industry, we envision a multi-pronged solution that consists of three major research objectives, as is illustrated in Fig \ref{Multi-pronged}. 
\begin{itemize}
    \item \textbf{Energy Efficiency Enhancement}. Data centers must meet a high standard for energy efficiency by optimizing their IT and physical infrastructure and the sophisticated coupling of the IT and physical system must be taken into account in order to achieve more aggressive PUE targets.  
    \item \textbf{Renewable Energy Integration}. Data centers must judiciously leverage the emergence of clean and renewable energy to meet their electricity demand gradually, optimizing their carbon usage effectiveness (CUE).
    \item \textbf{Circular Economy}. Data centers will set a high standard for circular economy practices, in a multi-fold manner, including the reuse of cold energy from the liquefied natural gas (LNG) regasification, waste heat from data centers, and electrical equipment.
\end{itemize}

We hope to reduce a data center's total energy consumption and carbon emissions with this multi-pronged solution. Furthermore, data centers may play as a critical carbon prosumer in the carbon market by touting their surplus renewable energy or waste heat to facilitate the decarbonization of other economic sectors.  
\begin{figure*}[t]
	\centering
	\includegraphics[width=.9\textwidth]{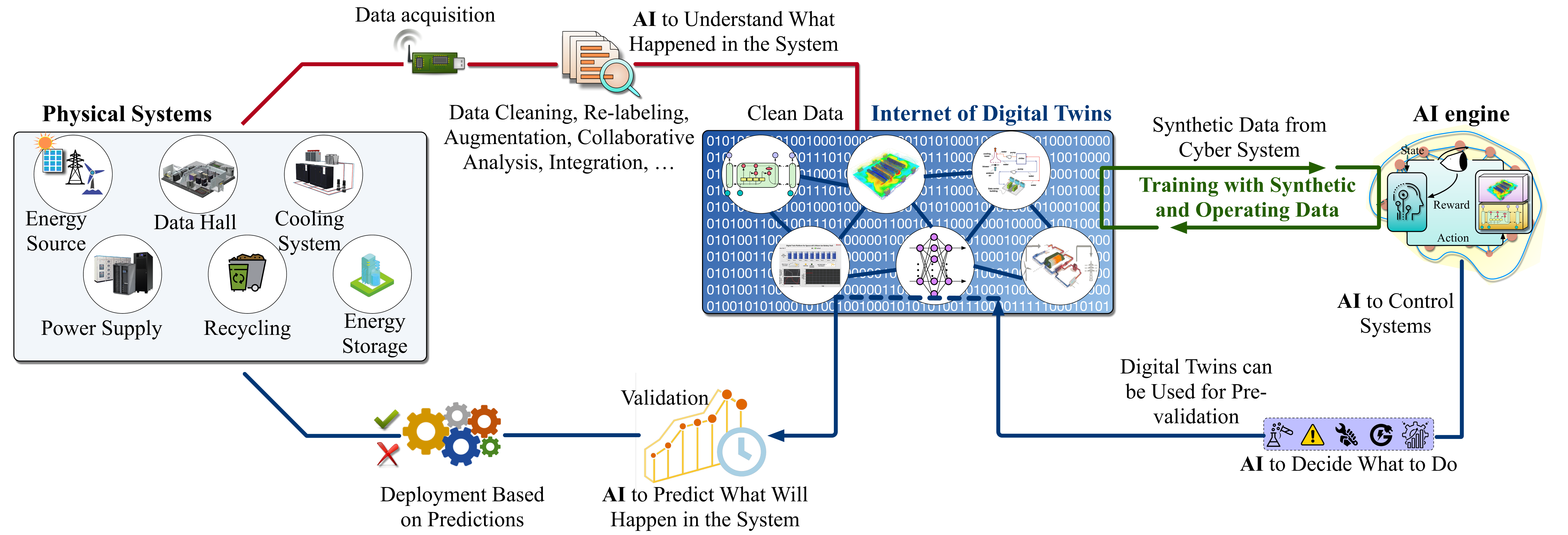}
	\caption{Digital twin-assisted industrial AI framework for carbon-neutral data centers. The framework includes three modules: a) the physical system; b) the digital twin; and c) the AI engine. The physical system is the real environment that is the target of our learning-based approach. The digital twin represents a digital counterpart of a physical system such as cooling system. Digital twins are intertwined due to the coupled underlying physical processes and  they are calibrated with real world data. The AI engine learns from the real data from physical systems and synthetic data from digital twins. It implements intelligent control or delivers recommendation to the operator after training. Digital twins can also predict future system states to prevent potential risks.}
	\label{DualControlLoop} 
\end{figure*}
\subsection{Framework Architecture}
To put the research framework into practice, a thorough understanding of the dynamic ICT workload, underlying thermal dynamics in a data hall, renewable energy generation process, and thermal energy circulation loop is indispensable. However, previous works often rely on simplified mathematical models and integrate them into a traditional optimization-based approach to derive control laws. Because the simplified models are inaccurate, and the practical optimality of the derived control laws cannot be guaranteed even if they are proved to be optimal mathematically. Furthermore, the nonlinearity in high-fidelity data center modelling makes mathematical programming-based solutions impractical to be solved efficiently. Meanwhile, AI-based data center controllers have emerged as a strong competitor since they can not only handle sophisticated nonlinearity in data center enclosures but also perform real-time inference once they are well trained offline.

Although the AI-based solution is intriguing, two major challenges must be addressed before it can be implemented in the real world.
\begin{itemize}
    \item \textbf{Data Scarcity}. Data collection in data centers is often a daunting task, consuming a huge amount of time and resources. In the events of system emergencies, anomalies, and failures, data capture and access may be prohibitively expensive due to safety compliance. AI algorithms' accuracy may suffer as a result of insufficient training data.
    \item \textbf{Risk-averse Mindset}. Since data centers are mission-critical infrastructure for the business continuity, DC operators tend to be extremely risk-averse in adopting new technologies. Moreover, even with better operational efficiency, the stochastic nature of AI algorithms makes widespread adoption difficult. 
\end{itemize}

Based on these considerations, we propose an industrial AI platform, assisted by an industry-scale digital twin and encapsulating cutting-edge AI algorithms that include three modules, as illustrated in Fig. \ref{DualControlLoop}. 

\begin{itemize}
    \item \textbf{Physical System} represents the real environment/system that is the target of the learning-based approach. The physical system is typically a collection of states that changes according to certain key factors (e.g., control action, weather, temperature, illumination, and time) and the physical laws. In this paper, the term ``physical system" refers to a variety of systems including the IT, cooling and power distribution system.
    \item \textbf{Digital Twin} represents the digitalized cyber models of the physical system. The digital twin has access to both physical and data-driven models, where the physical model is often built based on basic design data and coupled with physical laws, and the data-driven model is trained using massive historical data sampled from the physical system. Both models aim to accurately simulate the dynamics of the physical system. The digital twin will interact with the learning-based algorithm and synthesize additional datasets for system optimization. 
    \item \textbf{AI Engine} represents the implementation of learning-based approaches to optimize/diagnose/control the physical system operations. The AI engine can be trained through historical data or by directly interacting with the physical system or digital twin. To avoid the risks of deploying AI algorithms on the physical system, the proposed framework recommends that the AI engine interacts with the digital twin.
\end{itemize}

Multiple paths connect these three modules. The arrow from the physical system to the digital twin represents the digitalization of the physical system. Diverse raw data (e.g., design and operating data) is collected from the physical system. The AI model plays an important role in understanding inherent patterns behind the data and transforms them into a digital twin-acceptable form for modeling the digital twin. The digital twin is highly adaptable and can consist of a collection of physical formulations, simulators, or learning-based models. The cyclic path between the AI engine and digital twin represents the training of AI models on the digital twin in order to optimize the physical system. Instead of interacting with the physical system, the learning algorithm interacts directly with the digital twin to capture system behaviours and learn the complex patterns of the system. As the digital twin can synthesize the value of any attribute in a relatively short period of time (e.g., much less than a few years), the synthesized dataset will be more balanced and have a lower training cost, which benefits the training of the AI engine. In addition, the arrow from the AI engine to the digital twin represents the validation of AI engine's recommended control actions prior to the deployment on the physical system. This can significantly reduce data center operators' risk-averse attitude toward deploying learning-based approaches. The arrow from the digital twin to the physical system represents the prediction of future system states given specified inputs. The digital twin learns the system behaviours from the historical data by deploying AI algorithms. Without implementing the trail on the physical system, the trained AI model predicts future system states based on specified inputs. Based on the prediction results, appropriate actions can be implemented in the physical system to ensure system stability or improve system performance.

\subsection{Inherent Scientific Problems}
The proposed framework renders a dual-cycle control loop paradigm in operating physical infrastructure. The two cycles, i.e., the cyber and the physical cycle, are bridged by Internet of Digital Twins, as depicted in Fig. \ref{DualControlLoop}. Such a framework calls for solutions for several underlying scientific problems which will be discussed in this section.

\subsubsection{Intelligent Data Retrieval for AI Engine Training}
As digital twins are in charge of generating synthetic data to aid the training of the AI engine, the quality of synthetic data has a significant impact on the AI engine's generalization performance. However, synthetic data generated by digital twins may be noisy due to inaccurate calibration and real-world modelling. On the other hand, data from physical systems can serve as a complementary for AI engine training. Thus, a judicious data retrieval policy should be carefully designed to determine the best way of intermingling noisy but massive cyber data and accurate but costly physical data to improve the AI engine's generalization. 

\subsubsection{Policy Mapping Between Cyber and Physical World}
In the proposed framework, the AI engine implements control of the infrastructure or deliver recommendations on the operation of physical systems. However, as the AI engine is trained with a mixture of cyber and physical data, it is critical to find a proper mapping between the control law derived from the cyber world and that appropriate for the physical world. For example, control policy learned in the cyber world may cause a server's temperature to exceed its safe range. Such high temperatures are prohibited in the physical world to avoid potential disaster. Therefore, the control policy should be mapped via a policy transfer engine, which maintains excellent control performance of the cyber policy while guaranteeing safe operations of physical infrastructure. 

\subsubsection{Multi-Agent Equilibrium for Carbon Neutrality}
In the framework, each subsystem in a physical data center is incarnated with its digital twin in the cyber world and the subsystem is managed by an AI engine. Each digital twin has its own system dynamics, and they communicate with each other to form an Internet of digital twins. It is important to note that the dynamics of different systems differ greatly. For example, the response time for the cooling system and IT system control may differ by an order of magnitude. As a result, training these AI engines to work together is critical for achieving operational equilibrium for carbon neutrality. In this regard, a cooperative and interactive multi-agent reinforcement learning (MARL) approach could be adopted to train these AI engines. 

\subsection{Potential Applications}
In this section, we will envision some  potential applications of the proposed digital twin-assisted AI framework. These applications facilitates the actuation of our multi-pronged solution from a variety of perspectives.   

\subsubsection{Carbon Intelligent Computing Platform}
From the IT system stand point of view, carbon-aware intelligent computing is one key approach in realizing carbon-neutral data centers. The basic idea is to shift non-emergent computing tasks such as scientific computing jobs to the time when carbon-free energy such as wind and solar energy is adequate. Additionally, computing jobs can be migrated to locations with carbon-free energy as well. Although there have been some preliminary studies on carbon-aware  job scheduling, e.g., \cite{goiri2011greenslot,goiri2012greenhadoop}, these works do not consider the complicated interaction between the IT system and the physical infrastructure and the impact of such scheduling to the physical infrastructure is under exploration. In this regard, we envision that our data-driven digital-twin based solution will provide a comprehensive understanding of such impacts and will be beneficial for deriving optimal scheduling policy.

\subsubsection{Intelligent Management for Renewable Energy Portfolio}
Renewable energy plays a critical role in the transition towards carbon neutrality. However, renewable energy is difficult to manage and utilize due to its uncertainty. If we can accurately predict the yield of renewable energy, proactive control of workloads and the cooling system is able to reduce considerable carbon emissions, as shown in \cite{li2019thermal}. Previous works on renewable energy forecasting leveraged traditional time series analysis tools such as WCMA \cite{piorno2009prediction}. Nevertheless, these models typically utilize historical yields to predict future yields without further consideration of exogenous variables such as weather, locations, and surrounding environment, all of which have great impacts on renewable energy yields. Hence, a high-fidelity digital twin for renewable energy production should be established first, and AI-empowered data analysis tools can then be exploited to predict future renewable energy yields. Facilitated by these, data center operators are able to adjust their operating policies to maximize the utilization of renewable energy. Furthermore, if future renewable energy yields can be accurately predicted, cost-efficient carbon credit procurement is also in prospect. 

\subsubsection{Liquefied Natural Gas (LNG) Regasification}
As data centers generates a large amount of waste heat, it is urgent to find a proper way to utilize the waste heat in an efficient way because it can provide additional renewable energy \cite{araya2021study} or renewable cooling \cite{haywood2010sustainable}\cite{haywood2012AbsorptionChiller} to a data center while also alleviating the cooling load. However, as stated in Section \ref{Energy Circulation}, current technologies for recovering low-grade data center waste heat are inefficient. Nowadays, the combination of LNG regasification and data center waste heat is discussed in \cite{ayachi2019assessment}, where the authors discovered a lot of potential for the combination. LNG is produced when natural gas is cooled below its freezing point -160 $^\circ$C and a large amount of cold energy is encapsulated in it. The author proposed to regasify LNG with data center waste heat and cool the  data center with the cooled air after regasification. It was shown that over 250,000 metric tons of carbon emissions could be avoided when this technology was applied in Singapore. However, it is still in the infancy and many practical issues must be addressed before it can be used on a large scale. For example, since many LNG plants are located offshore, transporting the cold energy from offshore LNG plants to inland data centers with minimum thermal loss may be a great challenge. Hence, more efforts can be devoted to discussing the feasibility of its integration into data centers, as well as how data centers should modify its infrastructure to fully utilize the abundant cold energy embodied in LNG. 

\section{Summary}\label{Summary}
This survey paper systematically investigates the roadmap towards carbon-neutral data centers. We present data center carbon footprint at various time granularities, as well as the major carbon emission sources over the life of a data center. Definition and evaluation metrics of carbon-neutral data centers as well as industrial efforts towards carbon neutrality are discussed. As a policy instrument, a carbon market is introduced and various mechanisms that can be exploited by a data center operator to reduce carbon emissions are introduced. From a technological point of view, existing works on judicious management of a renewable energy-integrated data center, data center energy efficiency enhancement through coordinated control of the IT and the infrastructure, and data center waste heat recovery are summarized to provide a systematic view on carbon-neutral data centers. Based on these instruments, we envision a multi-pronged solution towards carbon-neutral data centers via energy efficiency improvement, renewable energy penetration and waste energy recycling simultaneously. To make it practical in the real world, a digital twin-based industrial AI framework is proposed, which incorporates digital twins and cutting-edge AI technologies into a dual cycle control loop. Finally, multiple potential applications for the proposed framework are presented to reveal its enormous potential as a supporting pillar for data center decarbonization.

% Reference
\bibliographystyle{IEEEtran} 
\balance
\bibliography{CNDC.bib}
\end{document}